\documentclass[onecolumn]{emulateapj}
\usepackage{lscape}
\newcommand{\be}{\begin{equation}}
\newcommand{\ee}{\end{equation}}
\newcommand{\bea}{\begin{eqnarray}}
\newcommand{\eea}{\end{eqnarray}}
\shortauthors{Yan \& Lazarian}
\begin{document}
\title{Polarization from aligned atoms as a diagnostics 
of 
circumstellar,
AGN and interstellar magnetic fields: II. Atoms with Hyperfine 
Structure}
\author{Huirong Yan\altaffilmark{1} \& A. Lazarian\altaffilmark{2}}
\altaffiltext{1}{Canadian Institute for Theoretical Astrophysics, 60 St. George, Toronto, ON M5S3H8, yanhr@cita.utoronto.ca}
\altaffiltext{2}{Department of Astronomy, University of Wisconsin, 475 N. Charter St., Madison, WI 53706, lazarian@astro.wisc.edu}

\begin{abstract}

We show that atomic alignment presents a reliable way to study topology of 
astrophysical magnetic fields. The effect of atomic alignment arises from
modulation of the relative population of the sublevels of atomic
ground state pumped by anisotropic radiation flux. 
As such aligned atoms precess in the external 
magnetic field and this affects the properties of the polarized radiation arising 
from both scattering and absorption by the atoms. As the result the polarizations of emission and absorption lines depend on the 3D geometry of the magnetic field as well as the direction and anisotropy of incident radiation. We consider 
a subset of astrophysically important atoms
 with hyperfine structure. For emission lines we obtain the dependencies of the
direction of linear polarization on the directions of magnetic field 
and the incident pumping radiation. For absorption lines
we establish when the polarization is perpendicular and parallel to magnetic field. For both emission and absorption lines we find the dependence on the
degree of polarization on the 3D geometry of magnetic field. We claim
that atomic alignment
 provides a unique tool to study magnetic fields in circumstellar
 regions, AGN, interplanetary and interstellar medium. This tool allows
studying of 3D topology of magnetic fields and establish other important
astrophysical parameters. We consider polarization arising from both
atoms in the steady state and also as they undergo individual
scattering of photons. We exemplify the utility of atomic alignment for 
studies of astrophysical magnetic fields by considering a case of Na alignment
in a comet wake.

\end{abstract}

\section{Introduction}

Magnetic fields play essential roles in many astrophysical circumstances. 
Incompatible with its importance, our knowledge about astrophysical magnetic 
field is very much limited. Exploring new tools to measure magnetic field 
is thus extremely important. 

In our previous paper (Yan \& Lazarian 2006, henceforth Paper I), we discussed
how the alignment of {\it fine} structure atoms and ions can be used to detect 
3D orientation of magnetic field in diffuse medium. As in Paper I, 
for the sake of simplicity, 
we shall term the alignment of both atoms and ions {\it atomic alignment}. 
In this paper, we shall discuss atomic alignment within hyperfine structures. 
In fact, historically optical pumping of atoms with hyperfine structure atoms, 
e.g., alkali atoms, He, Hg at al. (Happer 1972) were first studied in laboratory in relation with early day maser research. This effect was noticed and made use 
of for the interstellar case by Varshalovich (1968). And in a subsequent paper  
by Varshalovich (1971), it was first pointed out that the dependence of atomic 
alignment on the direction of magnetic field can be used to detect magnetic 
field in space. However, the study did not provide  either detailed 
treatment of the effect or quantitative predictions.

Atomic alignment has been addressed also  by solar researchers. 
The research into emission line polarimetry resulted in important
change of the views on solar chromosphere (see Landi Degl'Innocenti 1983, 1984, 1998, Stenflo \& Keller 1997, Trujillo
Bueno \& Landi Degl'Innocenti 1997, Trujillo
Bueno 1999, Trujillo Bueno et al. 2002, Manso Sainz \& Trujillo Bueno 2003). 
However,
they dealt with {\it emission} of 
atoms in a very {\it different} setting.
Similar to Paper I below we concentrate on the {\it weak} field regime, 
in which it is 
the atoms at ground level that are repopulated due to magnetic precession, 
while the Hanle effect that the aforementioned works deal with
is negligible. The closest to our study is regime discussed in 
the work by Landolfi \& Landi Degl'Innocenti
(1986), who considered an idealized two-level {\it fine} structure atom in a 
very restricted
geometry of observations, namely, when the magnetic field is along of line of
sight and perpendicular to the incident light. As we discussed
in Paper I, that dealt with {\it fine} structure atoms, these restrictions did not allow
to use this study to predict the directions of astrophysical magnetic fields
from polarimetric observations.

In this paper we consider atomic alignment for atoms
with {\it hyperfine} structure and provide quantitative
predictions for both absorption and emission lines. 
To exemplify the processes of alignment we perform calculations
for astrophysically important species.

Studies of magnetic field topology using atomic alignment are complementary
to the studies of magnetic fields using aligned dust.
Similar to the case of interstellar dust, the rapid precession of atoms
in magnetic field makes the direction of polarization sensitive to the
direction of underlying magnetic field (see Lazarian 2003 for a review).
However, as the precession of magnetic moments
of atoms is much faster than the precession of magnetic moments of grains, 
atoms can reflect much more rapid variations of magnetic field. More 
importantly, alignable atoms and ions can reflect magnetic fields in the
environments where either the properties of dust change or the dust cannot survive.
This opens wide avenues for the magnetic field research in circumstellar regions
interstellar medium, interplanetary medium, intracluster medium, AGN, etc. 
In addition, the polarization caused by atomic alignment is sensitive to the 
3D direction of magnetic
field. This information is not available by any other technique that are 
available for studies of magnetic field in diffuse gas.

In what follows we formulate the conditions for atomic alignment of
hyperfine species in \S2 and present our formalism for treating of atomic alignment  and optical pumping in \S3.  In \S4 we use NaI and KI to discuss the details of
practical calculations of polarization of emission lines
arising from atomic alignment of species with hyperfine
splitting. A nonequilibrium case is considered for Na I in \S5 and it is shown 
how alignment and polarization changes with the number of scattering events.
The results are then applied to turbulent comet wake in \S6. We show how the change
 of magnetic field direction influences the polarization of scattered 
Sodium D lines from the wake which can be used to ground-based studies
of interplanetary turbulence.
In \S7, we consider the alignment of neutral hydrogen, N V and P V and the 
resulting polarizations of Lyman $\alpha$. In addition, we also discuss 
their implications 
for HI 21cm and N V hyperfine radio lines. More complicated atomic species 
are considered in \S8, where we show that absorption from N I atoms is 
polarized even for unresolved hyperfine multiplet. We show how hyperfine structure changes its alignment compared to S II, which has the same electron configuration, but without nuclear spin. In \S9, we discuss how the average along the line of sight affects the results. The discussion and the summary are provided in, respectively, \S10 and \S11. 

\section{Conditions for atomic alignment in the presence of hyperfine structure}

\begin{figure}
\includegraphics[%
  width=0.45\textwidth,
  height=0.3\textheight]{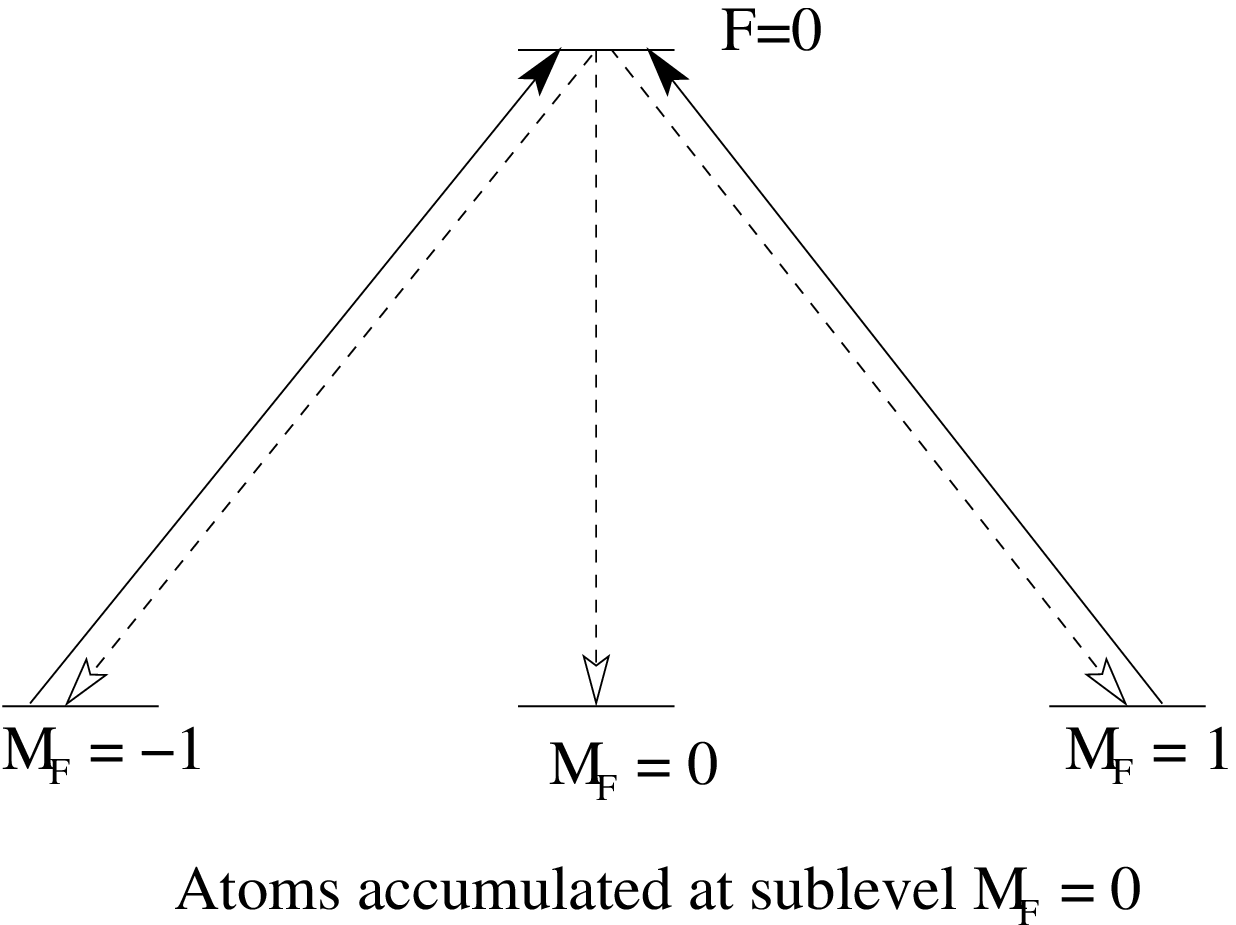}
\includegraphics[%
  width=0.45\textwidth,
  height=0.3\textheight]{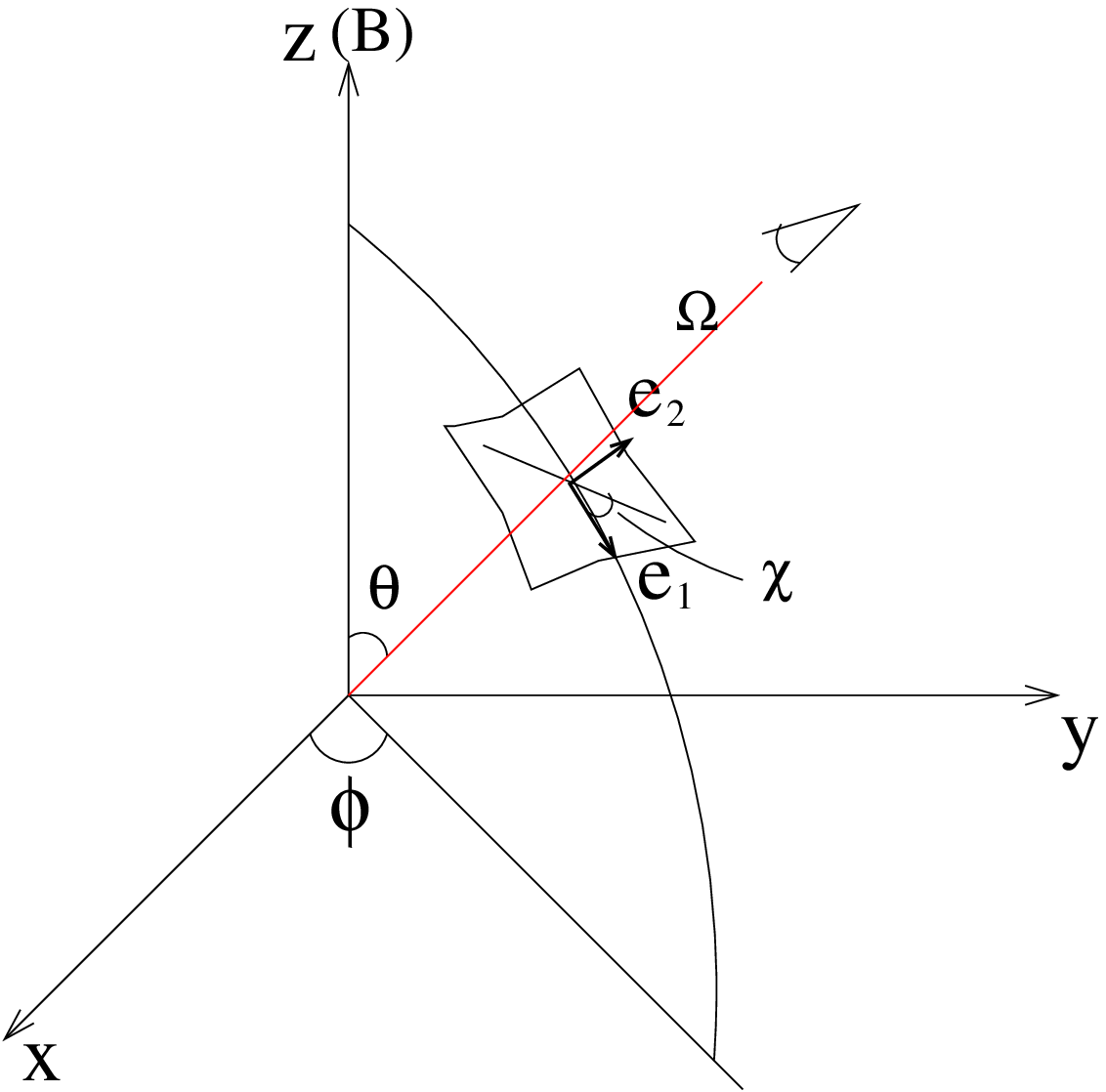}
\caption{{\it Left}: a toy model to illustrate how atoms are aligned by anisotropic light.
Atoms accumulate in the ground sublevel $M=0$ as radiation removes atoms from the ground states $M=1$ and $M=-1$; {\it Right}: Radiation geometry and the polarization vectors in a given coordinate system. ${\bf \Omega}$ is the direction of radiation, $\chi$ is the positional angle of a linear polarization.}
\label{nzplane}
\end{figure}

We discussed atomic alignment in Paper I. Atomic alignment is caused by
the anisotropic deposition of angular momentum from photons. As illustrated by the toy model shown in Fig.\ref{nzplane}{\it left}, absorption from $M_F = 0$ in the ground level is impossible owing to conservation of angular momentum. The differential absorptions for a realistic example of NaI will be provided later in \S4.1. As the result, atoms
scattering the radiation from a light beam are  aligned
in terms of their angular momentum.  To have the alignment of ground state, the atom should have non-zero ($\geq 1$) total angular
momentum on its ground state to enable various projection of
atomic angular momentum.

For atoms with nuclear spin, hyperfine structure need to be taken into account.
It is the total angular momentum ${\bf J+I=F}$, the vector summation of electron angular momentum $\bf J$ and nuclear spin $\bf I$, that should be considered. Alkali atoms thus are alignable with nuclear spins added. 
It is true that direct interaction between the nucleus and the radiation field is negligible. However, for resonant lines, the hyperfine interactions cause substantial 
precession of electron angular momentum ${\bf J}$ about
the total angular momentum ${\bf F}$ before spontaneous emission. Therefore 
total angular momentum should be considered and the $FM_{F}$
basis must be adopted (Walkup, Migdall \& Pritchard 1982). For alkali-like atoms,
 hyperfine structure should
be invoked to allow enough degrees of freedom to harbor the alignment and to 
induce the corresponding polarization.

In order for atoms to be aligned, the collisional rate should not be too high. In fact, as disalignment of the ground
state requires spin exchange (or flips), it is less efficient than one can naively
imagine. The calculations by Hawkins (1955) show that to disalign sodium
one requires more than 10 collisions with electrons and experimental
data by Kastler (1956) support this. This reduced sensitivity of aligned
atoms to disorienting collisions makes the effect important for various
astrophysical environments.

Magnetic field would mix up different $M$ states. However, it
is clear that the randomization in this situation is not complete
and {\it the residual alignment and resulting polarizations reflects the magnetic field direction}. Magnetic mixing happens if the angular
momentum precession rate is higher than the rate of the
excitation of atoms from the ground state, which is true
for many astrophysical conditions.

\begin{table}
\begin{tabular}{||c|c|c|c|c|c||}
\hline
 Atom&
 Nuclear spin&
 Lower state&
 Upper state&
 Wavl(\AA)&$P_{max}$
 \tabularnewline
\hline 
\hline 
H I&
$1/2$&
$1S_{1/2}$&
$2P_{1/2,3/2}$&
912-1216&26\%\\
\hline
{Na I}&
$3/2$&
$1S_{1/2}$&
$2P_{3/2}$&
5891.6&20\%)\\
\cline{4-6}
& &  &
$2P_{1/2}$&
5897.6&0\\
\hline
K I&
$3/2$&
$1S_{1/2}$&
$2P_{3/2}$&
7667,4045.3&21\%\\
\cline{4-6}
& &  &
$2P_{1/2}$&
7701.1,4048.4&0\\
 \hline
{N V}&
$1$&
  $1S_{1/2}$&
$2P_{3/2}$&
1238.8&22\%\\
\cline{4-5}
& & &
$2P_{1/2}$&1242.8&\\
 \hline
{P V}&
$1/2$&
  $1S_{1/2}$&
$2P_{3/2}$&
1117.977&27\%\\
\cline{4-6}
& & &
$2P_{1/2}$&1128&0 \\
\hline
 Al III&
$5/2$&
$1S_{1/2}$&
$2P_{3/2}$&
 1854.7&cna\\
\cline{4-5}
&& &
$2P_{1/2}$ & 1862.7&\\
\hline
N I&
1&
$4S^o_{3/2}$&
$4P_{1/2,3/2,5/2}$&
$1200$&5.5\%($J_u=\frac{1}{2}$)\\
\hline
N II&1&
$3P_{0,1,2}$&
$3D^o_{1,2,3}$&1083.99-1085.7&cna\\
\hline
N III&1&
$2P^o_{1/2,3/2}$&
$2D_{3/2,5/2}$&990-992&cna\\
\cline{1-2}\cline{5-5}
P III&1/2&&&1335-1345&\\
\hline 
Al I&
$5/2$&
$2P_{1/2,3/2}$&
$2S_{1/2}$&3945&cna\\
\cline{4-5}
& & &$2D_{3/2}$&3083&\\
\hline
Al II&
$\frac{5}{2}$&
$1S_0$&$1P^o_1$&1671&cna\\
\hline
Cl I& 3/2&
$2P^o_{3/2}$&$2D_{3/2,5/2}$&1097, 1189&cna\\
\cline{4-5}
&&&$2P_{1/2,3/2}$&1335.7,1347.2&\\
\hline
Cl II&3/2&
$3P_2$&$3P^o_2$&1064-1071&cna\\
\hline
Cl III&3/2&
$4S^o_{3/2}$&$4P_{1/2,3/2,5/2}$&1005-1015&cna\\
\hline
V II&7/2&
$a5D_{0,1,2,3,4}$&$z5D^o_{0,1,2,3,4}$&2672.8-2691.59&cna\\
\hline
V III&7/2&
$a4F_{3/2}$&$4D^o_{1/2,3/2,5/2}$&1121.158,1123.55,1124.298&cna\\
\cline{4-5}
&&&$z4F^o_{3/2}$&1153.179&\\
\hline
Mn II&5/2&$a7S_{3}$&$7P^o_{2,3,4}$&1162-2606.46&cna\\
\hline
Co II&7/2&$a3F_{4}$&$z3F^o_4$&2012&cna\\
\hline
Cu II&3/2&$1S_0$&$1P^o_1$&1358.773&cna\\
\hline
\hline
\end{tabular}
\caption{Selected alignable atomic species and corresponding transitions. Note only lines above $912 \AA$ are listed. Hyperfine structure should be considered for these elements with nuclear spins. Note that alkali-like species are not alignable without hyperfine structure taken into account (see text). Data are taken from the Atomic Line List http://www.pa.uky.edu/$\sim$peter/atomic/. The last column gives the maximum polarizations for emission by alkali species and absorption by NI; for other species, ``cna" stands for currently not available. }
\label{ch3t1}
\end{table}

\section{Physics of atomic alignment}
\label{physics}
We consider a realistic atomic system, which can have multiple upper levels $J_u\,,F_u$ and lower levels $F_l$.   When such an atomic system interacts with resonant radiation, there will be a photoexcitation followed by spontaneous emission. We describe the atomic occupation and radiation field  by irreducible density matrices $\rho^k_q$, ${\bar J}^K_Q$ (see App.\ref{density}). Unlike fine structure levels, the hyperfine separation $\nu_{F_u,F'_u}=[E({F_u})-E({F'_u})]/h$ is comparable to the natural line-width $A$, and therefore
the coherence between hyperfine levels must be taken into account on the upper level. The density matrix of the upper state is thus $\rho^k_q(F_u, F'_u)$. There is no coherence on the ground state, and its occupation thus can be characterized by $\rho^k_q(F_l)$. These density matrices are determined by the balance of the three processes: absorption (at a rate $\sim B_{lu}{\bar J}^0_0$), emission (at a rate $\sim$A) and magnetic precession (at a rate $\sim\nu_L$) among the sublevels of the state. The statistical equilibrium equations for hyperfine transitions can be extrapolated from the case of fine transitions (Paper I, see also Landi Degl'Innocenti \& Landolfi 2004). The formalism for hyperfine transitions can be obtained by replacing $J\,, M_J$ with $F\,, M_F$ and taking into account those additional factors in Eq.(\ref{hypfine}),
\bea
\dot{\rho^k_q}(F_u, F'_u)&+&2\pi i\nu_Lg_{u}q\rho^k_q(F_u, F'_u)+2\pi i \nu_{F_uF'_u}\rho^k_q(F_u, F'_u)  = -A(J_u\rightarrow J_l)\rho^k_q(F_u, F'_u)\nonumber\\
&+&[J_l]\sum_{F'_lk'q'}(\delta_{kk'}p_{k'}B_{lu}\bar{J}^0_0+r_{kk'}B_{lu}\bar{J}^2_0)  \rho^{k'}_{-q'}(F'_l)
\label{upevolution}
\eea
\bea
\label{lowevolution}
\dot{\rho^k_q}(F_l)&+&2\pi i \nu_Lg_lq \rho^k_q(F_l) = \sum_{J_u,F_u,F'_u}p_k[J_u]A(J_u\rightarrow J_l)\rho^k_q(F_u, F'_u) -\sum_{J_uF_uk'}(\delta_{kk'}B_{lu}\bar{J}^0_0+s_{kk'}B_{lu}\bar{J}^2_0 )\rho^{k'}_{-q'}(F_l),
\eea
where 
\bea
p_k=[F_l](-1)^{F'_u+F_l+k+1}\left\{\begin{array}{ccc}
F_l & F_l & k\\F_u & F'_u &1\end{array}\right\}[F_u, F'_u]^{1/2}\left\{\begin{array}{ccc} J_u& J_l &1\\F_l & F_u &I\end{array}\right\}\left\{\begin{array}{ccc} J_u& J_l &1\\F_l & F'_u& I\end{array}\right\},
\label{pcoe}
\eea

\bea
r_{kk'}=(-1)^{k'+q'}(3[k,k',2])^{1/2}[F_u, F'_u]^{1/2}[F'_l]\left(\begin{array}{ccc}
k & k' & 2\\ q & q' & Q \end{array}\right)\left\{\begin{array}{ccc} 
1 & F_u & F'_l\\1& F'_u & F'_l\\ 2 &k& k' \end{array}\right\}\left\{\begin{array}{ccc} J_u& J_l &1\\F'_l & F_u &I\end{array}\right\}\left\{\begin{array}{ccc} J_u& J_l &1\\F'_l & F'_u &I\end{array}\right\},
\label{rcoe}
\eea
\bea
s_{kk'}&=&(-1)^{J_u-I+F_l+q'+1}[J_l, F_l](3[k,k',2])^{1/2}\left(\begin{array}{ccc}
k & k' & 2\\ q & q'& Q\end{array}\right)\left\{\begin{array}{ccc} J_l& J_l &2\\1 & 1 &J_u\end{array}\right\}\left\{\begin{array}{ccc} J_l& J_l &2\\F_l & F_l &I\end{array}\right\}\nonumber\\
&&\left\{\begin{array}{ccc}
k & k' & 2\\ F_l & F_l& F_l\end{array}\right\}\frac{1}{2}[1+(-1)^{k+k'+2}].
\label{scoe}
\eea
In Eqs(\ref{pcoe}-\ref{scoe}), the matrices with big ``()" are 3-j symbol and the matrices with big ``\{ \}" represents the 6-j or 9-j symbol, depending on the size of the matrix. They are also known as Wigner coefficients (see Cowan 1981, Paper I). Throughout this paper, we define $[j]\equiv 2j+1$, which means $[F,F']=(2F+1)(2F'+1)$.
The evolution of upper state ($\rho^k_q(F_u, F'_u)$) is represented by Eq.(\ref{upevolution}) and the ground state ($\rho^k_q(F_l)$) is described by Eq.(\ref{lowevolution}). The second terms on the left side of Eq.(\ref{upevolution},\ref{lowevolution}) represent mixing by magnetic field, where $g_u$ and $g_l$ are the Land\'e factors for the upper and ground level. For the upper level, the mixing $\sim \nu_Lg_u\rho^k_q(F_u,F'_u)$ is much slower than the emission $\sim A\rho^k_q(F_u,F'_u)$ and is thus negligible as we consider a regime where the magnetic field is much weaker than Hanle field\footnote{For the Hanle effect to be dominant, magnetic splitting ought to be comparable to the energy width of the excited level.}. ($\nu_L\ll A/g_u$). The third term on the left side of Eq.(\ref{upevolution}) gives a measure of coherence of two hyperfine levels. It is easy to see if $\nu_{F_u,F'_u}\gg A$, the Einstein emission coefficient, the coherence component of the density matrix $\rho^k_q(F_u, F'_u)$) would be zeros. The two terms on the right side of Eq.(\ref{upevolution}, \ref{lowevolution}) are due to spontaneous emissions and the excitations from ground level. Transitions to all upper states are taken into account by summing over $J_u\,,F_u$ in Eq.(\ref{lowevolution}). Vice versa, for an upper level, transitions to all ground sublevels ($F_l$) are summed up in Eq.(\ref{upevolution}). The excitation is proportional to  
\be
\bar{J}^K_Q=\int d\nu \frac{\nu_0^2}{\nu^2}\xi(\nu-\nu_0)\oint \frac{d\Omega}{4\pi}\sum_{i=0}^3{\cal J}^K_Q(i,\Omega)S_i(\nu, \Omega),
\label{incidentrad}
\ee
which is the radiation tensor of the incoming light averaged over the whole solid angle and line profile $\xi(\nu-\nu_0)$. $S_i=[I,\, Q,\, U,\, V]$ represent Stokes parameters. The unit radiation tensors ${\cal J}^K_Q(i,\Omega)$ are given by:
\bea
{\cal J}^0_{0}(i,\Omega)&=&\left(\begin{array}{l}1\\0\\0\end{array}\right),\;~\;\; {\cal J}^2_{0}(i,\Omega)=\frac{1}{\sqrt {2}}\left[\begin{array}{l} ( 1-1.5\sin^2\theta )\\-3/2 \sin^2\theta\\0\end{array}\right],\nonumber\\
 {\cal J}^2_{\pm2}(i,\Omega)&=&\sqrt{3}e^{\pm 2i\phi}\left[\begin{array}{l}\sin^2\theta/4\\-( 1+\cos^2\theta )/4\\\mp i\cos \theta/2  \end{array}\right],\;~\;\;
{\cal J}^2_{\pm1}(i,\Omega)=\sqrt{3}e^{\pm i\phi}\left(\begin{array}{l}\mp\sin 2 \theta/4 \\\mp\sin2 \theta/4 \\-i\sin \theta/2 
\end{array}\right).
\label{unitrad}
\eea

 For unpolarized point source from ($\theta_r,\phi_r$), the radiation tensor is then:

\be
\bar{J}^0_0=I_*, \bar{J}^2_0=\frac{W_a}{2\sqrt{2}W}(2-3\sin^2\theta_r)I_*, \bar{J}^2_{\pm2}=\sqrt{3}\frac{W_a}{4W}\sin^2\theta_r e^{\pm i2\phi_r}I_*, \bar{J}^2_{\pm1}=\mp\sqrt{3}\frac{W_a}{4W}\sin2\theta_r e^{\pm i\phi_r}I_*
\label{irredradia}
\ee
where W is the dilution factor of the radiation field, which can be divided into anisotropic part $W_a$ and isotropic part $W_i$ (Bommier \& Sahal-Brechot 1978), $I_*$ is the solid-angle averaged intensity.  In the case of a point source, $W_i=0$. If $W_i\neq 0$, the degree of alignment and polarization will be reduced. The solid-angle averaged intensity for a black-body radiation source is 
\be
I_*=W\frac{2h\nu^3}{c^2}\frac{1}{e^{h\nu/k_BT}-1}.
\label{Idilu}
\ee
Since we are interested in the regime where for the
ground level the magnetic mixing is much faster than the optical pumping $\nu_L\gg \tau_R^{-1}=B\bar{J_0^0}$ for the ground state, magnetic coherence does not exist either. Thus there are only components $\rho^k_0(F_l)$ for the ground level. Taking into account this simplification, we obtain steady state solutions by setting the first terms of Eq.(\ref{upevolution},\ref{lowevolution}) on the left side to zeros,
\bea
&&\sum_{F_uF'_u}\frac{1}{1+2\pi i\nu_{F_uF'_u}/A}p_k[J_u,J_l]\sum_{F'_lk'}(\delta_{kk'}p_{k'}B_{lu}\bar{J}^0_0+r_{kk'}B_{lu}\bar{J}^2_0)  \rho^{k'}_{0}(F'_l)\nonumber\\
&-&\sum_{J_uk'}(\delta_{kk'}B_{lu}\bar{J}^0_0+s_{kk'}B_{lu}\bar{J}^2_0 )\rho^{k'}_{0}(F_l)=0
\label{hypground}
\eea

\bea
\rho^k_q(F_u, F'_u)&=&\frac{B_{lu}}{A+2\pi i\nu_{F_u,F'_u}}[J_l]\sum_{F_lk'}(\delta_{kk'}p_{k'}\bar{J}^0_0+r_{kk'}\bar{J}^2_0)  \rho^{k'}_{0}(F_l)
\label{hypupper}
\eea

It can be proved from Eq.(\ref{hypground}) that $\rho^2_0\propto {\bar J}^2_0$. Eq.(\ref{hypground}) represents a set of linear equations. Considering the equation with $k=0$, it only includes $\rho^{0,2}_0$ due to the triangular rule of the 3j symbol in the coefficient $r_{kk'}$. For $\rho^0_0$, the coefficient is $\propto {\bar J}^2_0$ as the coefficient $\propto {\bar J}^0_0$ is zero. Therefore $\rho^2_0\propto -\rho^0_0\propto {\bar J}^2_0$. As a result, the dipole component of density matrix changes its sign at Van Vleck angle as we shall show later. This is a generic feature of atomic alignment independent of their specific structures of atomic levels. The corresponding emission coefficient can be extrapolated from the one for fine structure atoms (see Landi Degl'Innocenti 1982) by replacing (L,S,J,M) with $(J, I,F,M_F)$,  
\bea
\epsilon_i(\nu, \Omega)&=&\frac{h\nu_0}{4\pi}An\xi(\nu-\nu_0)[J_u]\sum_{KQF_uF'_uF_l}[F_l]\sqrt{3[F_u,F'_u]}(-1)^{F_u+F_l+1}\left\{\begin{array}{ccc} J_u & J_l & 1\\F_l & F_u &I\end{array}\right\}\nonumber\\
&&\left\{\begin{array}{ccc} J_u & J_l &1\\F_l & F'_u &I\end{array}\right\}\left\{\begin{array}{ccc} F_u & F'_u & K\\1 & 1 & F_l\end{array}\right\}\rho^K_Q(F_u,F'_u){\cal J}^K_Q(i, \Omega),
\label{hfemissivity}
\eea
where $n$ is the total number density of the atoms. This is the expression if we can resolve the hyperfine components $F_l=1$ and $F_l=2$. For the D1 lines of alkali atomic species, polarization will be zero otherwise. The corresponding emissivities in unresolved case can be found in Landolfi and Landi Degl'Innocenti (1985). Since the separations among the hyperfine levels on the upper state is much smaller, the absorption coefficients in this case can be obtained by making the analogy with the emissivities of the unresolved case, 
\bea
\eta_i(\nu, \Omega)&=&\frac{h\nu_0}{4\pi}Bn\xi(\nu-\nu_0)[J_l]\sum_{KQF_l}[F_l]\sqrt{3}(-1)^{1-J_u+I+F_l}\left\{\begin{array}{ccc} J_l & J_l & K\\F_l & F_l &I\end{array}\right\}\nonumber\\
&&\left\{\begin{array}{ccc} 1 & 1 & K\\J_l & J_l & J_u\end{array}\right\}\rho^K_Q(F_l){\cal J}^K_Q(i, \Omega),
\label{hfmueller}
\eea
For optically thin case,  the linear polarization degree and the positional angle
\be
p=\sqrt{Q^2+U^2}/I=\sqrt{\epsilon_2^2+\epsilon_1^2}/\epsilon_0,\;
  \chi=\frac{1}{2}\tan^{-1}(U/Q)=\frac{1}{2}\tan^{-1}(\epsilon_2/\epsilon_1)
\label{pchi}
\ee
(see Fig.\ref{nzplane}); the polarization produced by absorption through optical depth $\tau=\eta_0d$ is
\be
\frac{Q}{I\tau}=\frac{-\eta_1d I_0}{(1-\eta_0d)I_0\eta_0d}\simeq-\frac{\eta_1}{\eta_0},\, U=0.
\label{genericabs}
\ee

The 6-j symbol in Eq.(\ref{hfmueller}) $\left\{\begin{array}{ccc} J_l & J_l & K\\F_l & F_l &I\end{array}\right\}=0$  for $K=2$ and $J_l<1$. This suggests that absorption is unpolarized for atoms with $J_l<1$. Alkali atoms can only produce polarized emissions therefore.

\section{Alignment of Na I, K I}
\label{alkali}
\subsection{D1 and D2 lines of Na I}
\begin{figure}
\includegraphics[%
  width=0.33\textwidth,
  height=0.25\textheight]{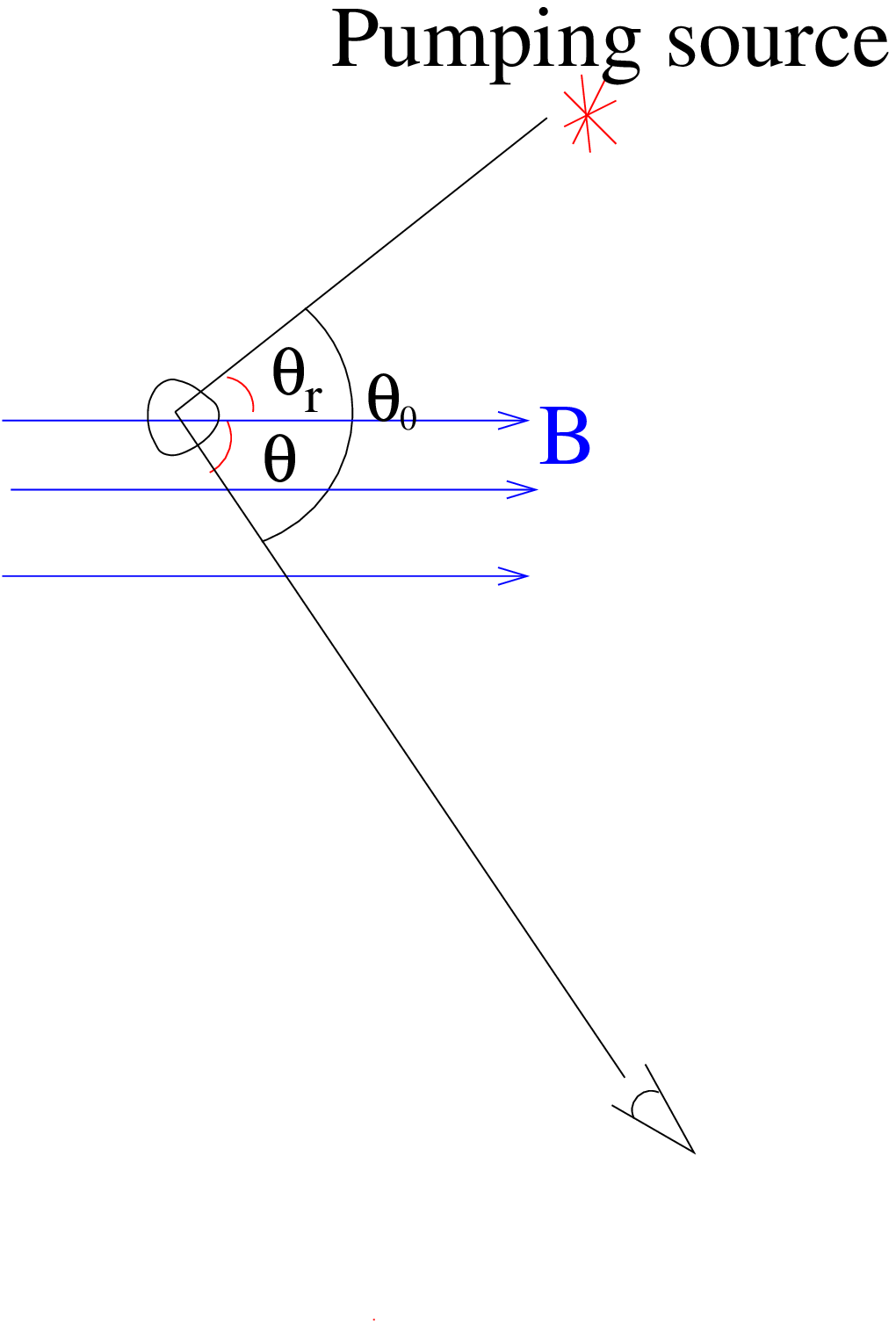}
\includegraphics[%
  width=0.33\textwidth,
  height=0.25\textheight]{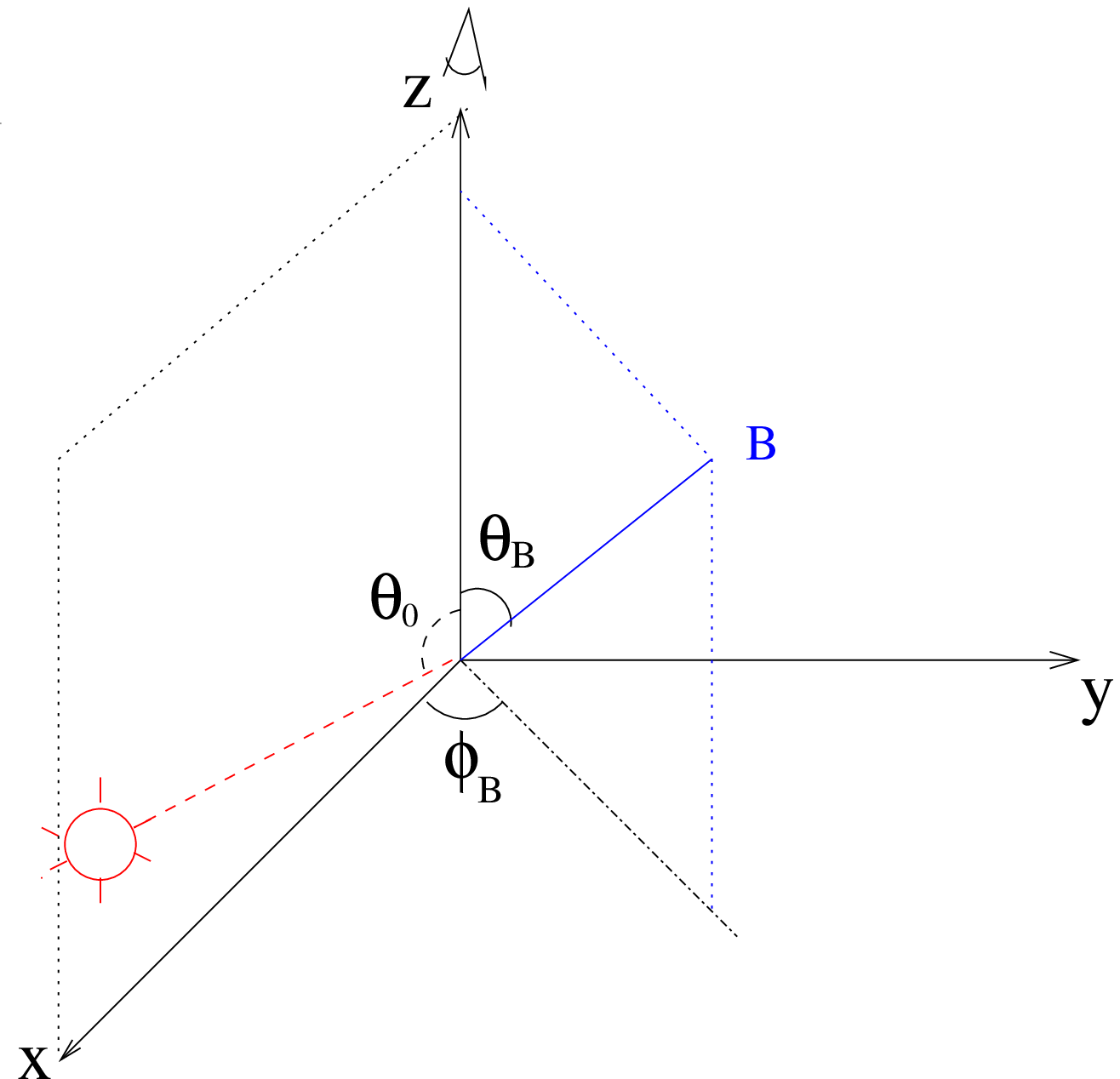}
\includegraphics[%
  width=0.33\textwidth,
  height=0.25\textheight]{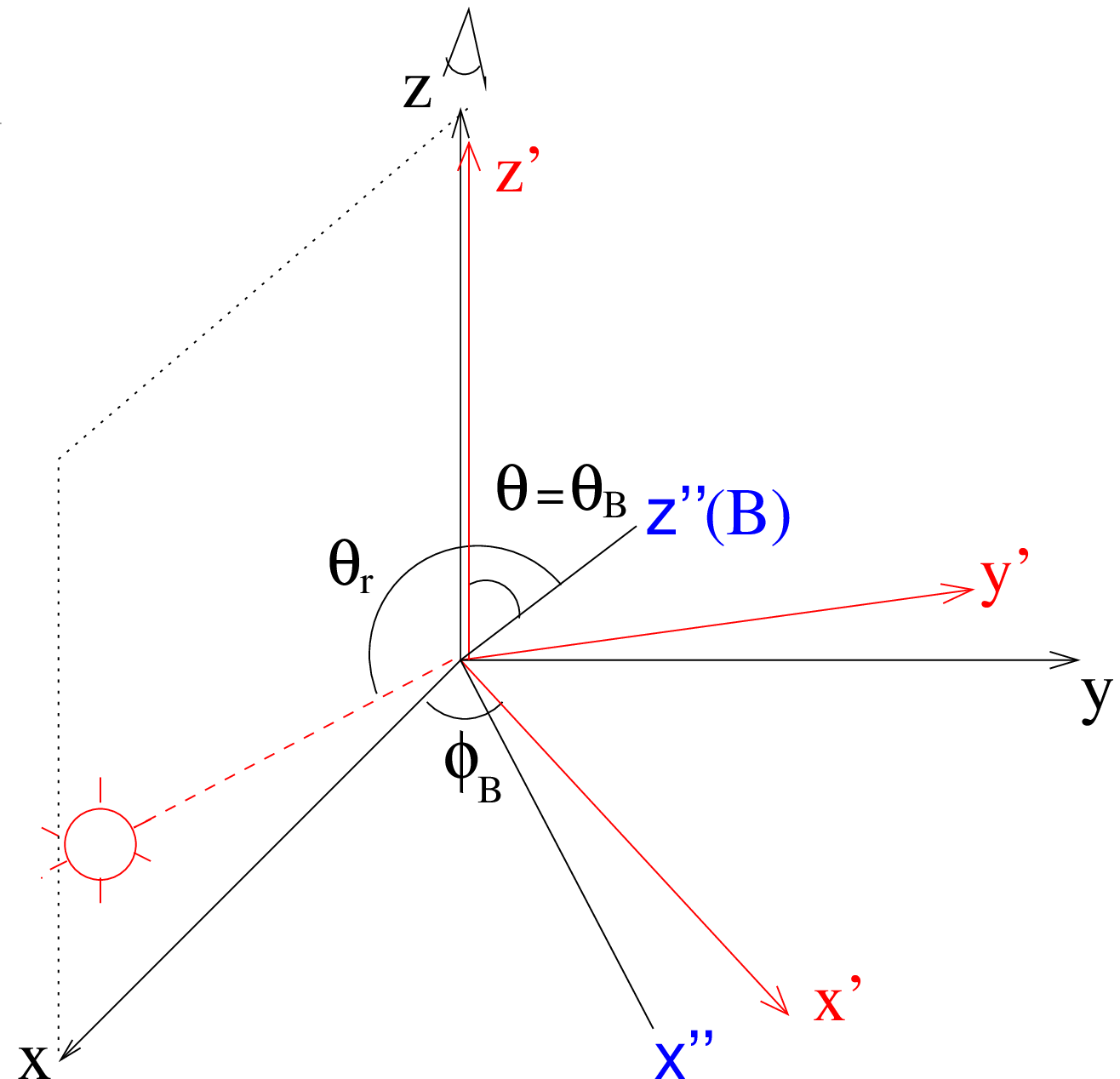}
\caption{{\it Left}: typical astrophysical environment where atomic alignment can happen. A pumping source deposits angular momentum to atoms in the direction of radiation and causes differential occupations on their ground states. In a magnetized medium where the Larmor precession rate $\nu_L$ is larger than the photon arrival rate $\tau_R^{-1}$, however, atoms are realigned with respect to magnetic field. Atomic alignment is then determined by $\theta_r$, the angle between the magnetic field and the pumping source. The polarization of scattered line also depends on the direction of line of sight, $\theta$ and $\theta_0$ (or $\phi_r$, defined afterward); {\it Middle}: geometry of the observational frame. In this frame, the line of sight is z axis, together with the incident light, they specify the x-z plane. Magnetic field is in ($\theta_B, \phi_B$) direction; {\it Right}: transformation to the ``theoretical frame" where magnetic field defines z" axis. This can be done by two successive rotations specified by Euler angles ($\phi_B, \theta_r$) (see App.\ref{euler} for details). The first rotation is from xyz coordinate system to x'y'z' coordinate system by an angle $\phi_B$ about the z-axis, the second is from x'y'z' coordinate system to x"y"z" coordinate system by an angle $\theta$  about the y'-axis. Atomic alignment and transitions are treated in the ``theoretical" frame where the line of sight is in ($\theta, \pi$) direction and the incident radiation is in ($\theta_r,\phi_r$) direction. }
\label{radiageometry}
\end{figure}

The geometry of the radiation system is illustrated by Fig.\ref{radiageometry}. The origin of this frame is defined as the location of the atomic cloud. The line of sight defines z axis, and together with direction of radiation, they specify x-z plane. The x-y plane is thus the plane of sky. In this frame, the incident radiation is coming from  ($\theta_0, 0$), and the magnetic field is in the  direction ($\theta_B, \phi_B$). 

The magnetic field is chosen as the quantization axis ($z"$) for the atoms. Alignment shall be treated in the frame $x"y"z"$ (see App.\ref{euler} how these two frames are related). In this ``theoretical'' frame, the line of sight is in ($\theta, \pi$) direction  (i.e., the x"-z" plane is defined by the magnetic field and the line of sight, see Fig.\ref{radiageometry}{\it right}), and the radiation source is directed along ($\theta_r, \phi_r$).

\begin{figure}
\includegraphics[%
  width=0.22\textwidth,
  height=0.25\textheight]{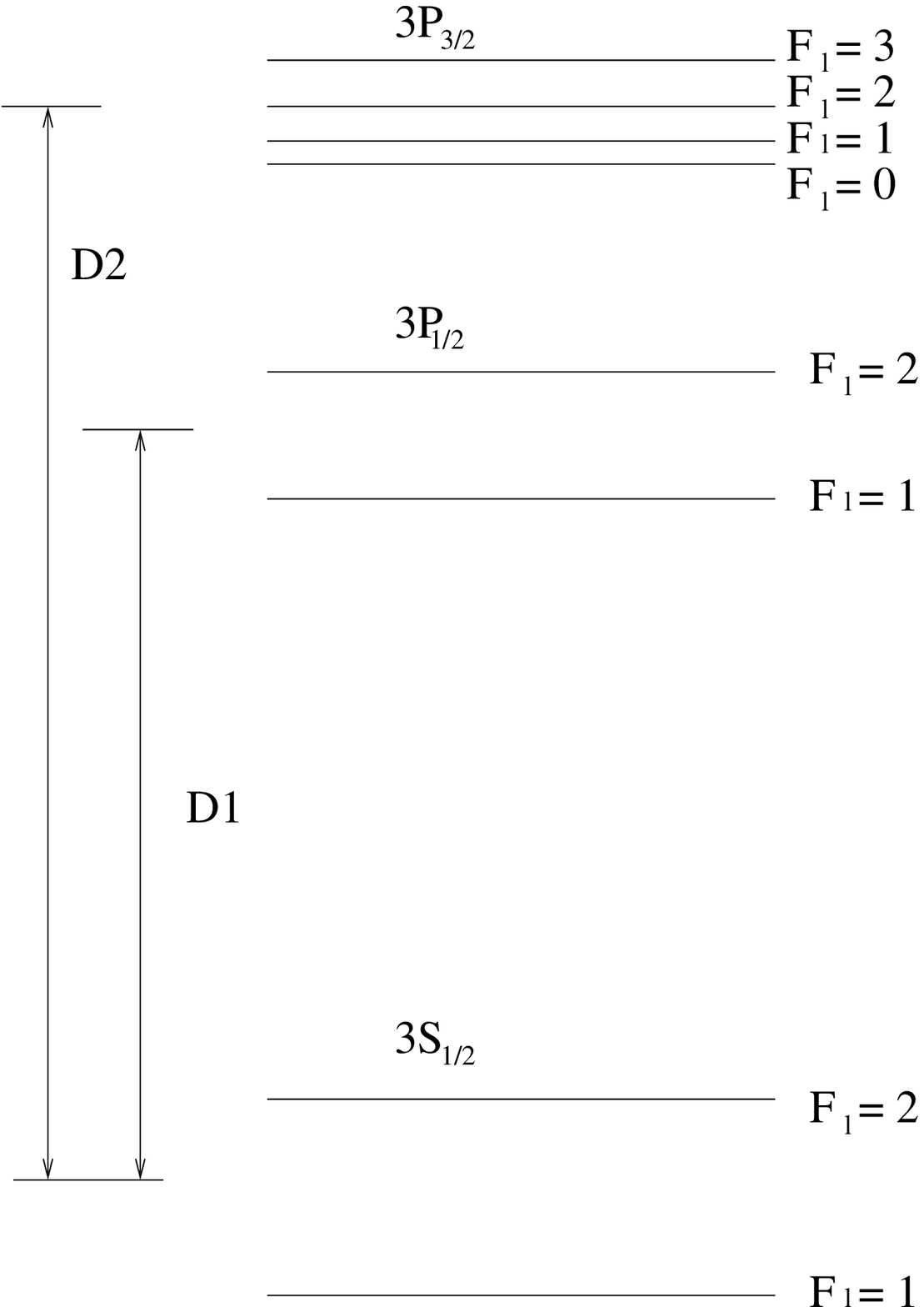}
\includegraphics[%
  width=0.38\textwidth,
  height=0.25\textheight]{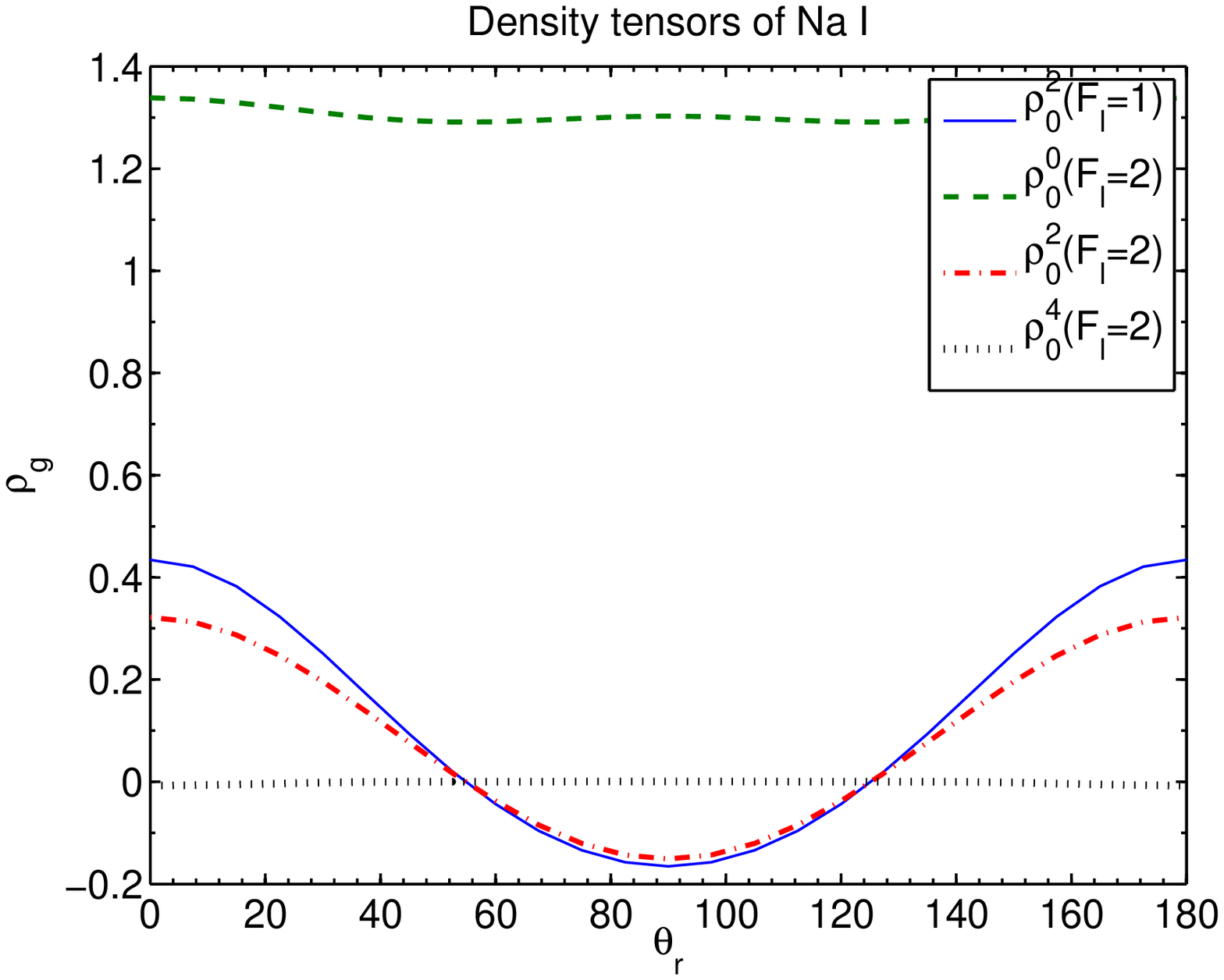}
\includegraphics[%
  width=0.38\textwidth,
  height=0.25\textheight]{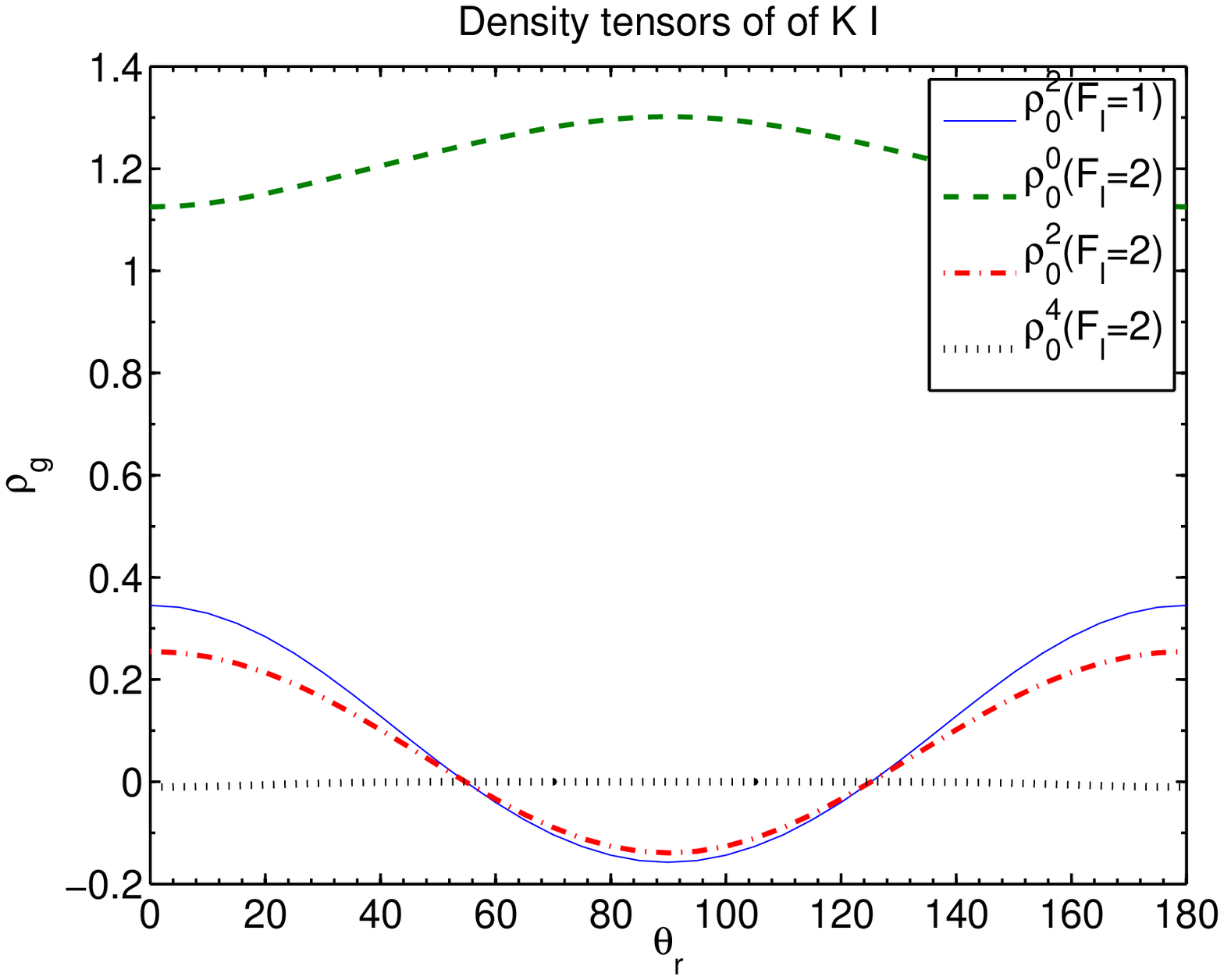}
\caption{{\it Left}: The schematic of Na and K hyperfine levels; {\it Middle}: Normalized density tensor components $\rho^k_0/\rho^0_0(F_l=1)$ for ground state of Na; {\it Right}: for ground state of K. The difference of Na and K alignments is due to different degrees of coherence on the excited state (see text).}
\label{Naocc}
\end{figure}

The ground state of Na is $2^{2}S_{1/2}$ and the first
excited states $2^{2}P_{1/2}$, $2^{2}P_{3/2}$ correspond to
D1 and D2 lines respectively. 
The nuclear spin of Na is $I=3/2$, its total angular momentum thus can be $F=I\pm J=1,2$ (see Fig.\ref{Naocc}{\it left}). According to the selection rule of the 3j symbol in Eq.(\ref{irreducerho}), the irreducible density tensor of the ground state $\rho^k_q(F_l=1)$ has components with $k=1-1, 1, 1+1=0, 1,2$, $\rho^k_q(F_l=2)$ has components with $k=2-2, 2-1,..., 2+2=0,1,2,3,4$. For unpolarized pumping light, we only need to consider even components. For the upper level $\rho^k_q(F_u)$, according to Eq.(\ref{hfemissivity}), only the components with $k=K\leq 2$ need to be counted. For Na D2 line, the hyperfine splittings of upper level $2^{2}P_{3/2}$ are comparable to their natural width: $\omega_{10}=2.6\gamma,\,\omega_{21}=5.2\gamma,\,\omega_{32}=7.7\gamma$. Thus the interference between levels (measured by the factor $1/(1+2\pi i\nu_{F_u, F'_u}/A)$ in Eq.(\ref{hypground},\ref{hypupper}) must be taken into account on the upper level. On ground level, there is no magnetic coherence term, namely, $\rho^k_{q\neq0}=0$. Owing to the triangle rule of "3j" symbols (see Eqs.\ref{hypground},\ref{rcoe},\ref{scoe}), only ${\bar J}^{0,2}_{Q=0}$ components appear and they are determined by the polar angle $\theta_r$ of the radiation (Eq.\ref{irredradia}, Fig.\ref{radiageometry}). As a result, $\rho^{2,4}_0$ are real quantities and they are independent of the azimuthal angle $\phi_r$ (see Eq.\ref{hypground}). Physically this results from fast procession around magnetic field. Their dependence on polar angle is shown in Fig.\ref{Naocc}. To calculate the alignment for the multiplet, all transitions should be counted according to their probabilities even if one is interested in only one particular line. This is because they all affect the ground populations and therefore the degree of alignment and polarization. This means that in practical calculations, a summation should be taken over all the hyperfine sublevels $F_u$ of both the upper levels $J_u=1/2,3/2$ in Eq.(\ref{hypground}). The key is the coefficients $p_k, r_{kk'}, s_{kk'}$ which are determined by the hyperfine structure of an atomic species. By inserting the values of $J_u,F_u,J_l,F_l$, $k,k'$, $K$ and ${\bar J}^K_Q$ (Eq.\ref{irredradia}) into Eqs.(\ref{pcoe})-(\ref{scoe}), we get the coefficients as given by Table~\ref{prs}. 

We see from Table~\ref{prs}, the coefficient $s_{kk'}\equiv 0$ and this is actually true for all the alkali species. Since $s_{kk'}$ represents the differential excitation from the ground level (see Eq.\ref{lowevolution}), this means alkali species are not aligned by the same mechanism\footnote{In fact, the absorptions from alkali species are not polarized for the same reason.} (so called depopulation pumping) as illustrated by the toy model (\S2). Instead they are aligned through another mechanism, repopulation pumping. Atoms are repopulated as a result of spontaneous decay from a polarized upper level (see Happer 1972). Upper level becomes polarized because of differential absorption rates to the levels (given by $r_{kk'}$, see Eq.\ref{upevolution}). For instance, $r_{20}=-0.1042$ for the absorption from $F_l=1$ to $F_u=1$. As the result, the density component of the upper level $F_u=1$, $\rho^2_0(u)<0$, indicating that atoms are accumulated in the sublevel $F_lM_F=1,0$ according to the definition of irreducible tensor $\rho^2_0=[\rho(1,1)-2\rho(1,0)+\rho(1,-1)]$ (see App.\ref{irreducerho}). 

With the coefficients $p_k, r_{kk'}, s_{kk'}$ known, one can then easily attain the coefficients matrix of Eq.(\ref{hypground}). For NaI, there are totally five linear equations for $\rho^{0,2}_0(F_l=1)$ and $\rho^{0,2,4}_0(F_l=2)$. By solving them, we obtain 
\bea
\left[\begin{array}{c}\rho^0_0(F_l=2)\\\rho^2_0\left(\begin{array}{c}F_l=1\\F_l=2\end{array}\right)\\\rho^4_0(F_l=2)\end{array}\right]=\left(\begin{array}{c}
34492 - 5511 \cos2\theta_r + 100 \cos4\theta_r\\2327 + 7289\cos2\theta_r - 157.4 \cos4\theta_r - 23.4 \cos6\theta_r\\
 1606 + 5823\cos2\theta_r - 435.1 \cos4\theta_r - 4.5\cos6\theta_r\\ 39.31+75.31 \cos2\theta_r + 35.56 \cos4\theta_r +15.94 \cos6\theta_r\end{array}\right)\varrho^0_0
\label{Naground}
\eea
where we have defined $\varrho^0_0=\rho^0_0(F_l=1)/(26416 - 4549\cos2\theta_r - 163.4 \cos4\theta_r + 24.5 \cos6\theta_r)$. The results are demonstrated in Fig.\ref{Naocc}. The triangle dependence is caused by the precession of the atoms around magnetic field. As we see, the alignment changes sign at the Van Vleck angle $\theta_r=54.7^o$ same as the case for atoms with only fine structures (Paper I). As explained in \S\ref{physics} and in Paper I, this is a generic feature of atomic alignment determined by the geometric relation of the pumping source and the magnetic field.
\begin{table}
\begin{tabular}{||c|c|ccc|cccc|c||}
\hline
\hline
&$F_l\rightarrow F_u,F'_u$&$p_0$&$p_2$&$p_4$&$r_{20}$&$r_{02}$&$r_{22}$&$r_{44}$&$s_{kk'}$\\
\hline
D1($F_u=F'_u$)&$1\rightarrow 1$&.0833&-.0417&0&-0.0417&-0.0417&-0.0589&0&0\\
\cline{2-9}
&$1\rightarrow 2$&.3277&.1909&0& 0.1909 &0.0323&0.0386& 0& \\
\cline{2-9}
&$2\rightarrow 1$&.3277&.1909&0& 0.0323&0.1909&0.0386& 0& \\
\cline{2-9}
&$2\rightarrow 2$&.25&.125&-.1667&-0.1479&-0.1479&-0.1263& 0&\\
\cline{1-8}
D2&$1\rightarrow 00$&0.1443&0&0&0&0&0& 0&\\
\cline{2-9}
&$1\rightarrow 11$&.2083&-0.1042&0 &-0.1042&-0.1042 &  -0.1473& 0&\\
\cline{2-9}
&$1\rightarrow 22$&0.1614&0.0955&0&0.0955&0.0161 &   0.0193& 0& \\
\cline{2-9}
&$1\rightarrow 12(21)$&0&0.1398&0&-0.1614& 0& 0.1141& 0& \\
\cline{2-9}
&$1\rightarrow 02(20)$&0&0.1021&0&0.0884&0& 0& 0& \\

\cline{2-9}
&$2\rightarrow 11$&0.0323&0.0191&0&0.0032&0.0191    &0.0039& 0&\\
\cline{2-9}
&$2\rightarrow 22$&0.125&0.0625&-0.0833& -0.0740&-0.0740 &-0.0631& 0& \\
\cline{2-9}
&$2\rightarrow 33$&0.2958&0.2449&0.1236&0.1449&0.0500&0.0495&0&\\
\cline{2-9}
&$2\rightarrow 12(21)$&0&0.0427&0&-0.0354&0&-0.003&0.0036&\\
\cline{2-9}
&$2\rightarrow 23(32)$&0&0.1118&0.1318&-0.1323&0&-0.0113&0.0133&\\
\cline{2-9}
&$2\rightarrow 13(31)$&0&0.0204&0.0791&0.0592&0&0.0051&-0.006&\\
\hline
\hline
\end{tabular}
\label{prs}
\caption{The numerical coefficients of Eqs.(\ref{upevolution},\ref{lowevolution},\ref{hypground}, \ref{hypupper}) calculated for NaI according to Eqs.(\ref{pcoe})-(\ref{scoe}).}
\end{table}

Scattering from such aligned atoms causes polarization in both D lines. Put Eq.(\ref{Naground}) into Eqs(\ref{hypupper}, \& \ref{hfemissivity}) and combine with Eq.(\ref{unitrad}), we obtain the expression for emission coefficients after some tedious calculations. For D2 line,
$F_l=1$,
{\scriptsize
\bea
\epsilon_0&=&\frac{3\sqrt{3}\lambda^2}{8\pi}AI_*n\varrho^0_0\xi(\nu-\nu_0)
\left\{62830-10084\cos 2\theta_r-412 \cos 4\theta_r+56
   \cos 6\theta_r+(0.8 \cos 8\theta_r-8.7 \cos 6 \theta_r+30.3 \cos4\theta_r+2236.6 \cos 2\theta_r\right.\nonumber\\
&+&776)\cos2\theta+\cos 2\phi_r
   \left[(395-513.4\cos 2\theta_r+107.4\cos 4\theta_r+12.3 \cos6\theta_r-1.3\cos 8\theta_r) \sin ^2\theta\right]\nonumber\\
&+&\cos \phi_r\left[104.3+1.95\cos 2\theta_r-212.19\cos4 \theta_r-1.78 \cos 6\theta_r+2.98 \cos8 \theta_r-0.18 \cos 10 \theta_r\right.\nonumber\\
&+&\left.\left.8.5(\cos 2\theta_r-1.4\cos 4\theta_r+0.1\cos 6\theta_r+48.5) \sin2\theta \sin 2\theta_r\right] \right\}\nonumber\\
\epsilon_1&=&-\frac{3\sqrt{3}\lambda^2}{8\pi}AI_*n\varrho^0_0\xi(\nu-\nu_0) \left\{(1552.7+4473.3 \cos 2\theta_r+60.6 \cos 4 \theta_r-17.4\cos 6\theta_r+1.5\cos 8\theta_r) \sin^2\theta+ \cos 2\phi_r \left[592.5\right.\right.\nonumber\\
&+&\left.\cos 2 \theta
   (197.5 -0.7 \cos8\theta_r+6.2 \cos 6\theta_r+53.7\cos 4\theta_r-256.7\cos 2 \theta_r)-770.1
   \cos 2\theta_r+161.1\cos 4\theta_r+18.5
   \cos 6\theta_r-2\cos 8\theta_r\right]\nonumber\\
&+& \left.\cos \phi_r (17.0\cos 2\theta_r-23.8\cos 4\theta_r+1.4\cos 6\theta_r) \sin 2\theta \sin 2\theta_r\right\}\nonumber\\
\epsilon_2&=&-\frac{3\sqrt{3}\lambda^2}{8\pi}AI_*n\varrho^0_0\xi(\nu-\nu_0) \left[\cos \theta \left(790-1026.9 \cos 2\theta_r+214.8\cos 4\theta_r+24.7 \cos 6\theta_r+2.6\cos 8\theta_r\right) \sin 2\phi_r+(34.0 \cos2\theta_r\right.\nonumber\\
&-&\left.47.6 \cos 4\theta_r+2.8\cos 6\theta_r+1649.9) \sin\phi_r \sin \theta \sin 2\theta_r\right] 
\label{D2F1}
\eea}
$F_l=2$,
{\scriptsize
\bea
\epsilon_0&=&\frac{3\sqrt{3}\lambda^2}{8\pi}AI_*n\varrho^0_0\xi(\nu-\nu_0)
 \left\{136670-28893\cos 2\theta_r+899 \cos 4
   \theta_r-2\cos 6\theta_r+\cos
   ^2\theta(20729 \cos 2\theta_r-1118 \cos 4
   \theta_r-19 \cos 6\theta_r\right.\nonumber\\
&-&2\cos 8
   \theta_r+6057) +\cos 2 \phi_r \left(5649.4-6278.5 \cos 2
   \theta_r+656.2\cos 4\theta_r-27.4\cos 6\theta_r+0.3\cos 8\theta_r\right) \sin^2\theta+\cos \phi_r \left[2522.6\right.\nonumber\\
&-&\left.\left.147.8\cos 2\theta_r -2521.1\cos 4
   \theta_r+147.6\cos 6\theta_r-1.6 \cos 8\theta_r+0.1\cos 10 \theta_r+(5090-591\cos 2\theta_r+6\cos 4\theta_r) \sin 2\theta \sin 2\theta_r\right]\right\}\nonumber\\
\epsilon_1&=&-\frac{3\sqrt{3}\lambda^2}{8\pi}AI_*n\varrho^0_0\xi(\nu-\nu_0)\left\{(20729 \cos 2\theta_r-1117.7\cos 4
   \theta_r-19.3 \cos 6\theta_r-1.9\cos 8\theta_r+6057.1) \sin^2\theta+\cos 2\phi_r
   \left[(-6278.5\cos 2\theta_r\right.\right.\nonumber\\
&+&\left.656.2\cos 4
   \theta_r-27.4\cos 6\theta_r+0.3 \cos 8\theta_r+5649.4) \cos
   ^2\theta-6479.6\cos 2\theta_r+656.2\cos 4\theta_r-27.4 \cos 6\theta_r+0.3 \cos 8
   \theta_r+5649.4\right]\nonumber\\
&+&\left.\cos \phi_r (10097-1183.0 \cos2\theta_r+12.4 \cos 4\theta_r-\cos 6
   \theta_r) \sin 2\theta \sin 2\theta_r+6057.1\right\} \nonumber\\
\epsilon_2&=&-\frac{3\sqrt{3}\lambda^2}{8\pi}AI_*n\varrho^0_0\xi(\nu-\nu_0) \left\{\cos \theta \left(20189-4818 \cos 2\theta_r+215
   \cos 4\theta_r-2\cos 6\theta_r\right) \sin ^2\theta_r \sin 2\phi_r+(20193-2366 \cos
   2\theta_r\right.\nonumber\\
&+&\left.25\cos 4\theta_r-2\cos 6\theta_r \cos 8\theta_r) \sin\phi_r \sin \theta \sin 2\theta_r\right\}
\label{D2F2}
\eea}

For D1 line,

$F_l=1$,
{\scriptsize
\bea
\epsilon_0&=&\frac{3\sqrt{3}\lambda^2}{8\pi}AI_*n\varrho^0_0\xi(\nu-\nu_0) \left\{36142-6350 \cos 2\theta_r-285 \cos 4\theta_r+35 \cos 6\theta_r-\cos 2\theta(40- 2\cos 6\theta_r+76  \cos4\theta_r+26 \cos2\theta_r)\right.\nonumber\\
&-&\cos   2\phi_r \left[(588-702\cos 2\theta_r+126\cos 4\theta_r-11 \cos 6\theta_r) \sin ^2\theta\right]-\cos \phi_r \left[268-1.19\cos 2\theta_r-267 \cos
   4\theta_r\right.\nonumber\\
&+&\left.\cos 6\theta_r+\left. ( 537.0-5.1\cos 2\theta_r+2.7\cos 4\theta_r-0.3\cos 6\theta_r) \sin2 \theta \sin2 \theta_r\right]\right\}\nonumber\\
\epsilon_1&=&-\frac{3\sqrt{3}\lambda^2}{8\pi}AI_*n\varrho^0_0\xi(\nu-\nu_0) \left\{(130.89 \cos 2\theta_r-153.45 \cos 4\theta_r+5.26 \cos 6\theta_r+1.82\cos 8\theta_r-2050.2)\sin^2\theta\right.\nonumber\\
&+&\cos 2\phi_r \left[\cos 2 \theta(-293.92-0.20\cos8 \theta_r+5.66\cos 6\theta_r-62.56\cos 4\theta_r+251.03 \cos 2\theta_r)+1053.08\cos 2\theta_r-187.674\cos 4\theta_r\right.\nonumber\\
&+&\left.16.975\cos 6\theta_r-0.606 \cos 8\theta_r-881.772\right]-\cos \phi_r \left[1073.92-10.216\cos 2\theta_r+5.472 \cos 4\theta_r-0.673 \cos 6\theta_r\right]\nonumber\\
&&\left. \sin 2 \theta \sin 2\theta_r+1969.6\right\} \nonumber\\
\epsilon_2&=&-\frac{3\sqrt{3}\lambda^2}{8\pi}AI_*n\varrho^0_0\xi(\nu-\nu_0)  \left[1404.1\cos \theta ( \cos 2\theta_r-0.1782\cos 4\theta_r+0.0161 \cos 6\theta_r-0.0006\cos 8\theta_r-0.8373) \sin 2\phi_r\right.\nonumber\\
&+&\left.(20.4 \cos 2\theta_r-10.9\cos 4\theta_r+1.3\cos 6\theta_r-2147.8) \sin \phi_r \sin \theta \sin 2\theta_r\right]
\label{D1F1}
\eea}
$F_l=2$,
{\scriptsize
\bea
\epsilon_0&=&\frac{3\sqrt{3}\lambda^2}{8\pi}AI_*n\varrho^0_0\xi(\nu-\nu_0) \left\{64488-10148\cos 2\theta_r+337\cos 4\theta_r-10\cos 6\theta_r-\cos 2 \theta(0.11\cos 8\theta_r+2.63\cos6\theta_r-66.73 \cos 4\theta_r+65.45 \cos 2\theta_r\right.\nonumber\\
&-&40.301)-\cos 2\phi_r   \left[(-587.8+702\cos 2\theta_r-125.2 \cos 4\theta_r+11\cos6\theta_r) \sin^2\theta\right]\nonumber\\
&+&\cos \phi_r\left[268-1.19\cos 2\theta_r-267.1 \cos 4\theta_r+1.1\cos 6\theta_r-0.7 \cos 8\theta_r+0.1 \cos 10\theta_r+\sin2\theta \sin 2\theta_r \right.\nonumber\\
&&\left.\left.(536.96-5.11 \cos 2\theta_r+2.74\cos 4\theta_r-0.34 \cos 6\theta_r)\right]\right\},\nonumber\\
\epsilon_1&=&-\epsilon_2(F_l=1),\,\epsilon_2=-\epsilon_2(F_l=1).
\label{D1F2}
\eea}

\begin{figure}
\includegraphics[%
  width=0.33\textwidth,
  height=0.25\textheight]{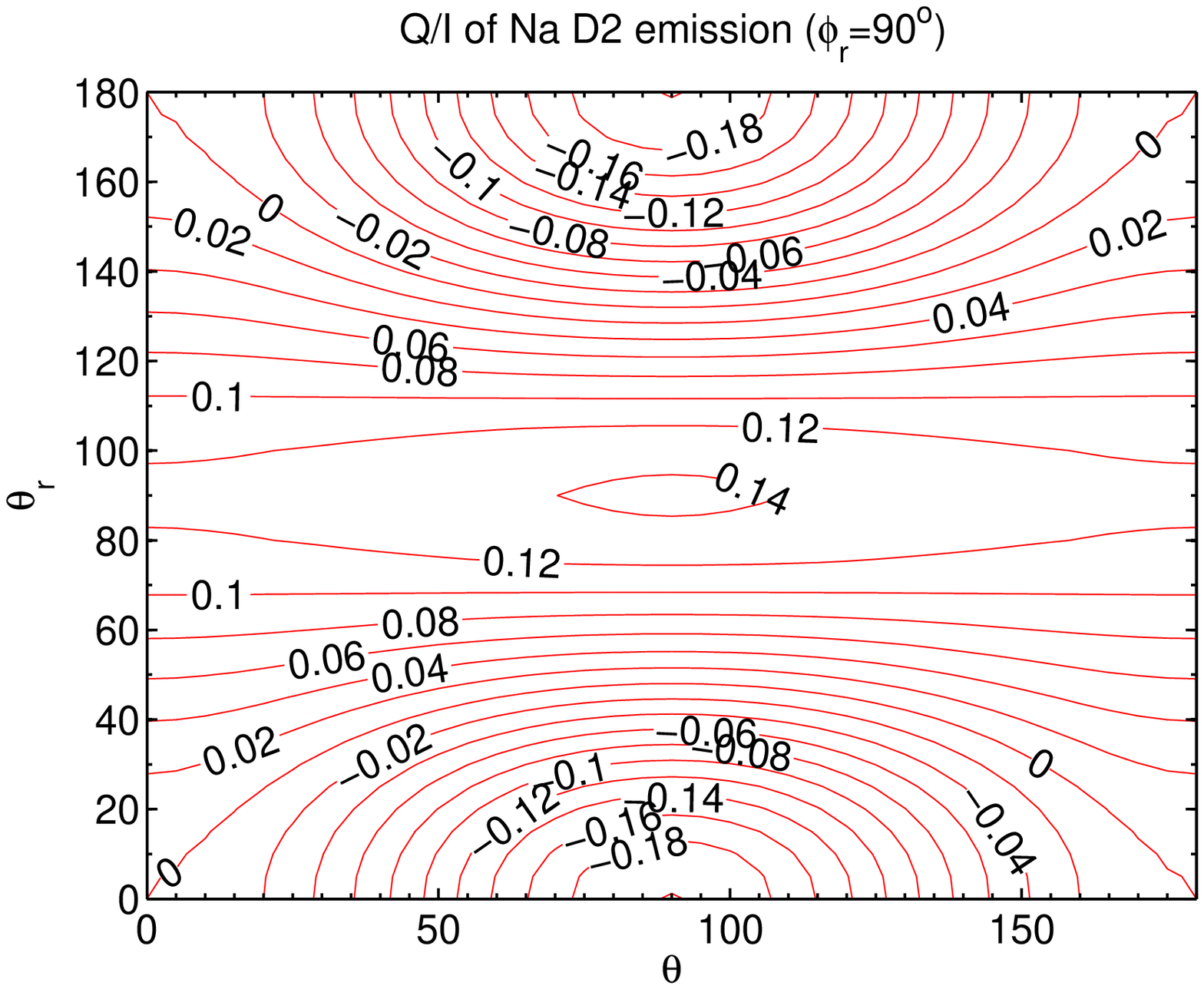}
\includegraphics[%
  width=0.33\textwidth,
  height=0.25\textheight]{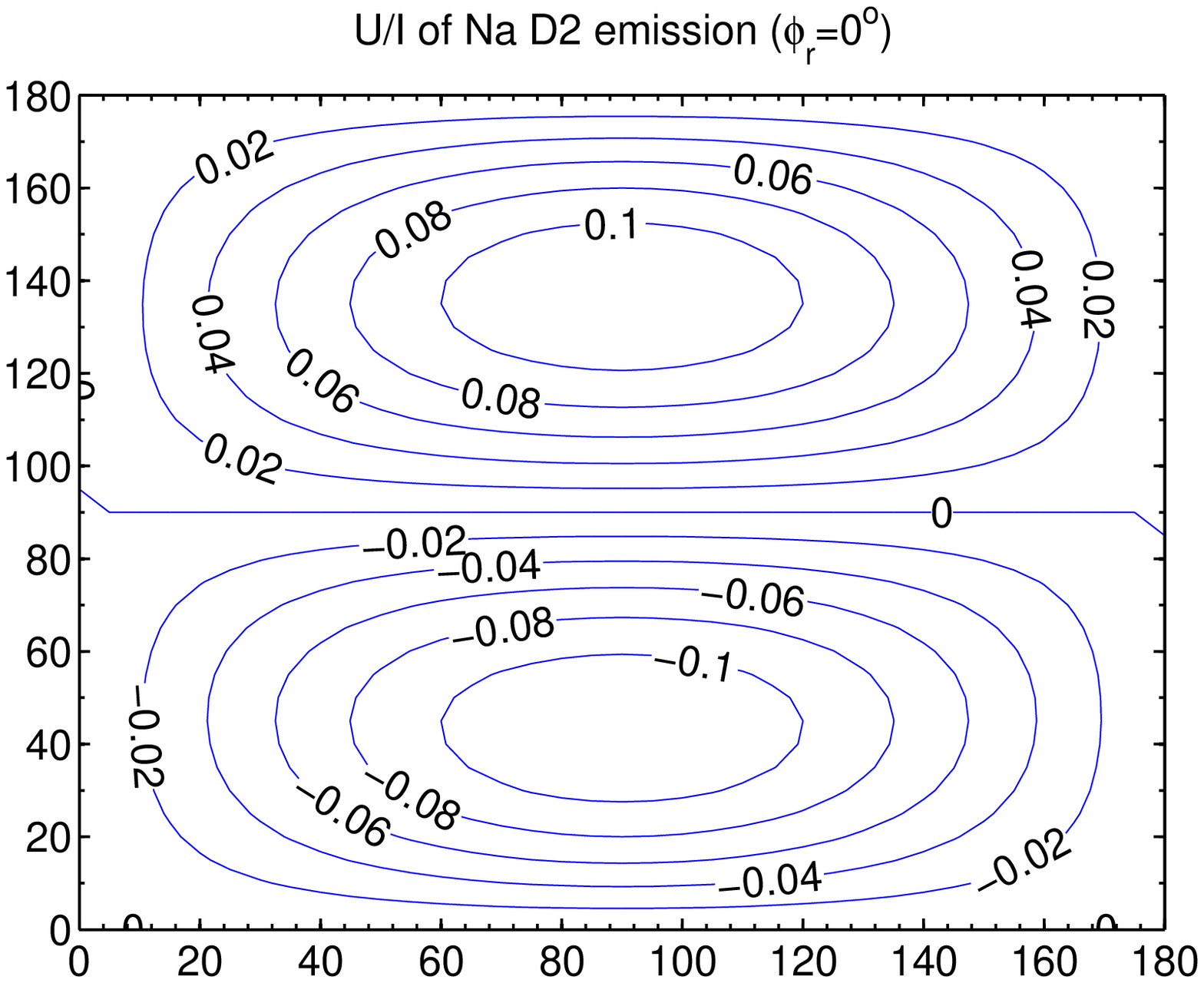}
\includegraphics[%
  width=0.33\textwidth,
  height=0.25\textheight]{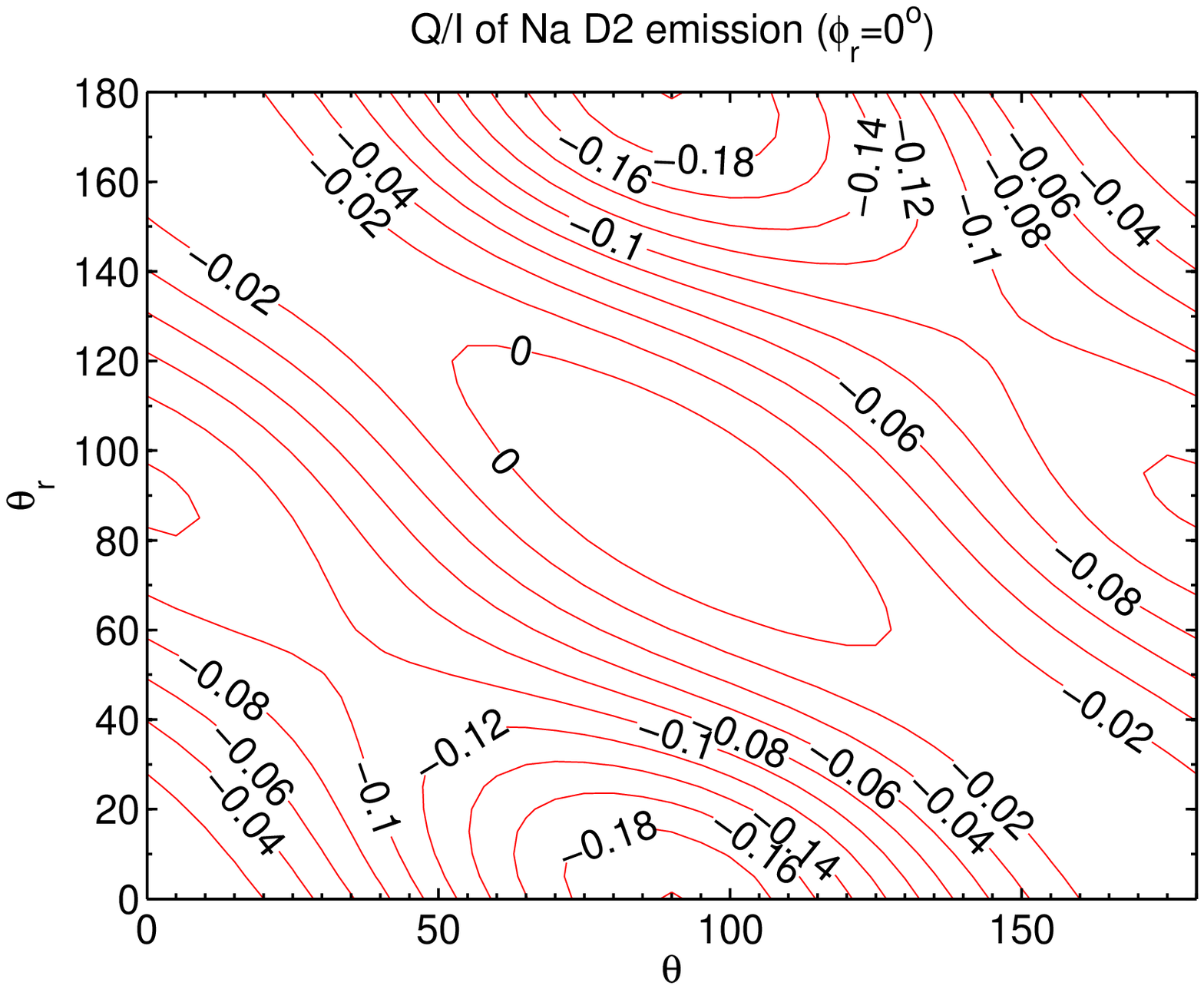}
\caption{Contour graphs of polarization signals of Na D2 emission line: {\it left, right}: Q/I, {\it middle}: U/I. Polarization depends on three angles: $\theta_r$, $\theta$ and $\phi_r$ (Fig.\ref{radiageometry}). $\phi_r$ is fixed to $\pi/2$ and 0. At $\phi_r=0$, U=0. The Stokes parameters Q represents the linear polarization along ${\bf e}_1$ minus the linear polarization along ${\bf e}_2$; U refers to the polarization along $({\bf e_1+e_2})/\sqrt{2}$ minus the linear polarization along $({\bf -e_1+e_2})/\sqrt{2}$ (see Fig.\ref{nzplane}{\it right}). }
\label{Napolcontour}
\end{figure}

\begin{figure}
\includegraphics[%
  width=0.45\textwidth,
  height=0.3\textheight]{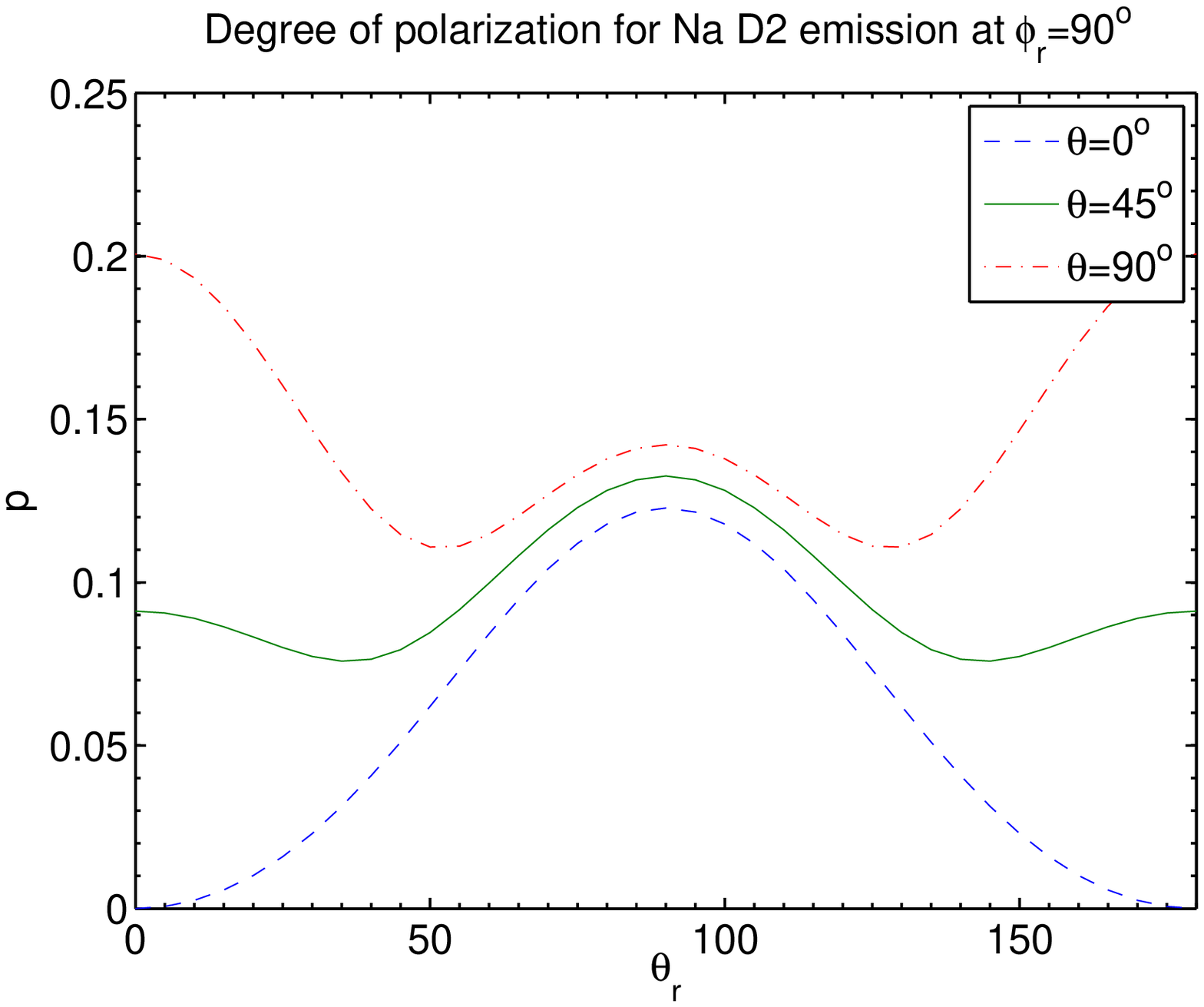}\includegraphics[%
  width=0.45\textwidth,
  height=0.3\textheight]{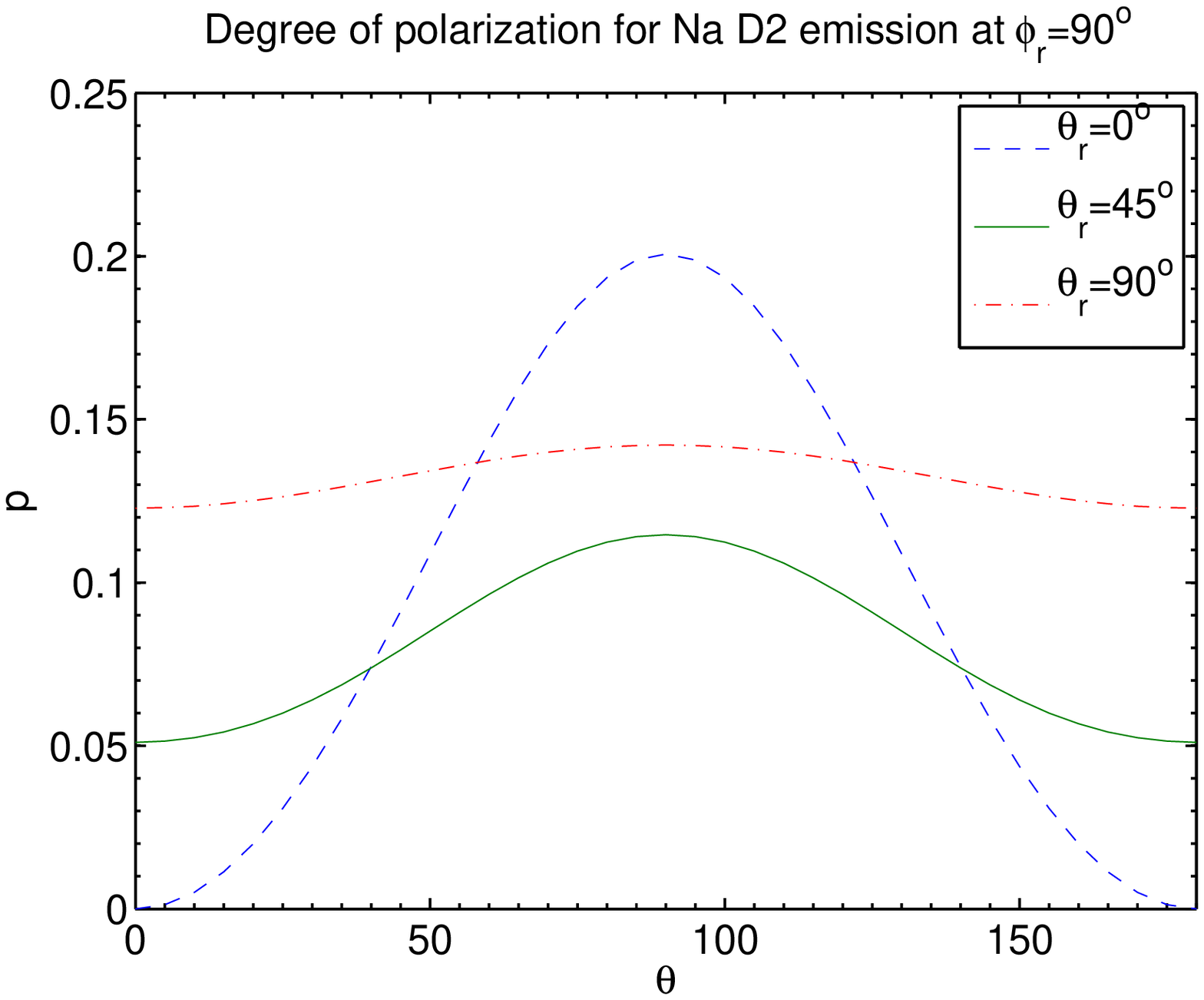}\\
\includegraphics[%
  width=0.45\textwidth,
  height=0.3\textheight]{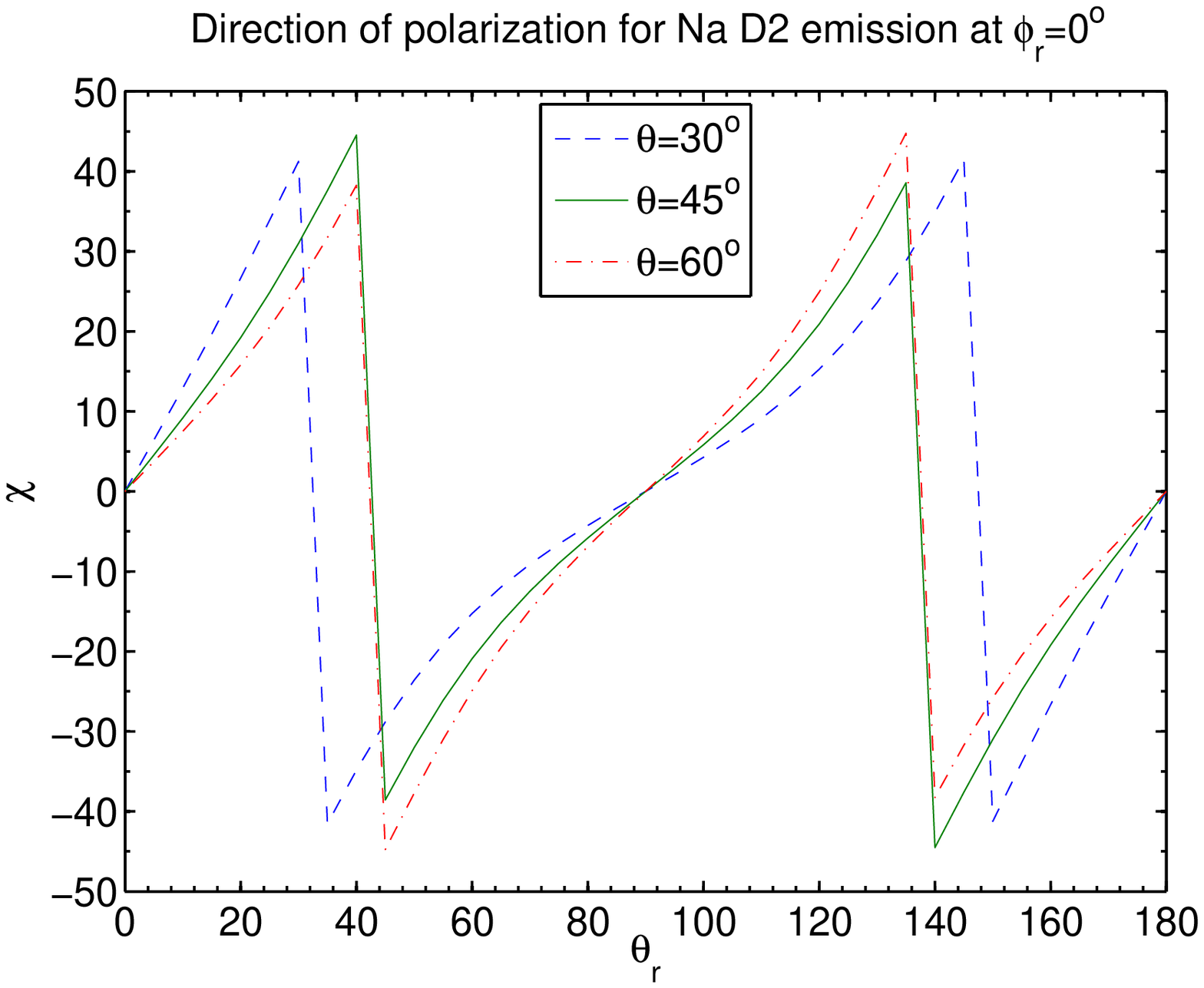}\includegraphics[%
  width=0.45\textwidth,
  height=0.3\textheight]{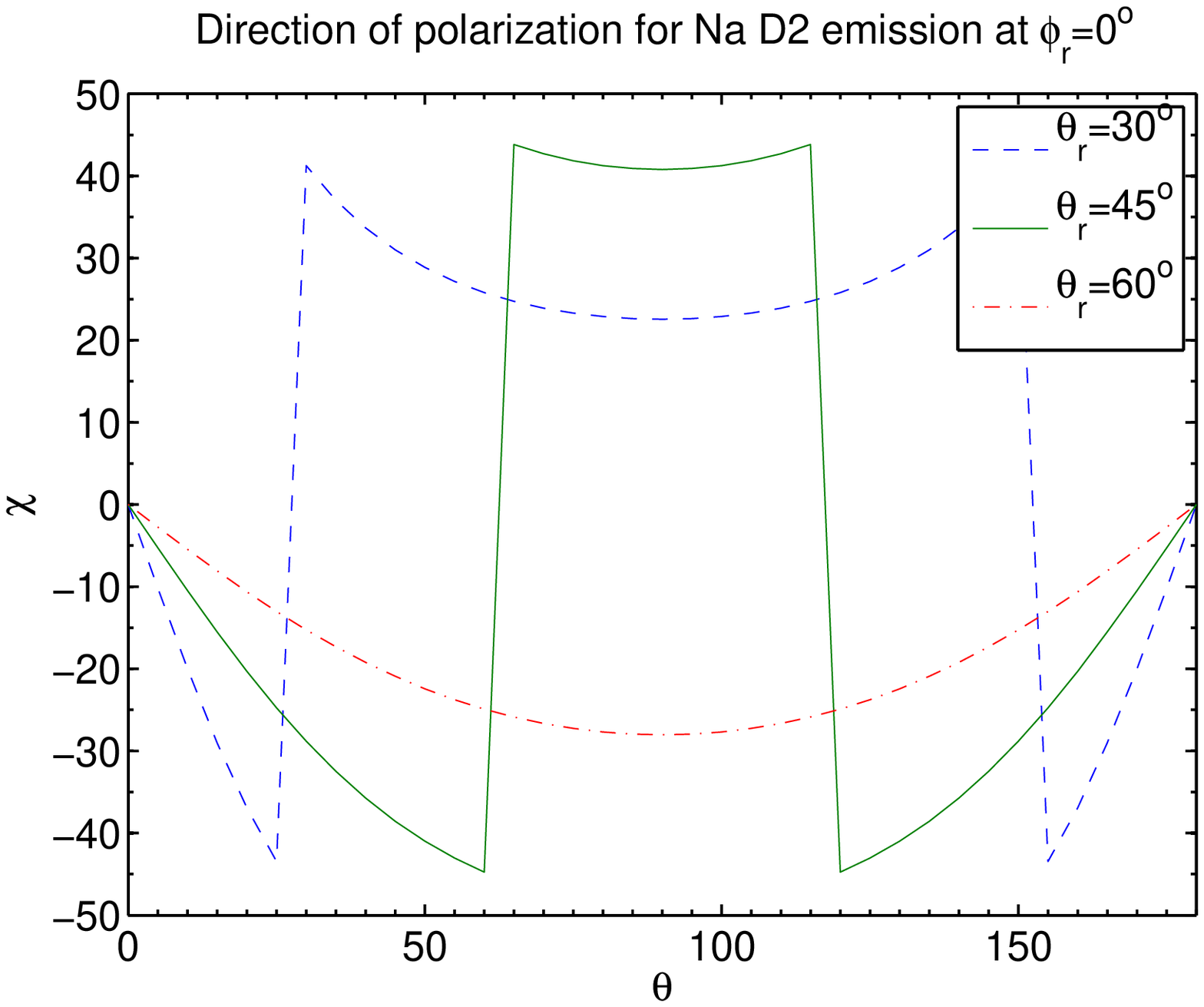}
\caption{Polarizations of Na D2 emission line and their dependence on ($\theta_r,\theta$) at $\phi_r=90^o$. Upper panels give the degree of polarization of emission line; Lower panels show the positional angle of polarization measured from the plane parallel to magnetic field (see Fig.\ref{nzplane}{\it right}).}
\label{NaD2}
\end{figure}

\begin{figure}
\includegraphics[%
  width=0.45\textwidth,
  height=0.3\textheight]{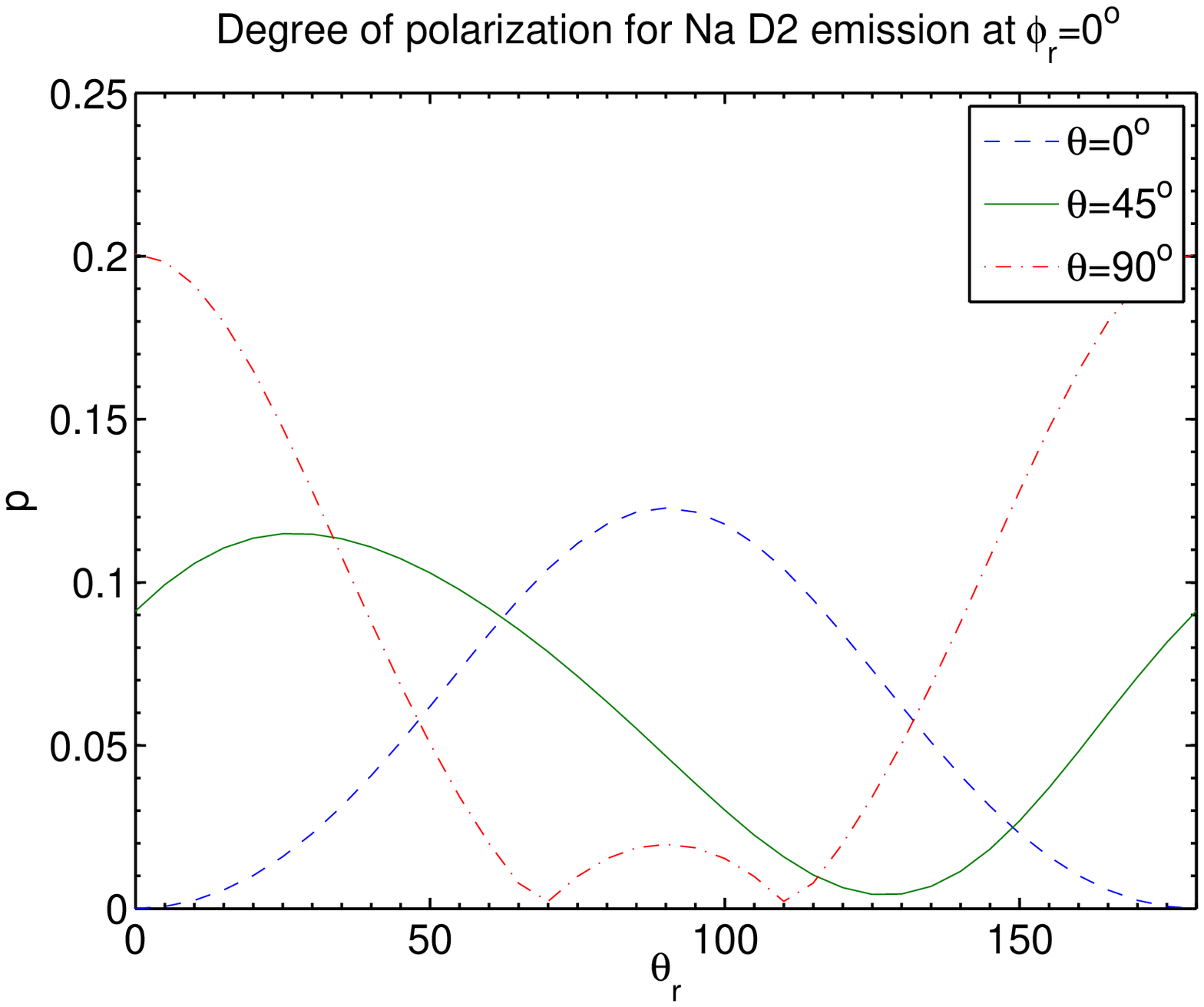}\includegraphics[%
  width=0.45\textwidth,
  height=0.3\textheight]{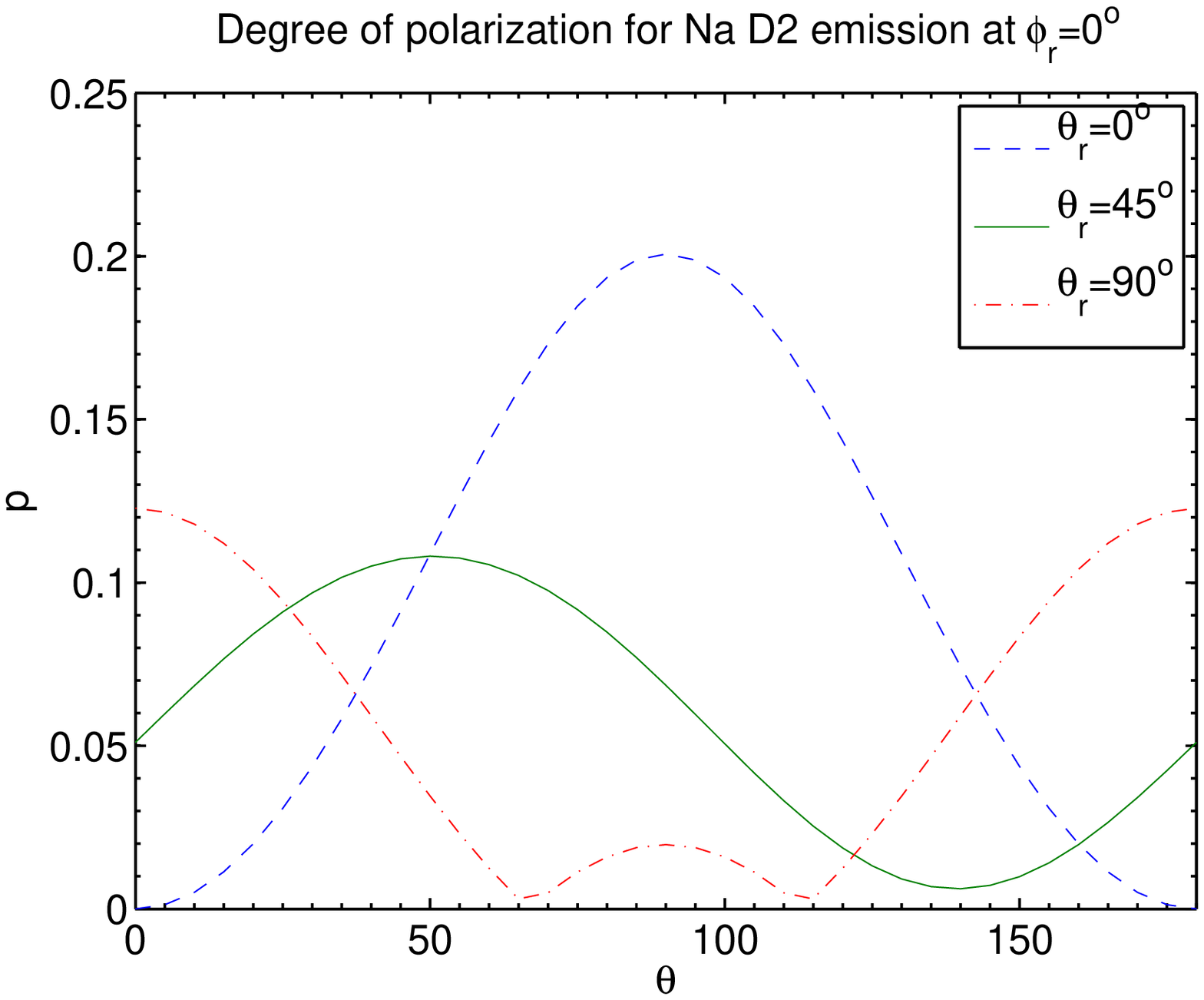}
\caption{Degree of polarizations of Na D2 emission line and their dependence on ($\theta_r,\theta$) at $\phi_r=0^o$. The positional angle of polarization is zero as the Stokes parameter U=0 (see Fig.\ref{Napolcontour} and the text).}
\label{NaD290}
\end{figure}

\begin{figure}
\includegraphics[%
  width=0.45\textwidth,
  height=0.3\textheight]{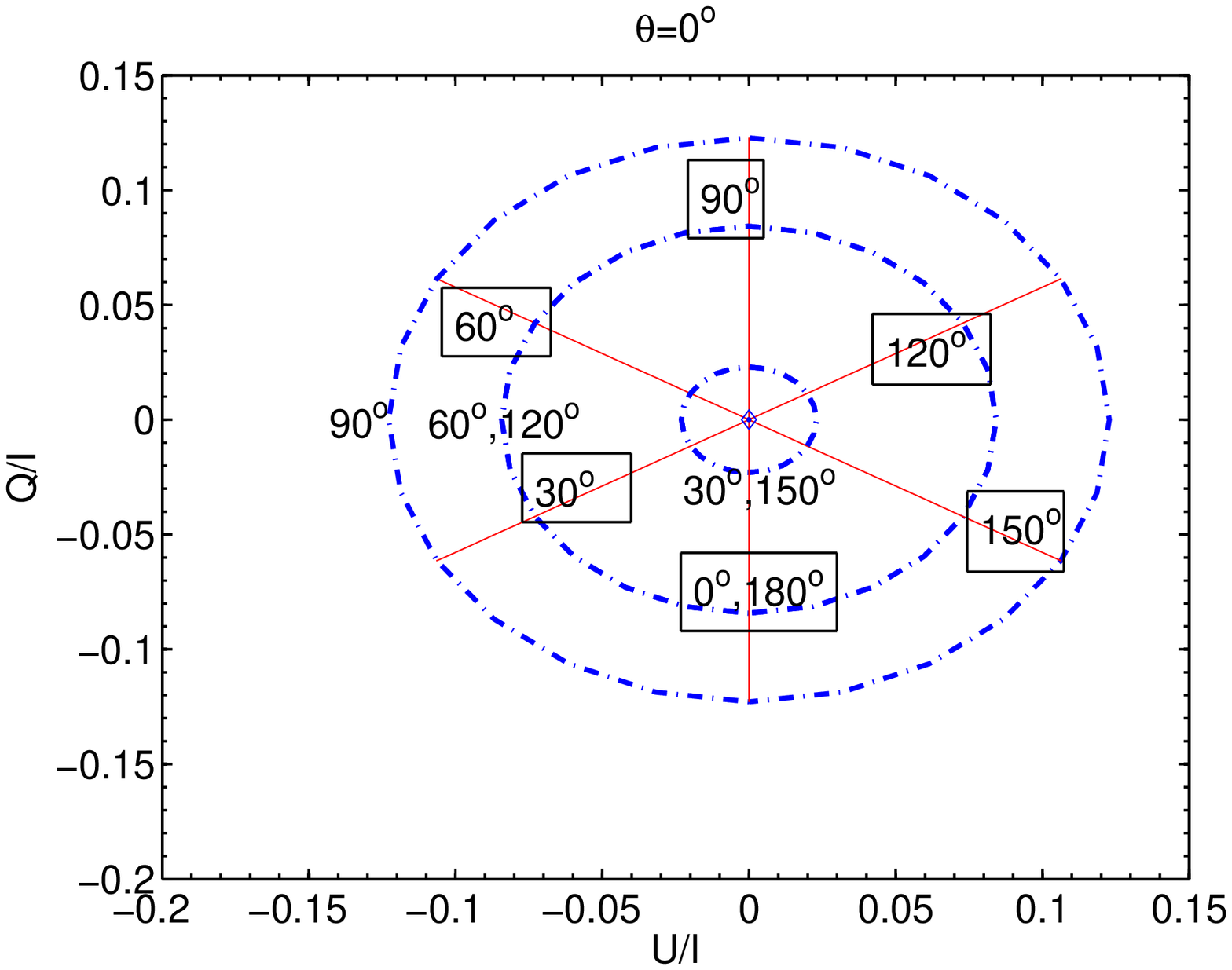}
\includegraphics[%
  width=0.45\textwidth,
  height=0.3\textheight]{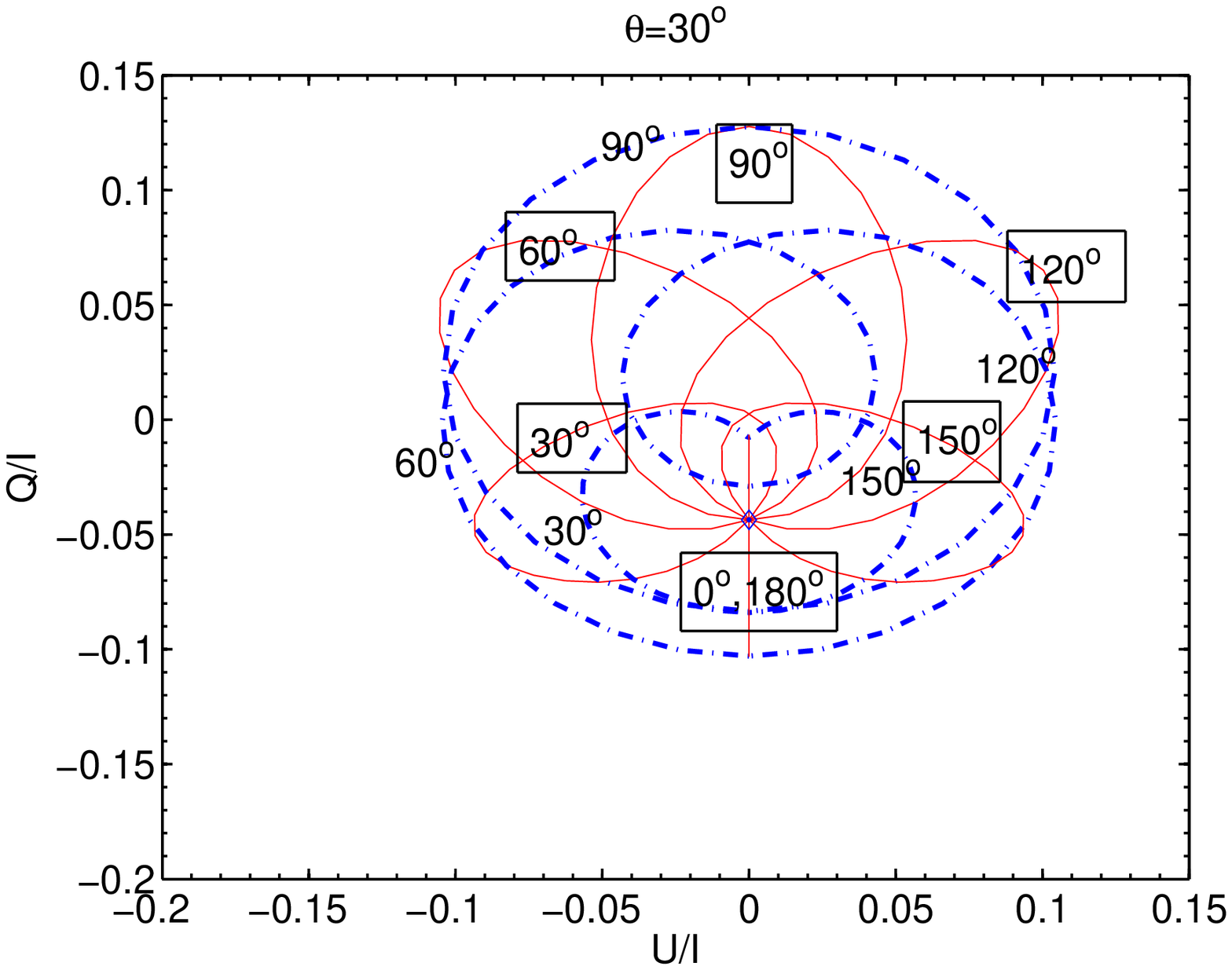}
\includegraphics[%
  width=0.45\textwidth,
  height=0.3\textheight]{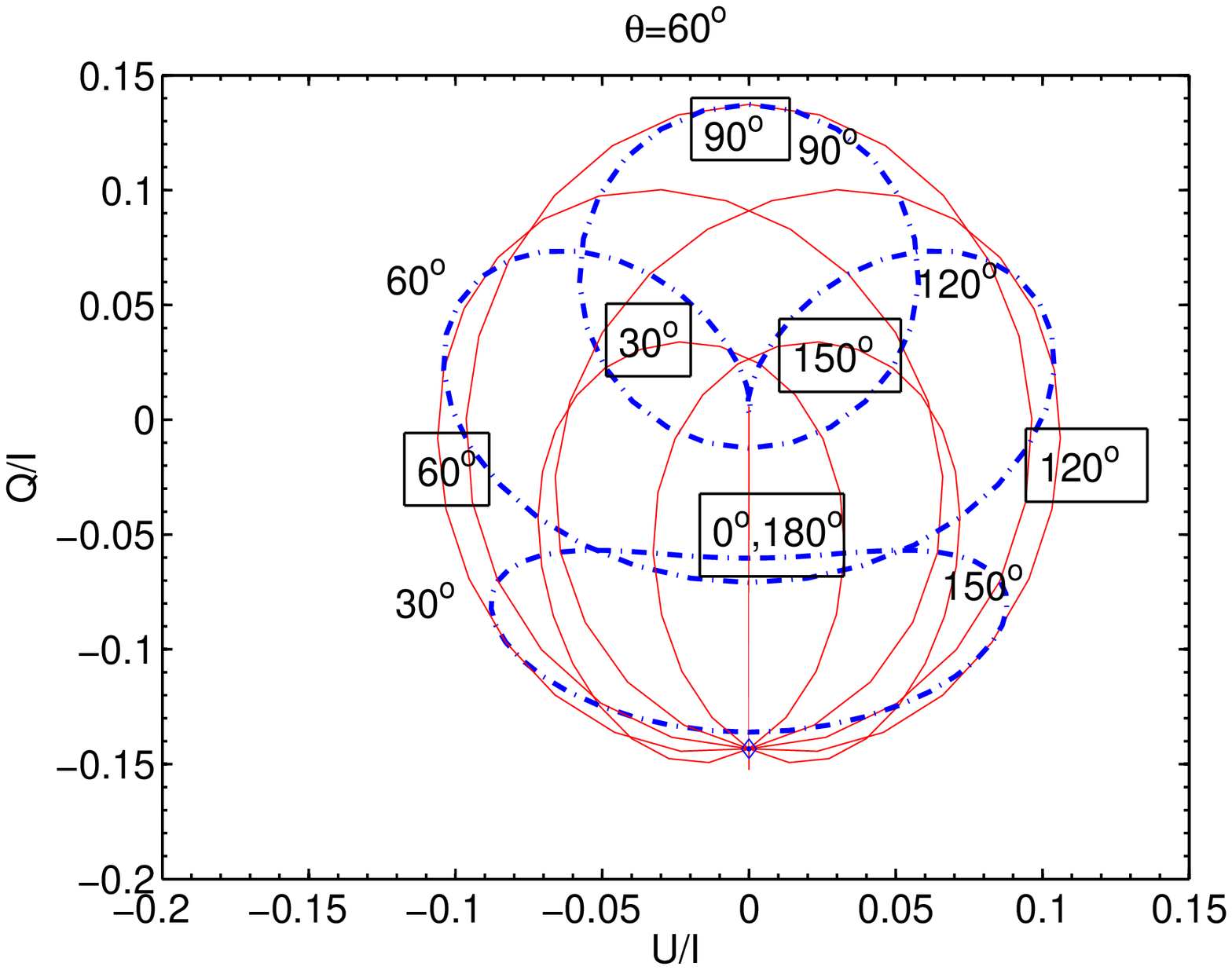}
\includegraphics[%
  width=0.45\textwidth,
  height=0.3\textheight]{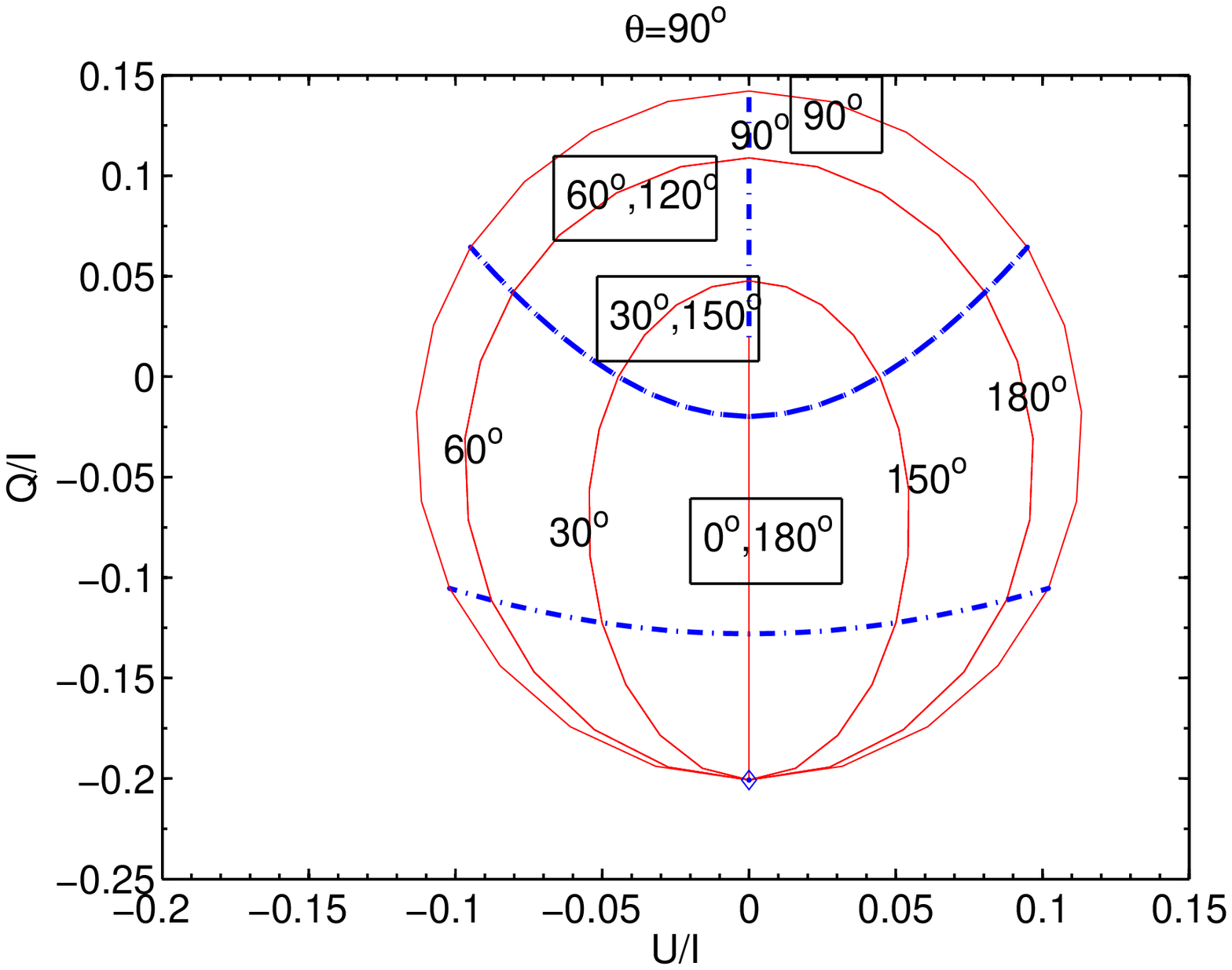}
\caption{Polarization diagrams (or Hanle diagrams) of Na D2 emission line observed at different angles $\theta$. In the diagram, solid lines represent contour of equal $\phi_r$, while dash-dot lines refer to contour of equal $\theta_r$ (see Fig.\ref{radiageometry}). The numbers in textbox mark the values of $\phi_r$, while the numbers without boxes give the values of $\theta_r$. Note that when $\theta_r=0^o$,\& $180^o$, the corresponding polarizations degenerate to one point in each diagram (marked by $\diamond$).}
\label{Hanle}
\end{figure}

\begin{figure}
\includegraphics[%
  width=0.33\textwidth,
  height=0.22\textheight]{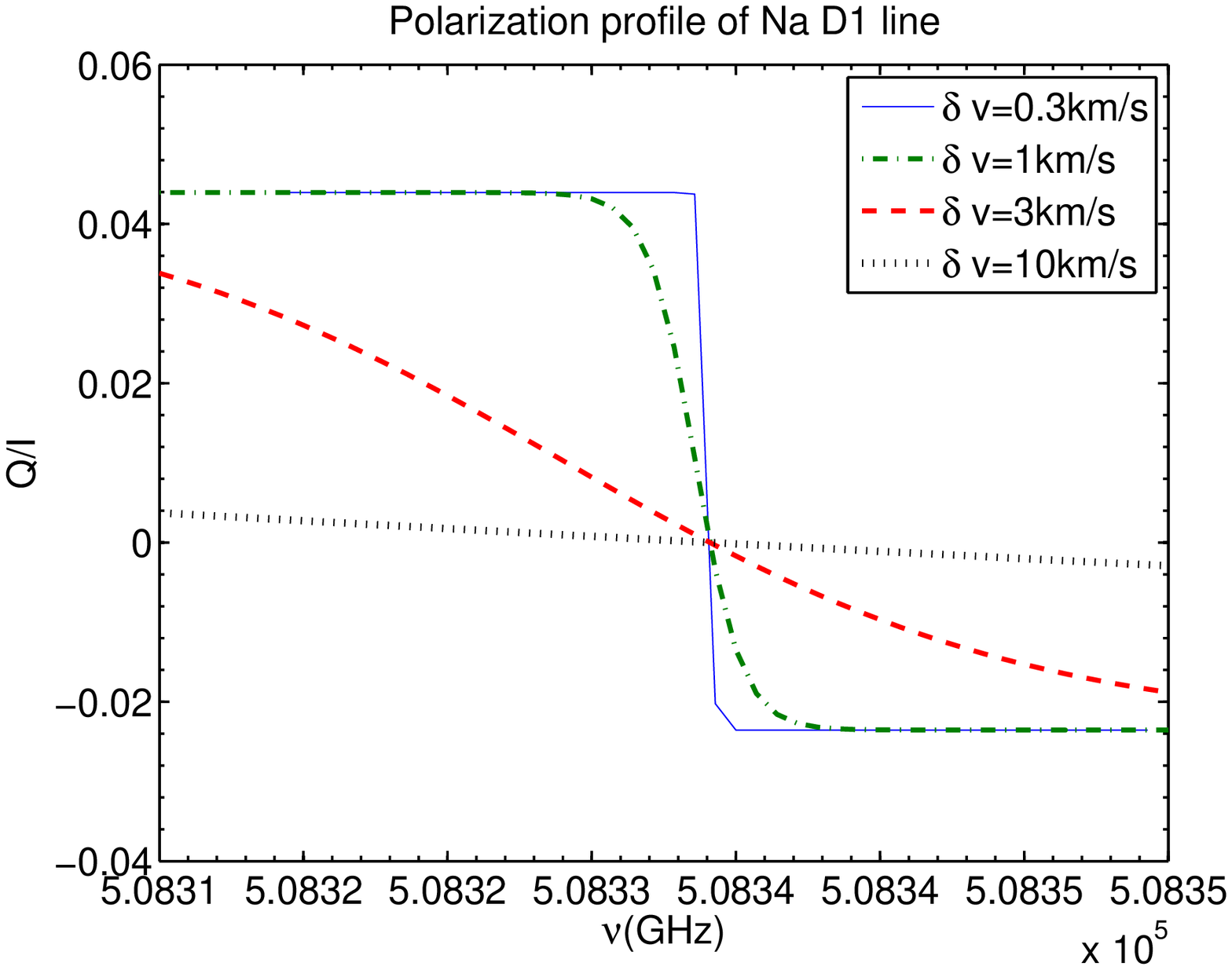}
\includegraphics[%
  width=0.33\textwidth,
  height=0.22\textheight]{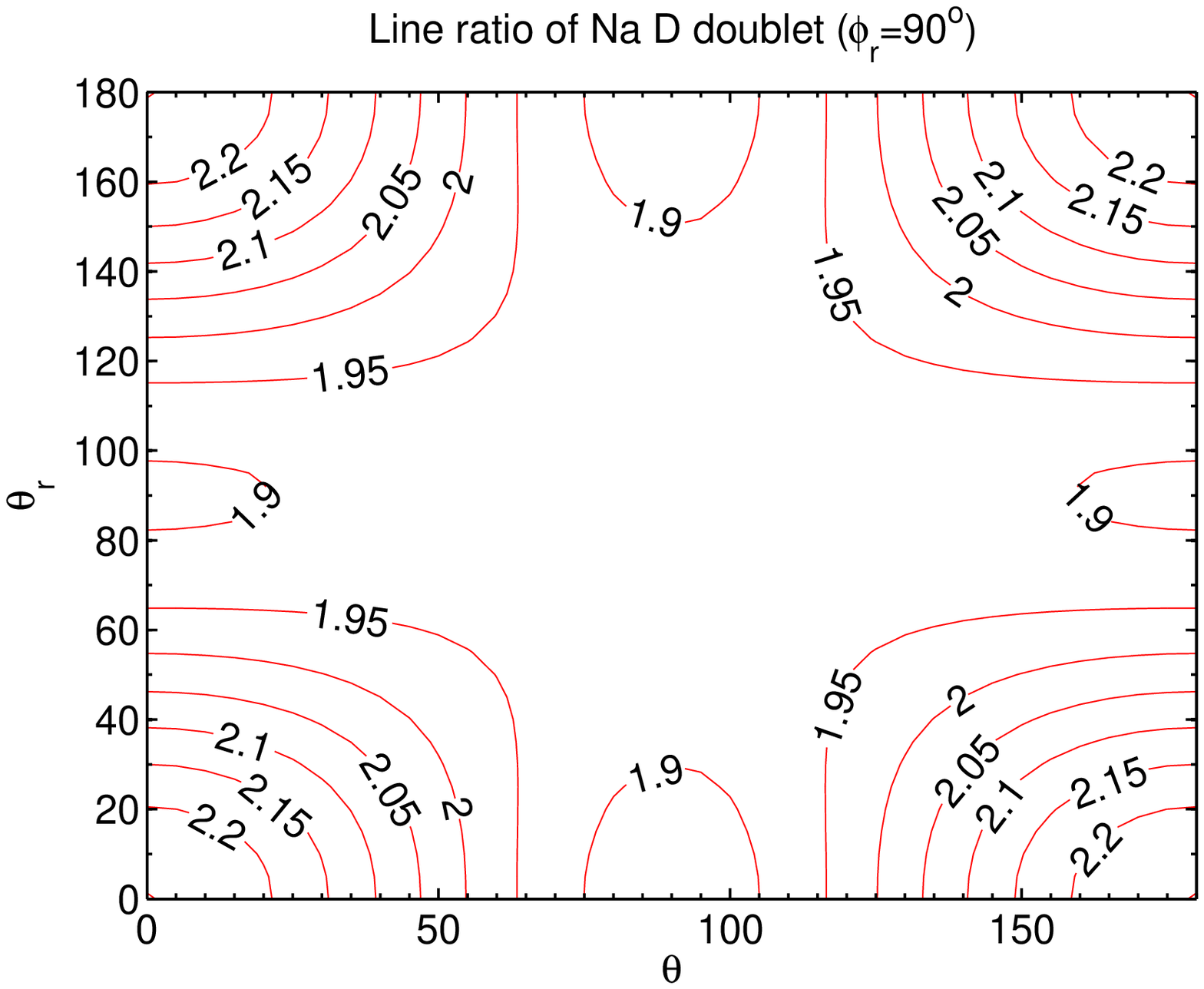}
\includegraphics[%
  width=0.33\textwidth,
  height=0.22\textheight]{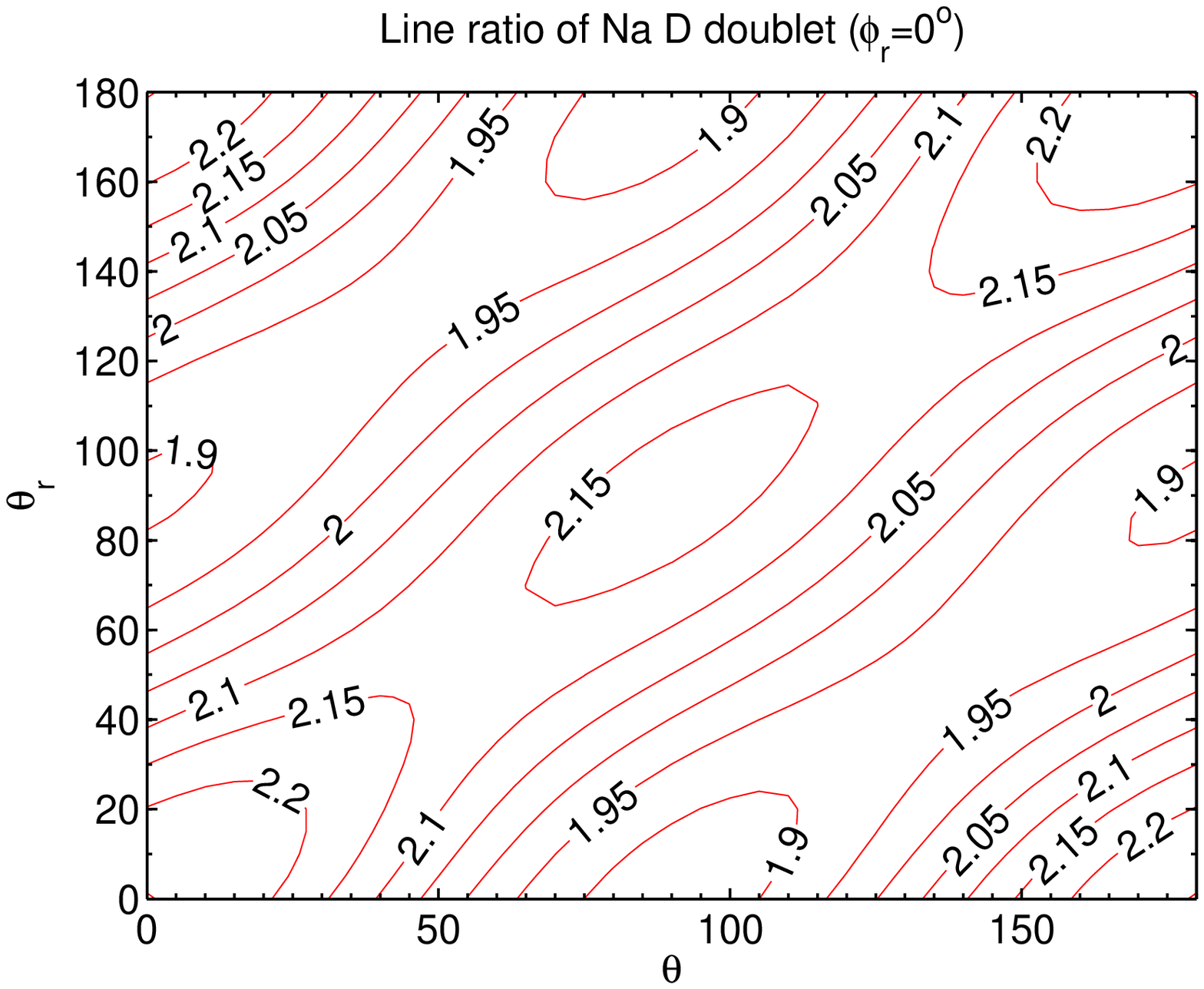}
\caption{{\it Left}: Polarization of Na D1 line at $\theta_r=90^o, \theta=90^o, \phi_r=0^o$. As we see, the polarization degree and the profile of D1 line depends on the line-width. The stokes parameters of the two hyperfine components ($F_l=1,2$, see Fig.\ref{Naocc}{\it left}) are reversal to each other. Thus if the line-width is much wider than their separation ($\sim 1$km/s), their polarizations cancel each other out; {\it Middle \& right}: Contour graphs of the emissivity ratio of Na D doublet. $\phi_r$ is fixed to $\pi/2$ and 0.}
\label{NaD1}
\end{figure}

For unresolved D2 line, the result can be obtained by
 the summation of the two hyperfine components $F_l=1$ and $F_l=2$. The angular dependence of the polarization comes from both $\rho^K_Q(F_u, F'_u)$, the density matrix of the upper level and ${\cal J^K_Q}(i,\Omega)$ (see Eq.\ref{hfemissivity}). While the former one reflects the direction of incident radiation seen in the theoretical frame, the latter is an observational effect which solely determined by line of sight in the theoretical frame.
Fig.\ref{Napolcontour} is a contour graph showing the dependence of polarization on $\theta_r$ and $\theta$ (see Fig.\ref{radiageometry}) with fixed $\phi_r=0^o, 90^o$. Along $\theta_r$ axis the principle harmonic is of order 2 dependence appears as expected from Eq.(\ref{D2F1}, \ref{D2F2}). At $\phi_r=90^o$, $Q/I$ and $U/I$ are shifted with respect to each other in $\theta_r$ by a quarter of their period $180^o$. At $\phi_r=0^o$, U=0, and Q/I is distorted as the phase dependence of $\theta_r$ and $\theta$ is entangled. Since U=0, the polarization always lies in the single plane formed by the radiation source, magnetic field and observer, just as expected from the symmetry of the system (see Fig.\ref{radiageometry}). Fig.\ref{NaD2},\ref{NaD290} are the corresponding plots for degree of polarization $p$ and the positional angle of polarization $\chi$ (see Fig.\ref{nzplane}{\it right}) calculated according to Eq.\ref{pchi}. 

Fig.\ref{Hanle} is the polarization diagram (or Hanle diagram) of Na D2 emission line. Solid lines are the contour of equal $\phi_r$, while dash-dot lines are contour of equal $\theta_r$. For any pair of $\theta_r, \phi_r$, the polarization diagram gives the polarization $Q/I, U/I$. The actual diagram is three-dimensional and its shape depend on the perspective we observe (the angle $\theta$, see Fig.\ref{radiageometry}). We present here its projection at four directions $\theta=0^o, 30^o, 60^o, 90^o$. For $\theta>90^o$, the contour of diagram $\phi_r$ at $\theta$, is the same as that for $180^o-\phi_r$ at $180^o-\theta$; and the contour of diagram of $\theta_r$ is the same as that for $180^o-\theta_r$ at $180^o-\theta$. This  At $\theta=0^o$, the polarization is symmetric about $\theta_r=90^o$, i.e. contour of $\theta_r$ coincides with that of $180^o-\theta_r$. At $\theta=90^o$, the polarization is symmetric about $\theta_r=90^o$. In other cases, the equal-value lines are always increasing clock-wise with anti-symmetry of U about $90^o$. At $\theta_r=0^o, 180^o$, the lines degenerated to a point in every diagram (marked by 'x') as expected. At $\phi_r=0^o,180^o$, $U=0$ and the direction of polarization traces magnetic field in pictorial plane with $90^o$ uncertainty. The lines at $\phi_r=180^o$ are anti-symmetric in respect to $\theta_r=90^o$ with that of $\phi_r=0^o$. These symmetric features are generic and independent of specific species as they are solely determined by the scattering geometry. 

The polarizations of the two hyperfine components ($F_l=1,2$, see Fig.\ref{Naocc}{\it left}) of D1 line are reversal to each other (see Eq.\ref{D1F1},\ref{D1F2}). The polarization of D1 line is thus dependent on the ratio of its line-width and the separation of the two hyperfine components $\sim 1$km/s. As an illustration, Fig.\ref{NaD1}{\it left} shows how the polarization of D1 line with a fixed radiation and magnetic geometry changes with the line-width.

Measurements of the polarization degree of both D lines can constrain up to four parameters, from which we may extract both magnetic field ($\theta_B, \phi_B$) direction and information of radiation source ($\theta_0, W_a/W$). In future, in the situations when we can resolve the hyperfine components of D lines, we can cross-check and make the detection of magnetic field even more accurate. 

\begin{figure}
\includegraphics[%
  width=0.45\textwidth,
  height=0.3\textheight]{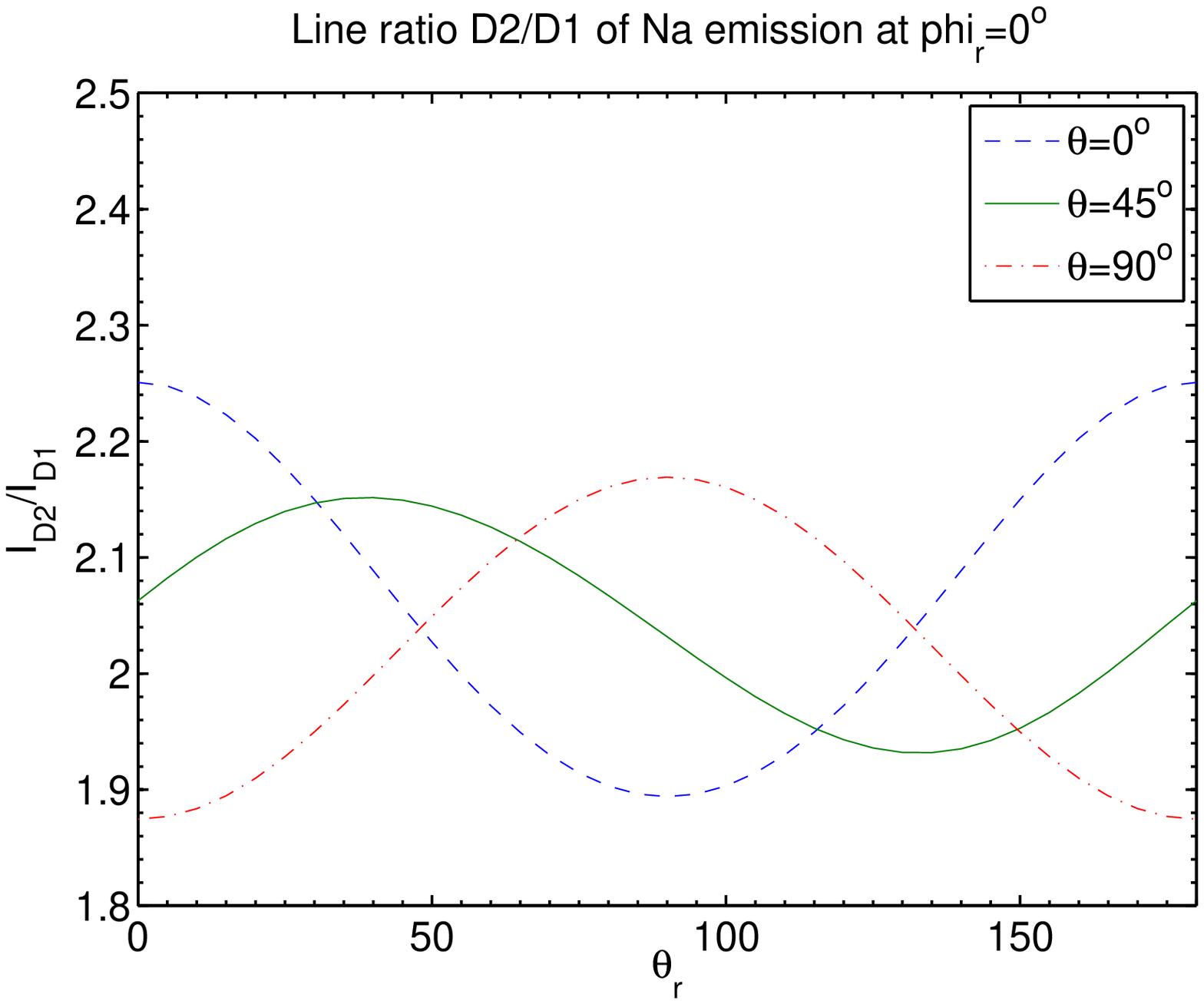}
\includegraphics[%
  width=0.45\textwidth,
  height=0.3\textheight]{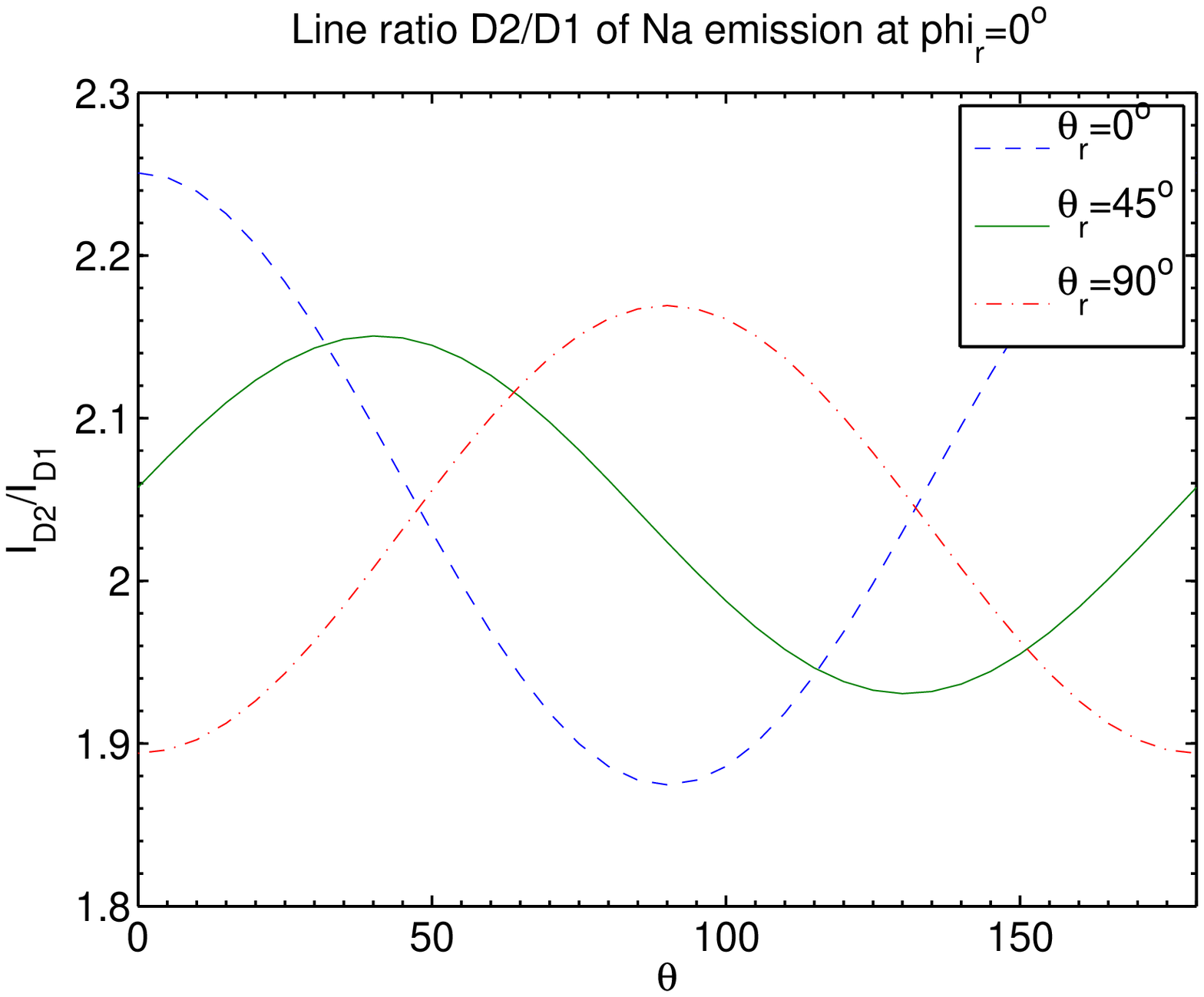}
\includegraphics[%
  width=0.45\textwidth,
  height=0.3\textheight]{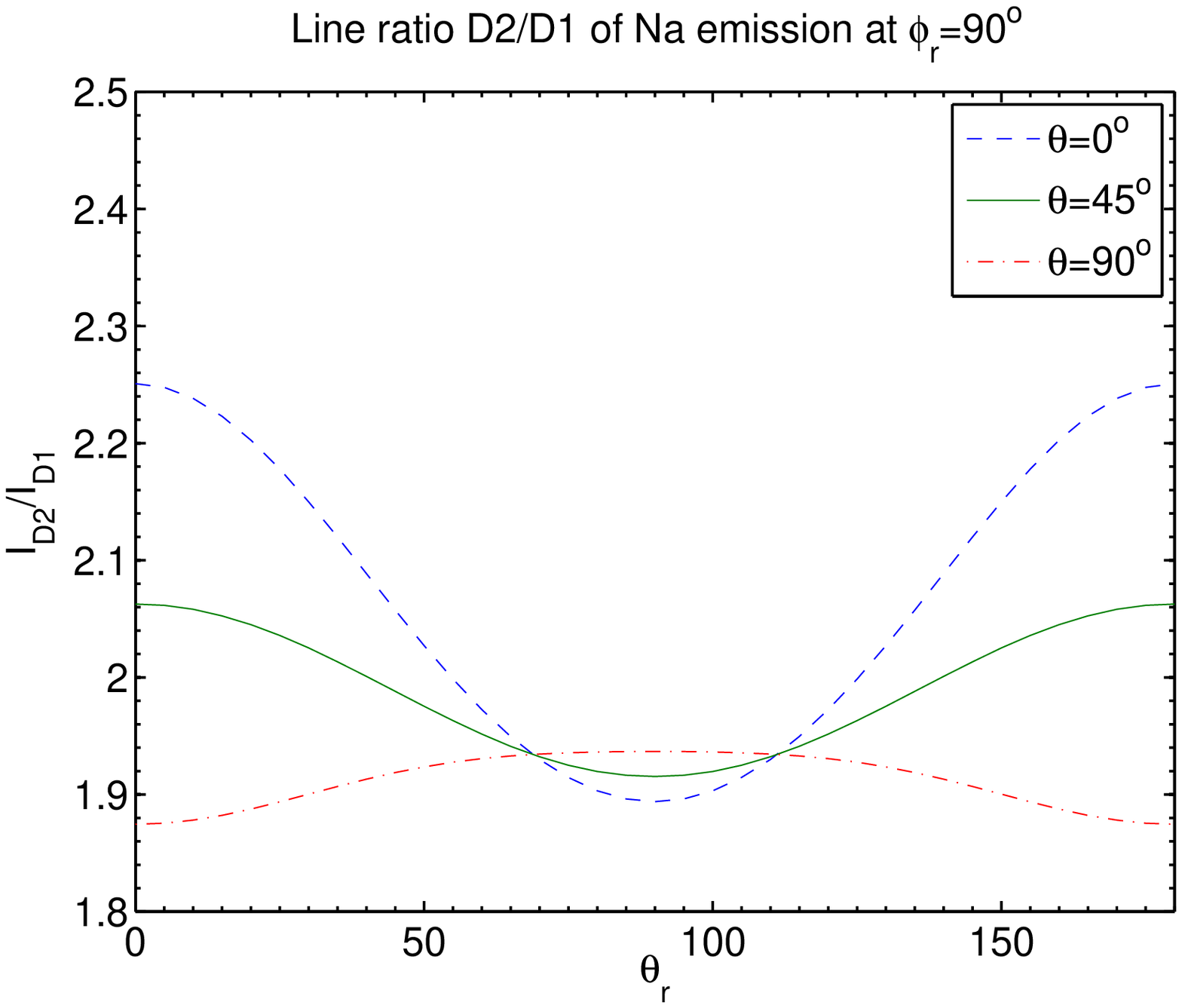}
\includegraphics[%
  width=0.45\textwidth,
  height=0.3\textheight]{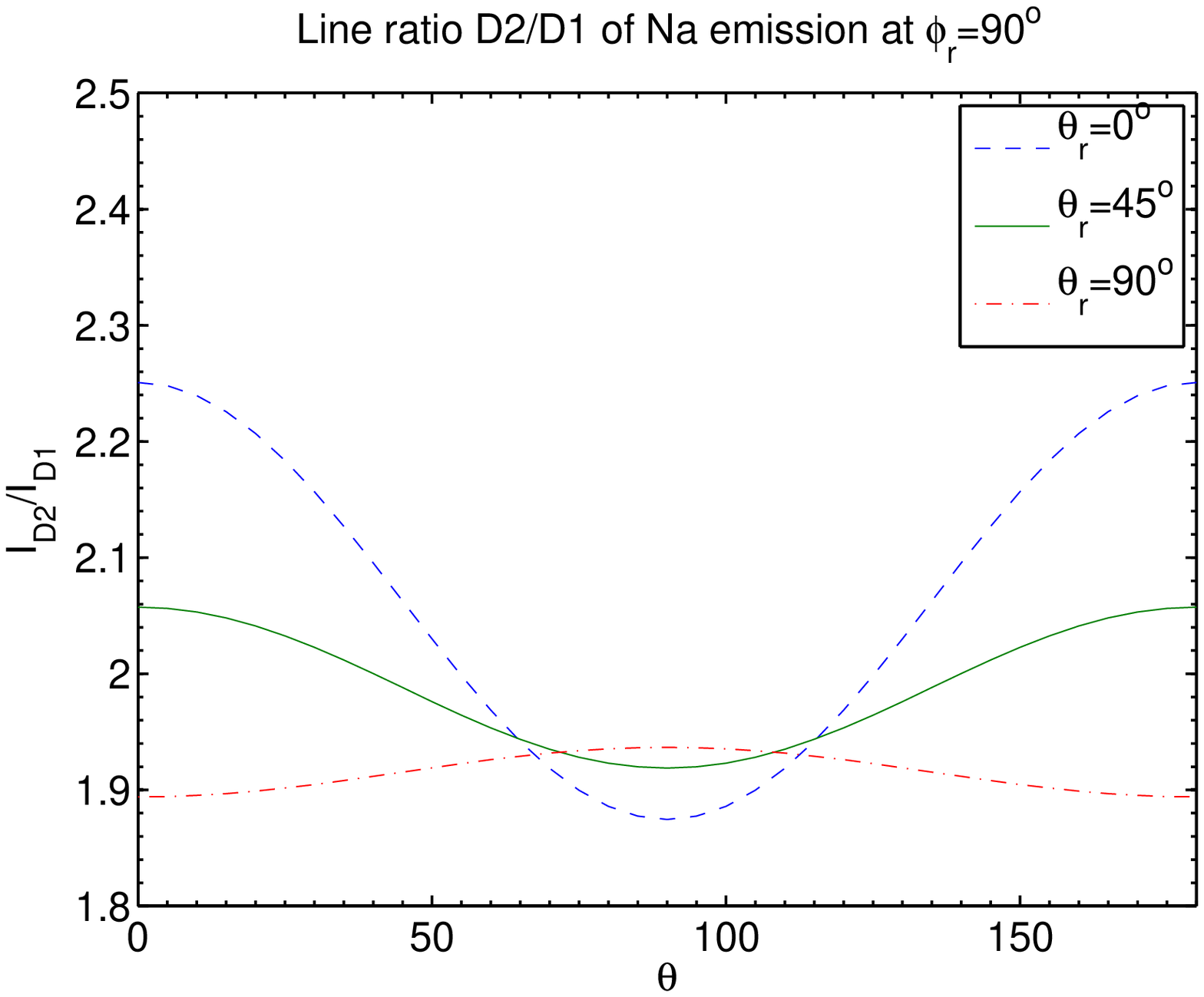}
\caption{The line ratios of Na D2 and D1 lines varies with the orientation of magnetic field. {\it Upper}: at $\phi_r=0^o$; {\it Lower} $\phi_r=90^o$. The corresponding ratios without magnetic realignment would be the curve of $\theta_r=0^o$ (see Fig.\ref{radiageometry}).}
\label{iratio}
\end{figure}

The intensity of the scattered light is also modulated by magnetic realignment. For comparison, we plot the D2/D1 line ratios of their intensities (see Fig.\ref{NaD1}{\it middle \& right}, Fig.\ref{iratio}). The corresponding line ratio without magnetic realignment are equal to those values at $\theta_r=0^o$, where magnetic field is parallel to the incident radiation.

\subsection{Results for KI}

{\bf K I} has the same electron configuration and nuclear spin as Na I does (see Fig.\ref{Naocc}{\it left}). The only difference between them is the coherence between hyperfine sublevels on the excited state $2P_{3/2}$. $\omega_{10}=0.51\gamma,\,\omega_{21}=1.5\gamma,\,\omega_{32}=3.5\gamma$. However, the result for K I only differs from that of Na I by $\lesssim 5\%$. Here we only provide the density tensors of the ground state.  

\bea
\left[\begin{array}{c}
\rho^2_0(F_l=1)\\
\rho^{0,2}_0(F_l=2)
\end{array}\right]&=&\left(\begin{array}{c}0.9 \cos
   ^{10}\theta_r+8.3 \cos
   ^8\theta_r-44.89\cos
   ^6\theta_r-135.8 \cos
   ^4\theta_r+535.8 \cos
   ^2\theta_r-161.9,\\-2.5 \cos ^8\theta_r+15.5\cos
   ^6\theta_r+67.0\cos ^4\theta_r-758.4\cos
   ^2\theta_r+1337.0,\\
-0.01 \cos
   ^{12}\theta_r+0.16 \cos ^{10}\theta_r+6.08
   \cos ^8\theta_r+2.54\cos ^6\theta_r-218.68\cos ^4\theta_r+502.59 \cos
   ^2\theta_r-143.33\end{array}\right)/\nonumber\\
&/&(-1.0    \cos ^{10}\theta_r-4.2 \cos
   ^8\theta_r+58.9 \cos ^6\theta_r-57.1 \cos
   ^4\theta_r-438.0 \cos
   ^2\theta_r+1026.9)
\eea
\be
\rho^{4}_0(F_l=2)\frac{-7.4893 \cos
   ^8\theta_r+91.6742\cos ^6\theta_r-46.534 \cos^4\theta_r+1.5742 \cos
   ^2\theta_r+1.3429}{2.8\cos
   ^8\theta_r-53.3 \cos ^6\theta_r+422.2 \cos
   ^4\theta_r+1320.0\cos ^2\theta_r-5521.9}
\ee
{\it There are a number of parameters that determine the polarization}: direction of magnetic field ($\theta_B,\phi_B$), direction of radiation field ($\theta_0,\phi_0$) and the percentage of radiation anisotropy $W_a/W$. For those cases the radiation source is known, $\phi_0$ can be easily obtained. If we know the distances of the source and atomic cloud, we can also determine $\theta_0$ and $W_a/W$ . Thus we have 2-4 unknown numbers depending on the specific situation. The list of observables we have are degree and direction of polarization, line intensity ratio of the doublet. If the hyperfine separation of the sodium doublet ($\sim 20m\AA$) 
is resolved the four line components are available. In this
case one has enough observables to constrain the 3D direction of magnetic field and the environment in situ.

\subsection{Comparison with earlier works}
 
As we mentioned in the Introduction, the optical pumping of atoms have been studied in laboratory in relation with early day maser research (see Hawkins 1955, Happer 1972). For instance, magnetic mixing of the occupations of Sodium atoms was extensively studied by Hawkins (1955). Their quantitative results, however, is incorrect. This is because of the classical approach they adopted for radiation field. As pointed out in Paper I, because conservation and transfer of angular momentum is the essence of the problem, it is necessary to quantize the radiation field as the atomic states are quantized. Hawkins (1955) claims that the alignment is zero at $\theta_r=90^o$ ($\theta_r$ is the angle between magnetic field and the pumping light). Our results show that the alignment is zero at the Van Vleck angle $54.7^o$. And this is a generic feature for optical pumping of any atoms regardless of their structures (see also Paper I).    

\section{Time-dependent alignment}
\label{noneq}
In this paper we mostly deal with alignment in equilibrium states. 
Astrophysical environments may present us with the cases when there are not enough scattering events
to reach equilibrium. This can happen if either the mean free path of the atom is comparable or larger than the system 
dimension so that atoms are leaving the system before acquiring the steady-state alignment, or if 
the rate of randomizing collisions is comparable to the optical pumping rate. In this case, a collisional term should be added to the statistical equilibrium equations (see App.\ref{collision}). Below we study the alignment of Na I under a given number of scattering events. 

The calculations are straightforward. Initially when no pumping has occurred, the ground state occupation is isotropic. There is only $\rho^0_0$ in this case. Since the energy splitting between the two hyperfine sublevels is negligible, atoms are distributed according to their level statistical weight, 
\bea
\left(\begin{array}{l}\rho_0^0(F_l=1)\\\rho_0^0(F_l=2)\\\rho^2_0(F_l=1)\\\rho^{2}_0(F_l=2)\\\rho^{4}_0(F_l=2)\end{array}\right)\propto\left(\begin{array}{l}\sqrt{3}\\\sqrt{5}\\0\\0\\0\end{array}\right).
\label{iniocc}
\eea
We simply need to find the explicit expression for scattering matrix and then multiply the initial density matrix by it as many times as scattering events. For such purpose, we need to go back to the statistical equilibrium equations (\ref{upevolution},\ref{lowevolution}). The first step, the stimulated emission is described by the second term on the right side of Eq.(\ref{upevolution}). There are two routes then for the excited atoms, either they spontaneously decay (first term on the right side of the same equation) to the ground level or precess under the magnetic field\footnote{Note, that in this paper we discuss a regime for which the
external magnetic field is not strong enough to affect the upper level population.} produced by nuclear spin (the 3rd term on the left side). The probability of spontaneous emission out of the two routes is thus easily obtained by multiplying the first term on the left side of Eq.(\ref{lowevolution}) by $A/(A+2\pi i\nu_{F_uF'_u})$. It turned out that the scattering matrix is the same as the first term in Eq.(\ref{hypground}) apart from the Einstein coefficient B which should be removed as we are concerned about the probability rather than rates. The effect of magnetic field is only to remove the magnetic coherence as explained earlier. We assume for the incident radiation the same intensity at the resonant frequencies of D lines. Then averaging over both D lines, we obtain the scattering matrix 

\begin{figure}
\includegraphics[%
  width=0.45\textwidth,
  height=0.3\textheight]{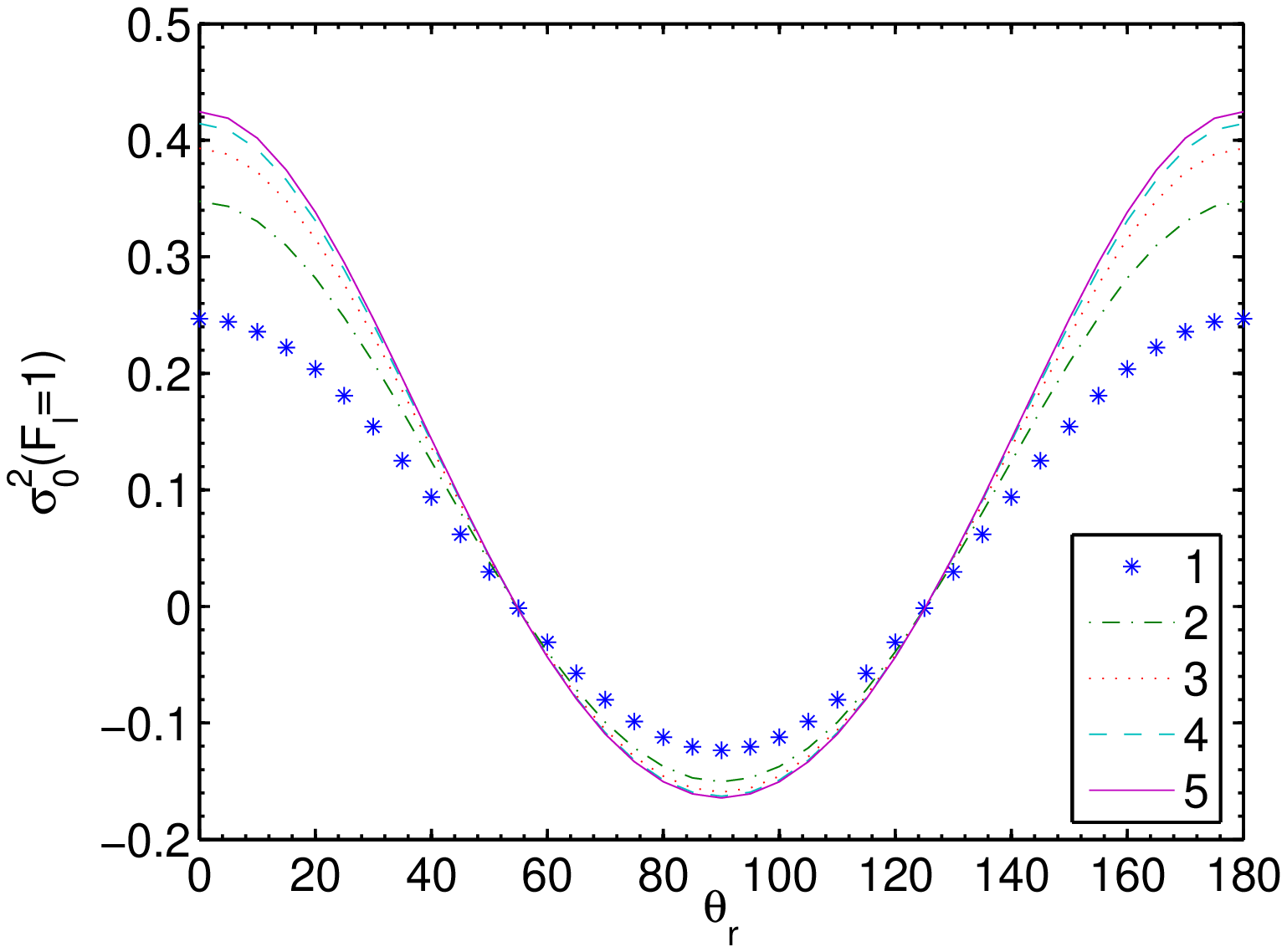}
\includegraphics[%
  width=0.45\textwidth,
  height=0.3\textheight]{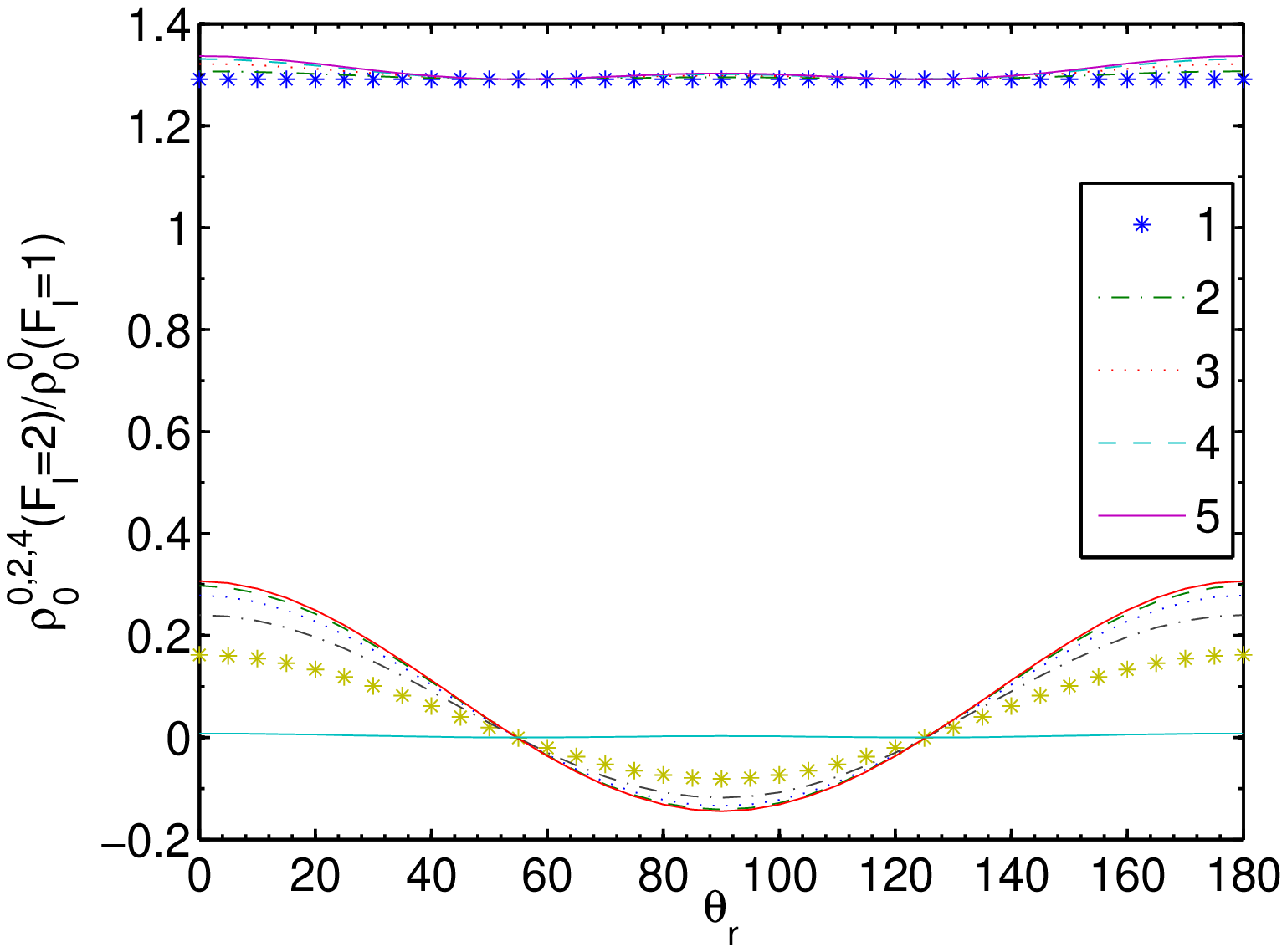}
\caption{The density tensors of Na ground state for nonequilibrium cases. The degree of alignment $\rho^2_0$ increases with the number of scattering events. Every scattering event is described by an statistically averaged scattering matrix (see text).}
\label{nonsatocc}
\end{figure}

{\scriptsize
\bea
\left(
\begin{array}{lllll}
 1.09& 0.497 &
   0.0340 \cos
   ^2\theta_r-0.0113&
   0.052-0.156 \cos
   ^2\theta_r & 0 \\
 0.497& 1.35 &
   0.00878-0.0264 \cos
   ^2\theta_r & 0.121\cos
   ^2\theta_r-0.0403 & 0 \\
 0.279 \cos
   ^2\theta_r-0.0929&
   0.115 \cos
   ^2\theta_r-0.0383 &
   0.190 \cos
   ^2\theta_r+0.225 &
   0.105\cos
   ^2\theta_r+0.206 &
   0.0219-0.0656 \cos
   ^2\theta_r \\
 0.0738 \cos
   ^2\theta_r-0.0246&
   0.269\cos
   ^2\theta_r-0.0896&
   0.0879-0.0315 \cos
   ^2\theta_r & 0.0602 \cos
   ^2\theta_r+0.699&
   0.136\cos
   ^2\theta_r-0.0453\\
 0 & 0 & 0.0845-0.254 \cos
   ^2\theta_r & 0.294\cos
   ^2\theta_r-0.0981&
   0.165 \cos
   ^2\theta_r+0.22
\end{array}
\right)
\label{scattmx}
\eea}
For a given number of scattering events, the density matrix of the ground level can be obtained by multiplying Eq.\ref{iniocc} the scattering matrix Eq.\ref{scattmx} as many times as the number  of scattering events. The results are shown in Fig.\ref{nonsatocc}. We see after $\sim 5$ scattering events, the system approaches its equilibrium. The scattering time scale is $B_{lu}{\bar J}^0_0\sim 5\times 10^{-10}[R(R_\sun)/r(pc)]^2$. In contrast, the collisional transition rate is $\sim 10^{-14}n_{e}{\rm cm}^2$ (Happer 1972). In this sense, collisions can be neglected if $r\ll 1pc$ for a solar type star. In the case where collisions are not negligible, the effect of collisions would be to limit the number of scattering events.

We also obtained the polarization of emission and absorption from atoms scattered one photon. The shape of polarization curves for all the lines are very similar to the equilibrium cases. In fact, the curves for positional angle of polarization are the same for all the cases. In this sense, direction of polarization is a more {\it robust measure} for the detection of magnetic field. Only the amplitude of degree of polarization is decreased from the equilibrium case by $\sim \%10$.

\section{Example: synthetic observation of comet wake}
\label{cometwake}
As an illustration, we discuss here a synthetic observation of a comet wake. 
Though the abundance of sodium in comets is very low, its high efficiency of
 scattering sunlight makes it a good tracer 
(Thomas 1992). It was suggested by Cremonese \& Fulle (1999) there are two categories of sodium tails. Apart from the {\it diffuse} sodium tail superimposed on dust tail, there is also a third {\it narrow} tail composed of only neutral sodium and well separated from dust and ion tails. This neutral sodium tail is characterized by fast moving atoms from a source inside the nuclear region and accelerated by radiation pressure through resonant D line scattering. While for the diffuse tail, sodium are considered to be released in situ by dust, it is less clear for the second case. Possibly the fast narrow tail may also originate from the rapidly fragmenting dust in the inner coma (Cremonese et al. 2002). 

\begin{figure}
\plottwo{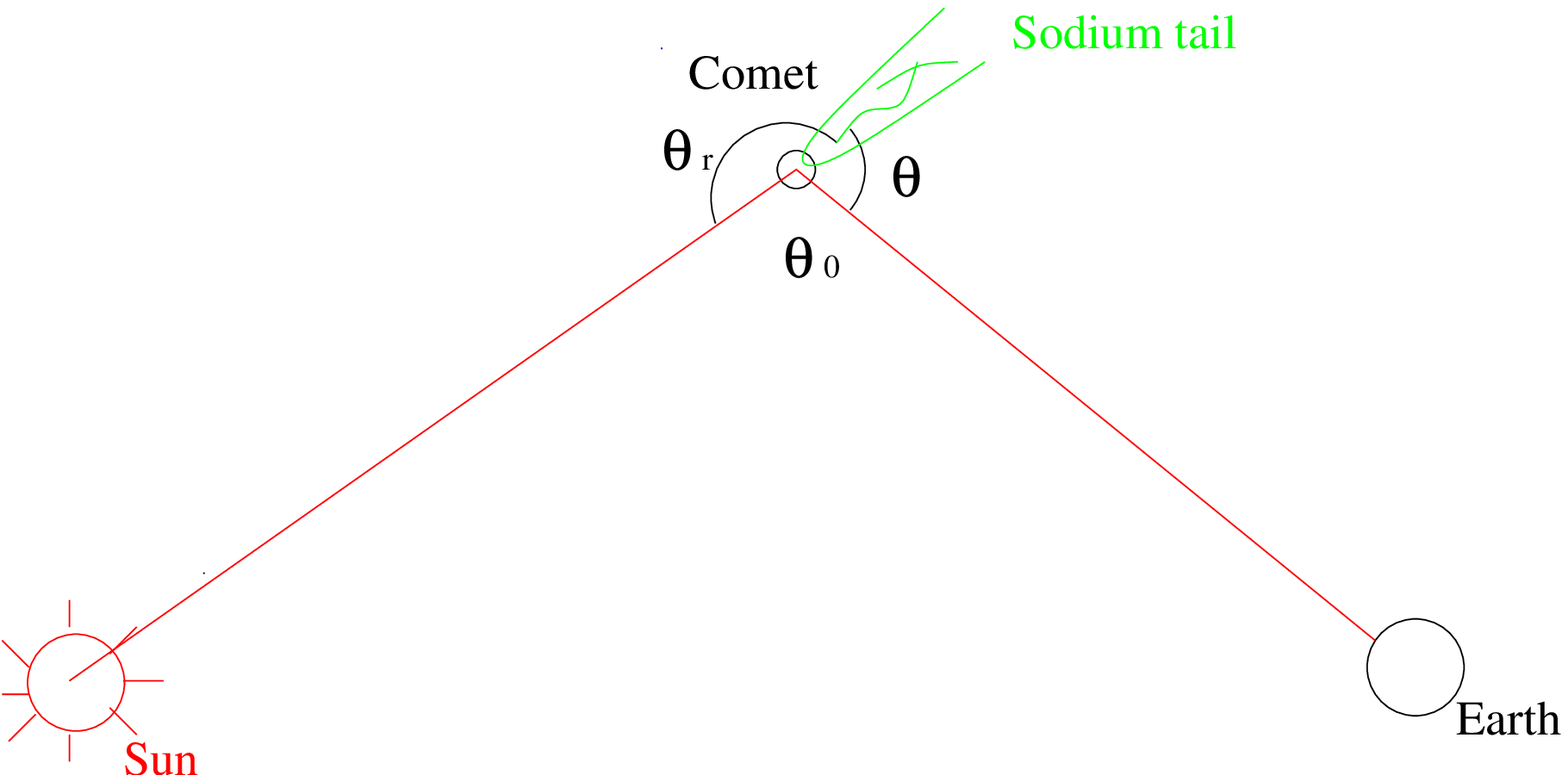}{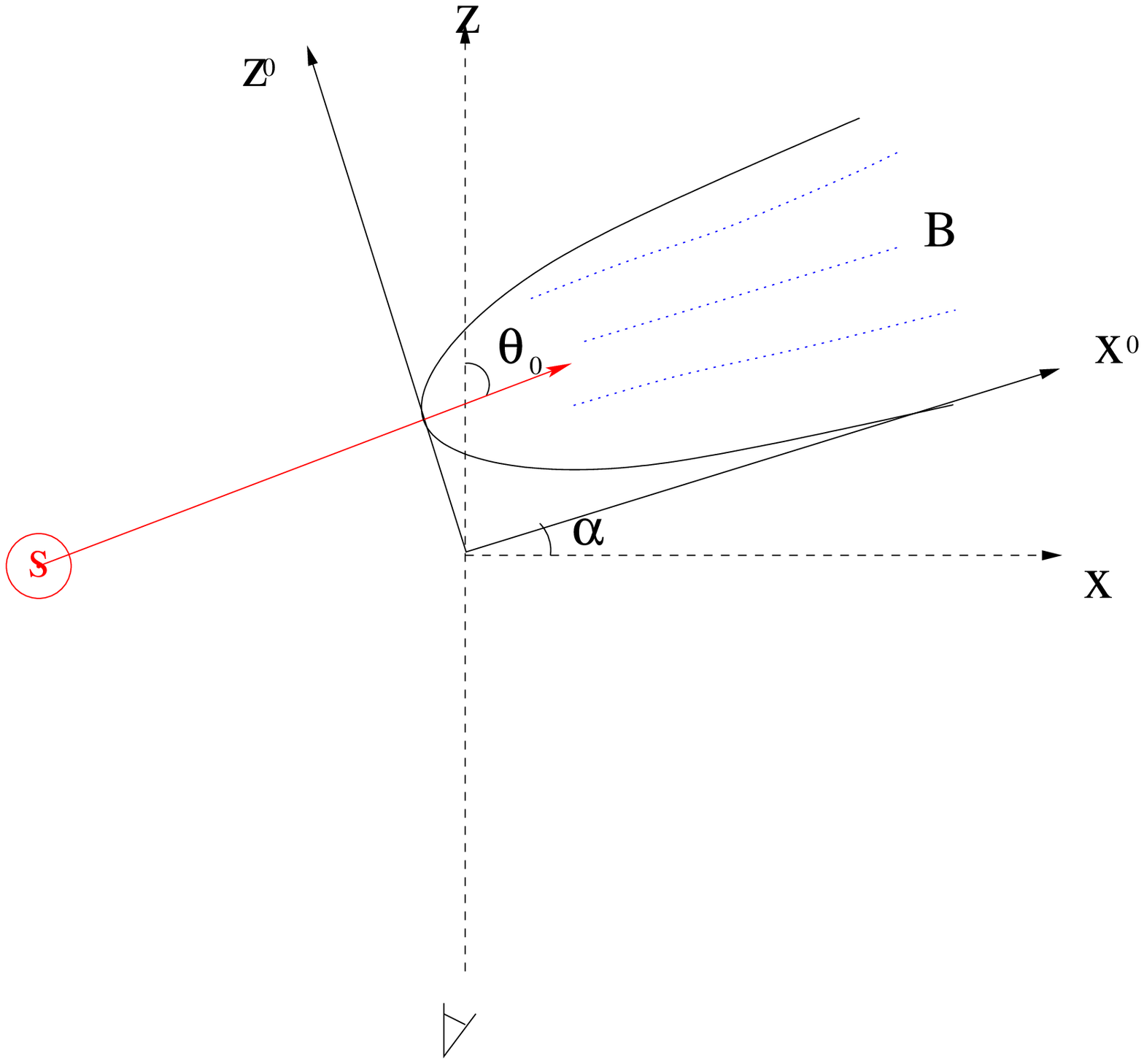}
\caption{{\it Left}: Resonance scattering of solar light by sodium tail from comet;{\it Right}: geometry relation of the observational system.} 
\label{comet}
\end{figure}

\begin{figure}
\includegraphics[%
  width=0.45\textwidth,
  height=0.3\textheight]{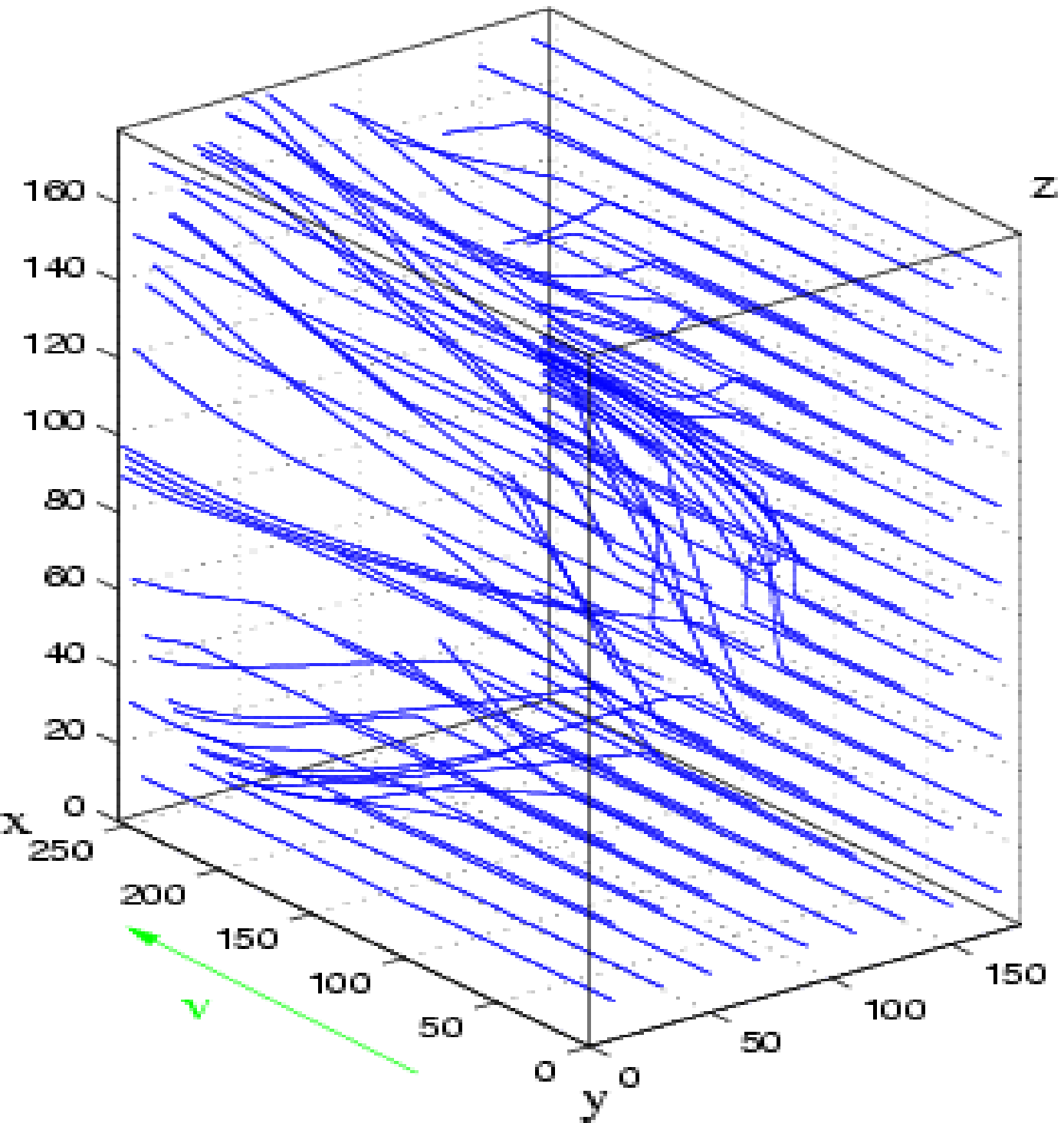}
\includegraphics[%
  width=0.45\textwidth,
  height=0.3\textheight]{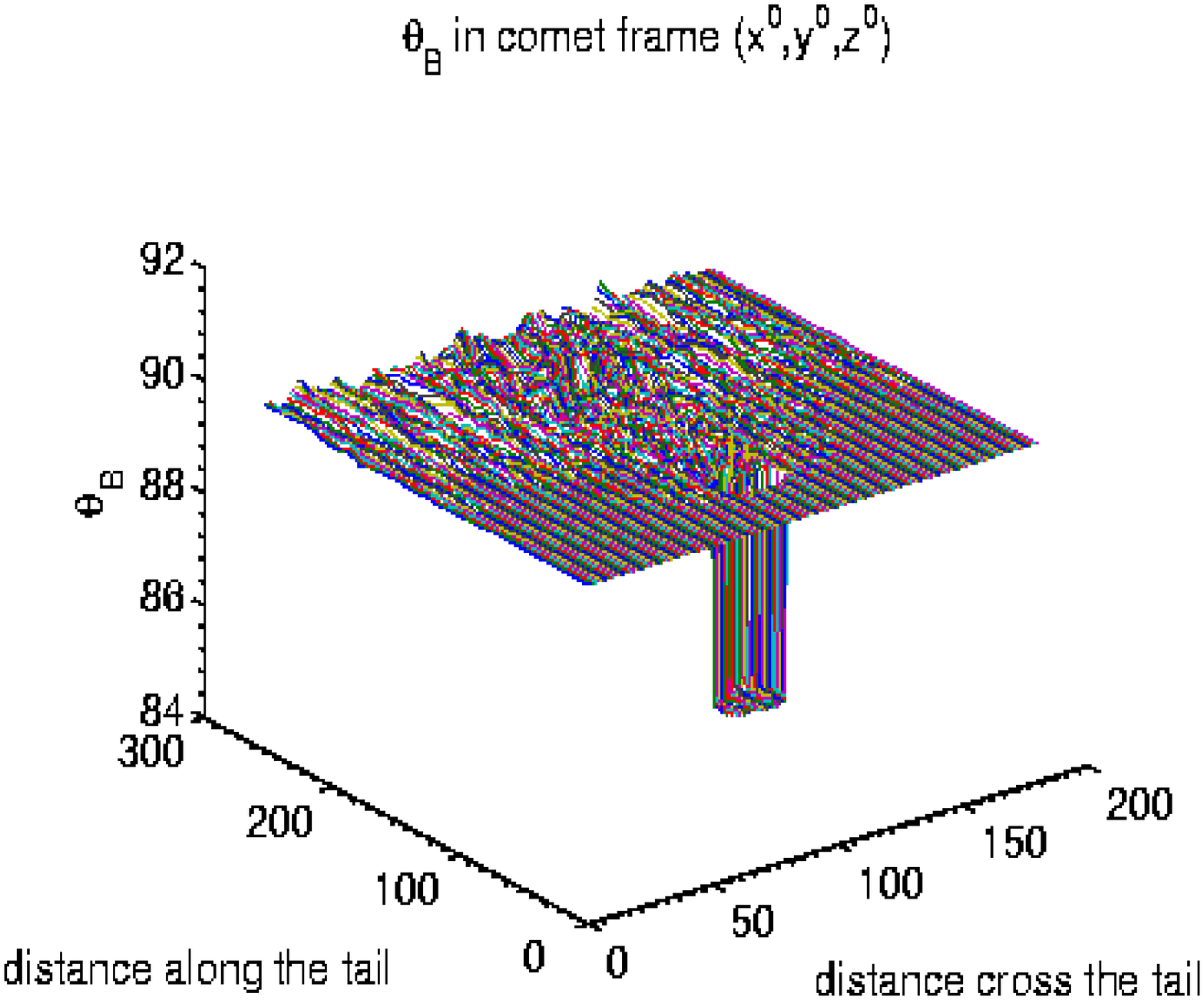}
\caption{{\it Left}: Simulated magnetic field distribution in the comet wake; {\it right}: the direction of magnetic field seen in the comet frame $x^0,y^0,z^0$ (see Fig.\ref{comet}).}
\label{MHD}
\end{figure}

\begin{figure}
\includegraphics[%
  width=0.45\textwidth]{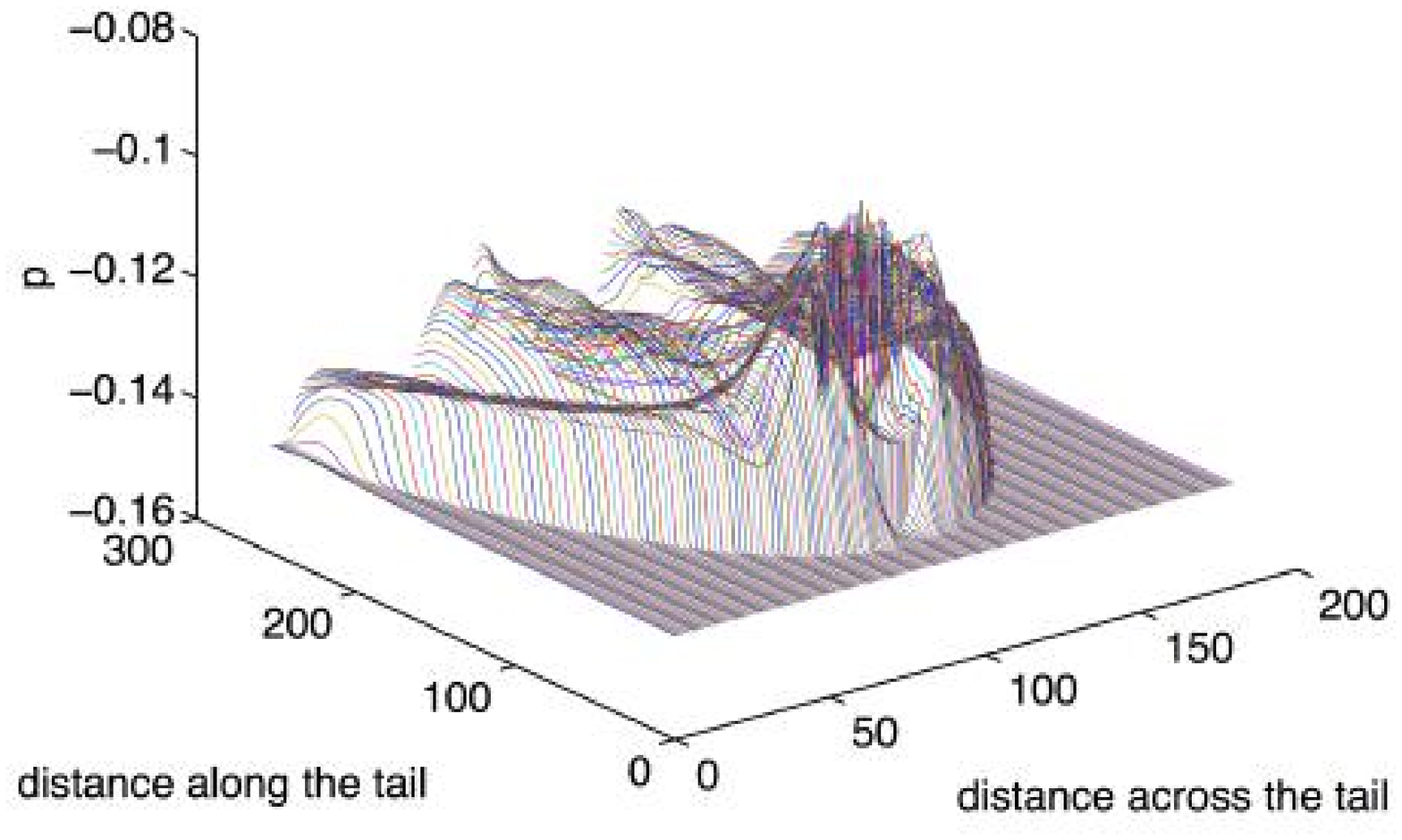}
\includegraphics[%
  width=0.45\textwidth]{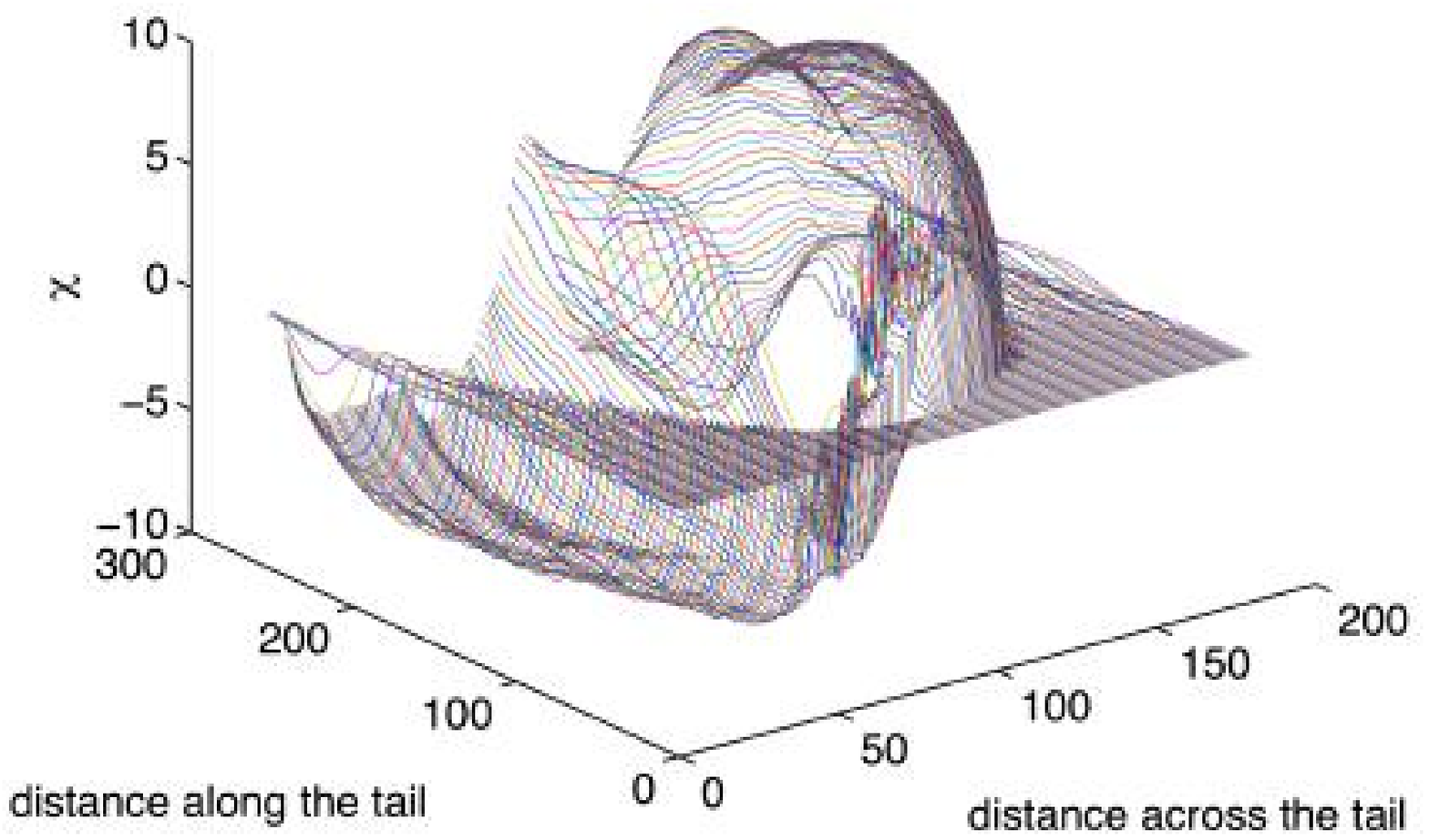}
\caption{Polarization caused by sodium aligned in the comet wake ($\alpha=0^o$, Fig.\ref{comet} {\it right}). Spatial and temporal fluctuations of polarization carry the information on MHD turbulence. Degree and direction of polarization for D2 emission and absorption lines. }
\label{norm}
\end{figure}

\begin{figure}
\includegraphics[%
  width=0.45\textwidth]{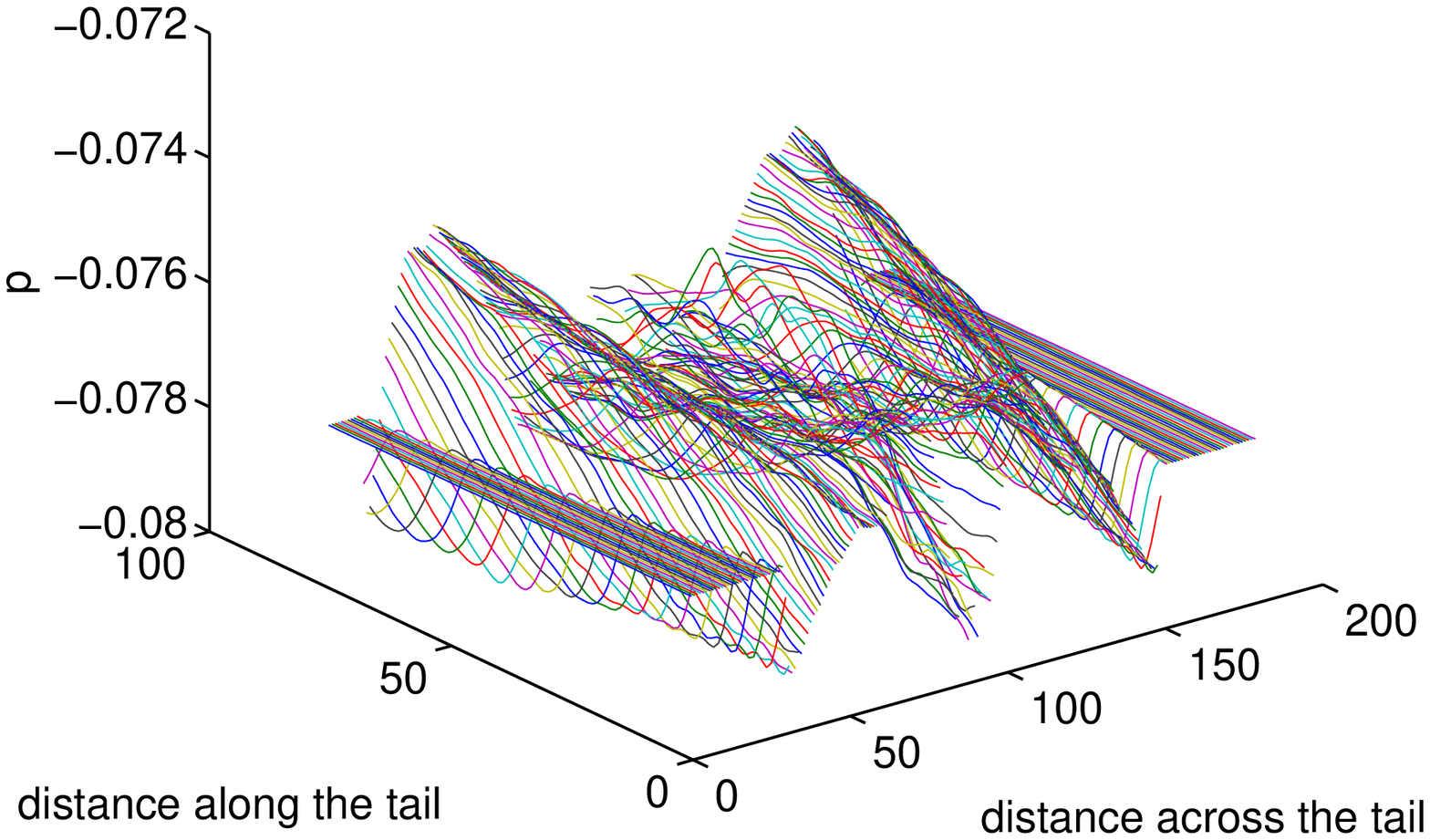}
\includegraphics[%
  width=0.45\textwidth]{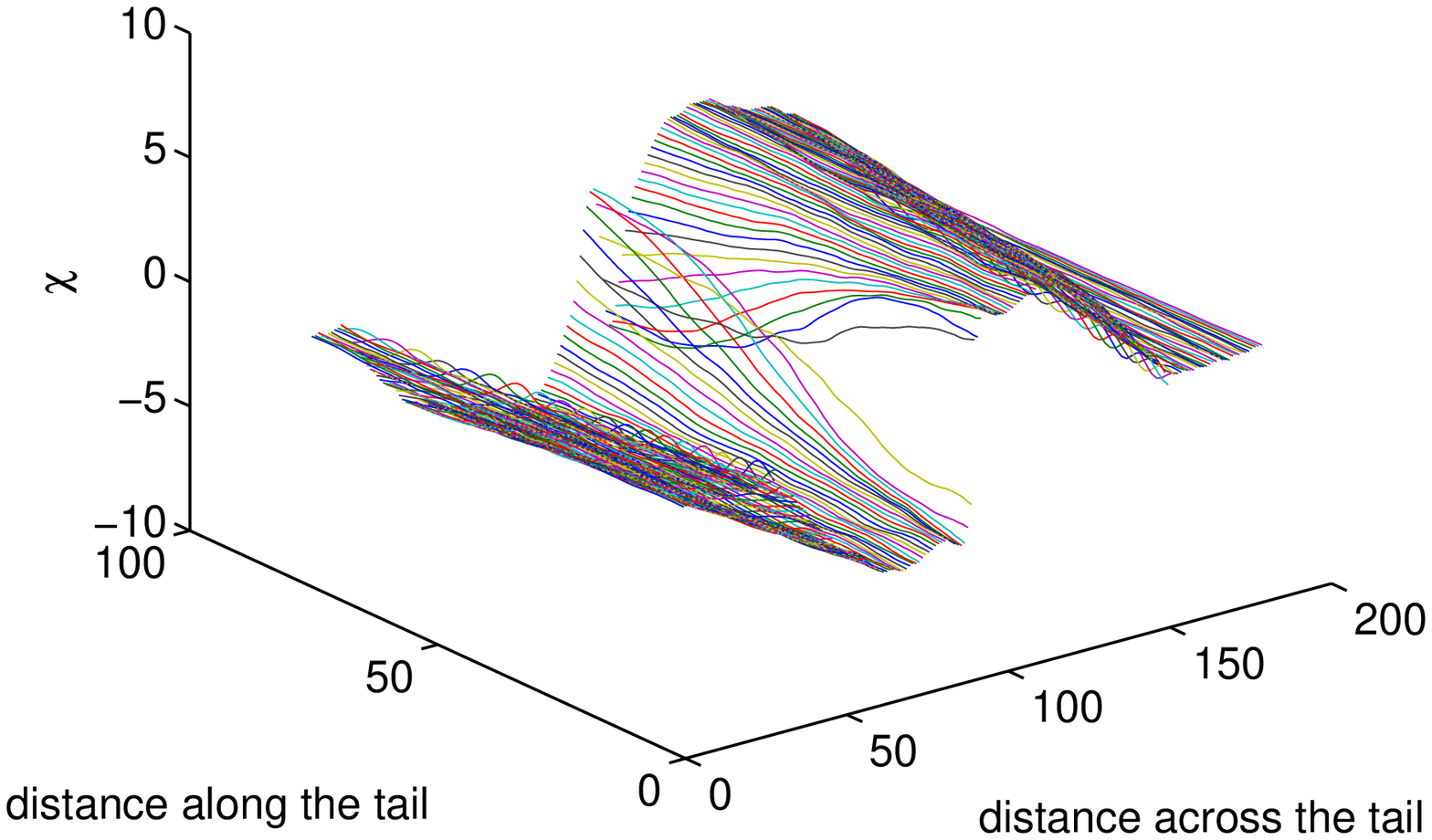}
\caption{Same as Fig.\ref{norm}, but for the case when comet tail has an inclination ($\alpha=45^o$, Fig.\ref{comet} {\it right}) with the plane of sky. As we see, the pattern is completely different for a different inclination.}
\label{inclined}
\end{figure}

The gaseous sodium atoms in the comet tail acquires not only momentum, but also angular momentum from the solar radiation, i.e. they are aligned.  Distant from comets, the Sun can be 
considered a point source. As shown in Fig.\ref{comet}, the geometry of the 
scattering is well defined, i.e., the scattering angle $\theta_0$ is known. 
The polarization of the sodium emission thus provides an exclusive 
information of the magnetic field in the comet wake.  Embedded in Solar 
wind, the magnetic field is turbulent in a comet wake. We take a 
data cube (Fig.\ref{MHD}) from MHD simulations of a comet wake. Depending on its 
direction, the embedded magnetic field alter the degree of 
alignment and therefore polarization of the light scattered by the aligned atoms. Therefore, fluctuations in the linear polarization are expected from 
such a turbulent field (see Fig.\ref{norm},\ref{inclined}). The calculation is done for the equilibrium case. If otherwise, the result for degree of polarization will be slightly different ($\lesssim \%10$) depending on the number of scattering events experienced by atoms as presented in \S\ref{noneq}. The direction of polarization, nevertheless, should be the same as in the equilibrium cases. Except from polarization, intensity can also be used as a diagnostic. Note that the result also depends on the inclination with the plane of sky $\alpha=90^o-\theta_0$. Fig.\ref{inclined} shows that the patterns are completely different at $\alpha=45^o$ from the ones with no inclination (Fig.\ref{norm}). By comparing observations with it, we can determine whether magnetic field exists and their directions. For interplanetary studies,
one can investigate not only spatial, but also temporal variations
of magnetic fields. Since alignment happens at a time scale $\tau_R$, magnetic field variations on this time scale will be reflected. This can allow cost effective way of studying
interplanetary magnetic turbulence at different scales.

\section{Alignment of H I, PV, \& NV}

\subsection{Aligned atomic hydrogen}

\begin{figure}
\includegraphics[%
  width=0.28\textwidth,
  height=0.25\textheight]{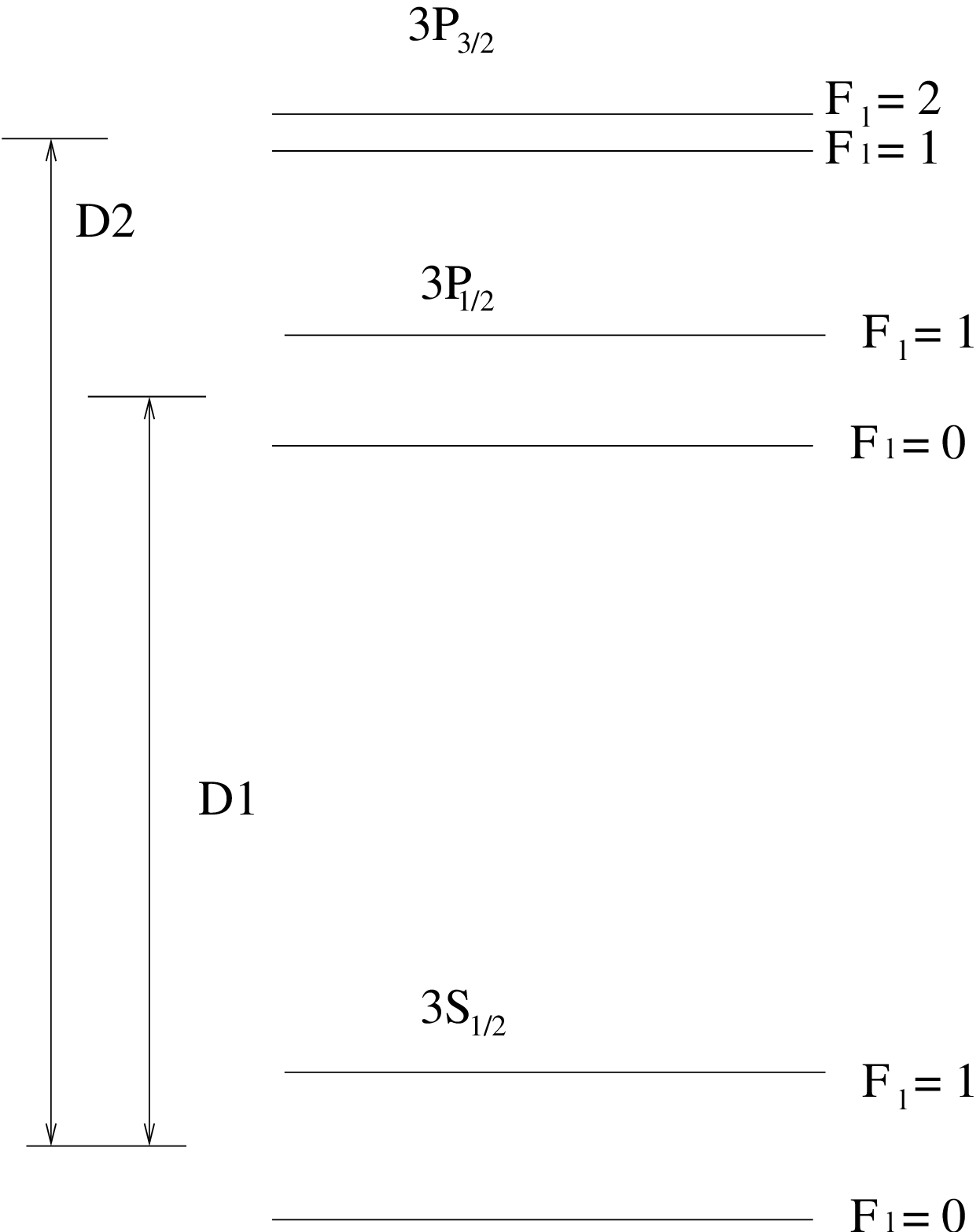}
\includegraphics[%
  width=0.28\textwidth,
  height=0.25\textheight]{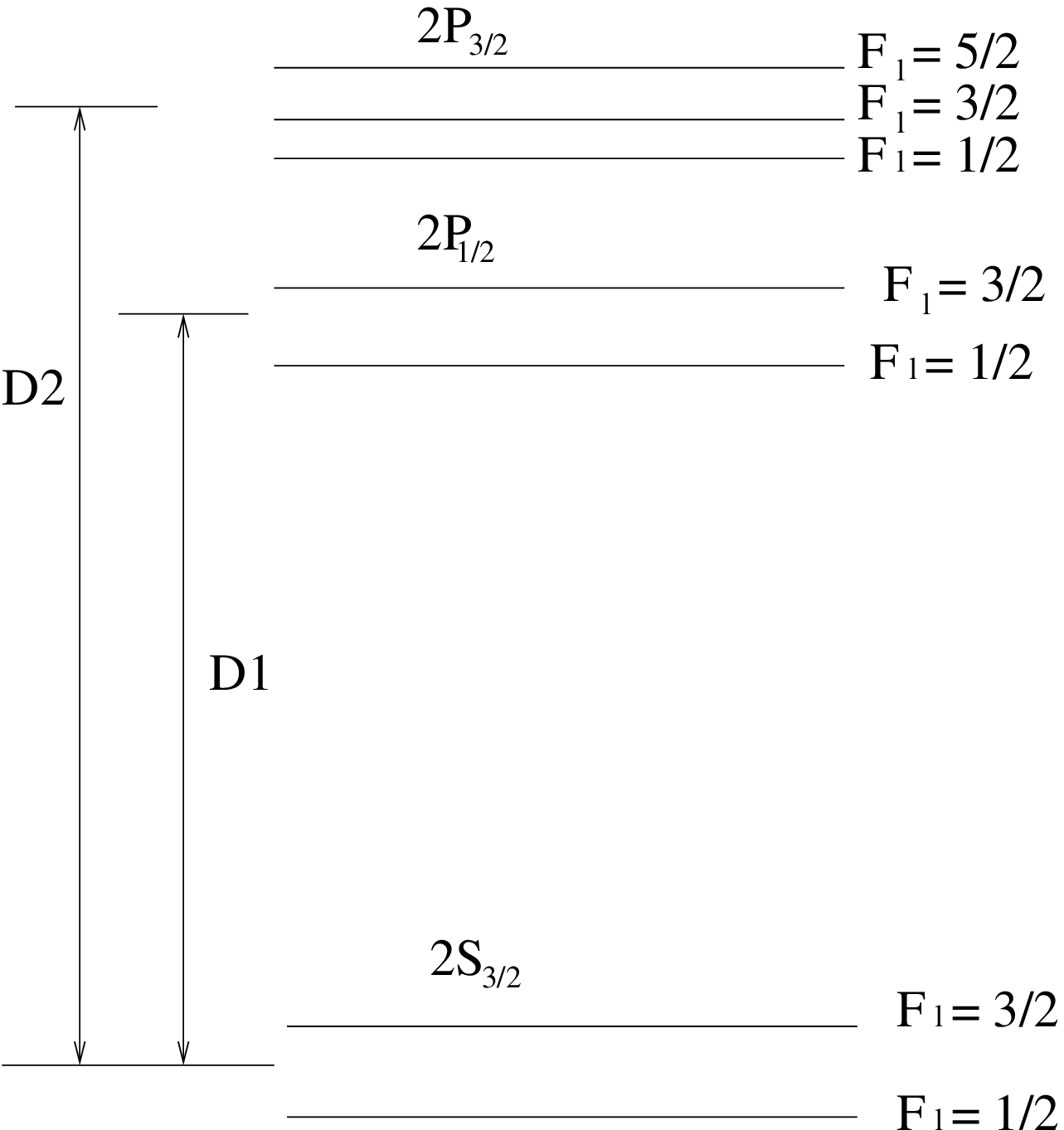}
\includegraphics[%
  width=0.40\textwidth,
  height=0.25\textheight]{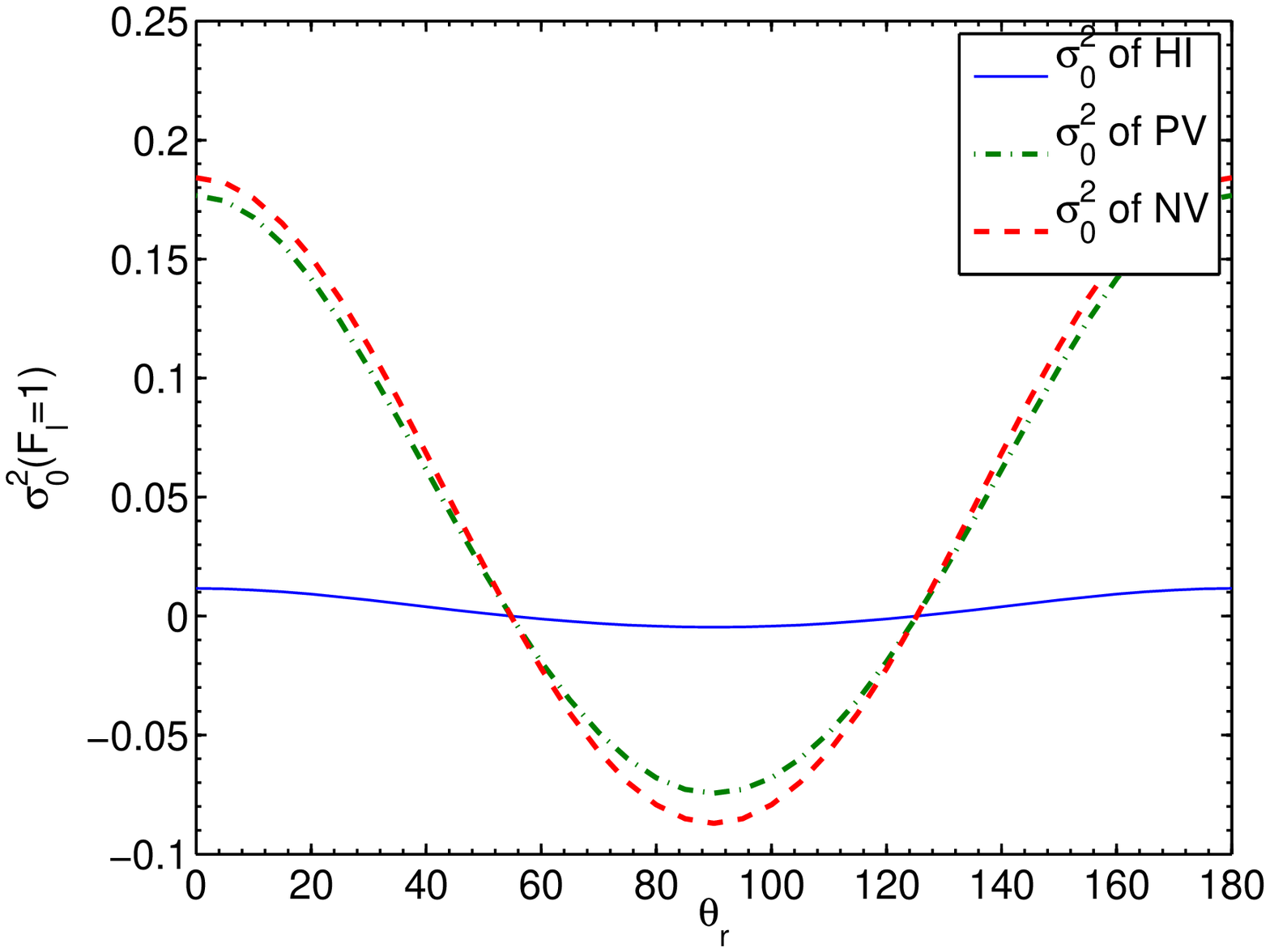}
\caption{{\it Left}: The schematic of H I and PV hyperfine levels; {\it Middle}: schematic of NV hyperfine levels; {\it Right}: The normalized density tensor components $\rho^k_0(F_l=1)/\rho^0_0(F_l=1)$ of ground state of HI (solid lines), PV (dash-dot lines) and NV (dotted lines). As we see, HI is less aligned compared to PV and NV. This is due to the high degree overlap of the hyperfine levels on the upper state of H I.}
\label{HIPV}
\end{figure}

Hydrogen has a similar structure as sodium does. The nuclear spin of hydrogen is $I=1/2$. The total angular momentum of the ground state can be $F_l=1/2\pm 1/2=0,1$ (see Fig.\ref{HIPV}{\it left}). Only the sublevel of $F_l=1$ can accommodate alignment.  The hyperfine splittings $\nu_{F_u,F'_u}$ of $nP_{1/2}, nP_{3/2}$ are smaller than their natural line width $\gamma$, $\nu_{21}(nP_{3/2})=0.229\gamma$, and $\nu_{10}(nP_{1/2})=0.258\gamma$. Thus coherence on both levels should be taken into account. We obtain from Eq.(\ref{hypground}),
\bea
\left[\begin{array}{c}\rho^0_0(F_l=1)\\\rho^2_0(F_l=1)\end{array}\right]=\left[\begin{array}{c}
\left(1.557\cos^4\theta_r+3.114\cos^2\theta_r-23.665\right)\\0.11008-0.33023\cos^2\theta_r\end{array}\right]\varrho^0_0,
\eea
where $\varrho^0_0=\rho^0_0(F_l=0)/(\cos ^4\theta_r+1.730 \cos^2\theta_r-13.652) $. 
Insert the ground density matrix $\rho^{0,2}_0(F_l)$ into Eqs.(\ref{hfemissivity},\ref{hfmueller}), we obtain their emissivities. For D2 line,
{\scriptsize
\bea
\epsilon_0&=& \frac{3\sqrt{3}\lambda^2}{8\pi}A\varrho^0_0\xi(\nu-\nu_0) \left\{
 \cos^2\theta(1.092\cos^6\theta_r+1.762\cos^4\theta_r-17.148\cos^2\theta_r+5.481)+\cos2\phi_r\sin ^2\theta(6.131\cos^2\theta_r-.3446\cos^4\theta_r)\right.\nonumber\\
&-&54.412+12.611 \cos^2\theta_r+2.9112 \cos^4\theta_r-.3641\cos^6\theta_r+\cos \phi_r\left[0.1190\cos2 \theta_r+2.2114\cos4 \theta_r-0.1190\cos6
   \theta_r-0.010\cos8
   \theta_r-2.2014\right.\nonumber\\
&+&\left.\left.0.3207\left(\cos ^4\theta_r+1.970\cos^2\theta_r-15.026\right) \sin 2\theta \sin2\theta_r-(5.422+0.3641\cos^6\theta_r)\sin^2\theta\right]\right\} \nonumber\\
\epsilon_1&=& \frac{3\sqrt{3}\lambda^2}{8\pi}AI_*n\varrho^0_0\xi(\nu-\nu_0) \left\{-1.0923 \cos ^6\theta_r-1.7621 \cos^4\theta_r+17.1512 \cos^2\theta_r-5.4808-0.6414\cos \phi_r 
   \left(\cos^4\theta_r+1.970\cos^2\theta_r\nonumber\right.\right.\\
&-&\left.15.026\right) \sin 2\theta \sin2\theta_r +1.0923 \cos ^2\theta \left(\cos ^6\theta_r+6.930\cos^4\theta_r- 15.702\cos ^2\theta_r+5.018\right)+\cos 2\phi_r\left[0.3641\cos^6\theta_r+0.3446 \cos ^4\theta_r-6.1312 \cos ^2\theta_r\right.\nonumber\\
&+&\left.\left.5.4225+\cos^2\theta \left(0.3641 \cos^6\theta_r+0.3446 \cos^4\theta_r-6.1312 \cos^2\theta_r+5.4225\right)\right]\right\}\\
\epsilon_2&=& \frac{3\sqrt{3}\lambda^2}{8\pi}AI_*n\varrho^0_0\xi(\nu-\nu_0) \left[3.4651\sin2\phi_r \cos \theta \left(0.2101\cos ^6\theta_r+0.1989\cos
   ^4\theta_r-3.5388 \cos^2\theta_r+3.1297\right)\right.\nonumber\\
&-&\left.1.2829\sin\phi_r\left(\cos^4\theta_r+1.970\cos^2\theta_r-15.026 \right) \sin \theta \sin2\theta_r\right].
\eea}

For D1 line,
\be
\epsilon_0=\frac{3\sqrt{3}\lambda^2}{8\pi}AI_*n\varrho^0_0\xi(\nu-\nu_0)
\left(1.749\cos^4\theta_r+3.447 \cos^2\theta_r-26.292\right),\,\epsilon_1=0,\,\epsilon_2=0.
\ee

\begin{figure}
\includegraphics[%
  width=0.33\textwidth,
  height=0.22\textheight]{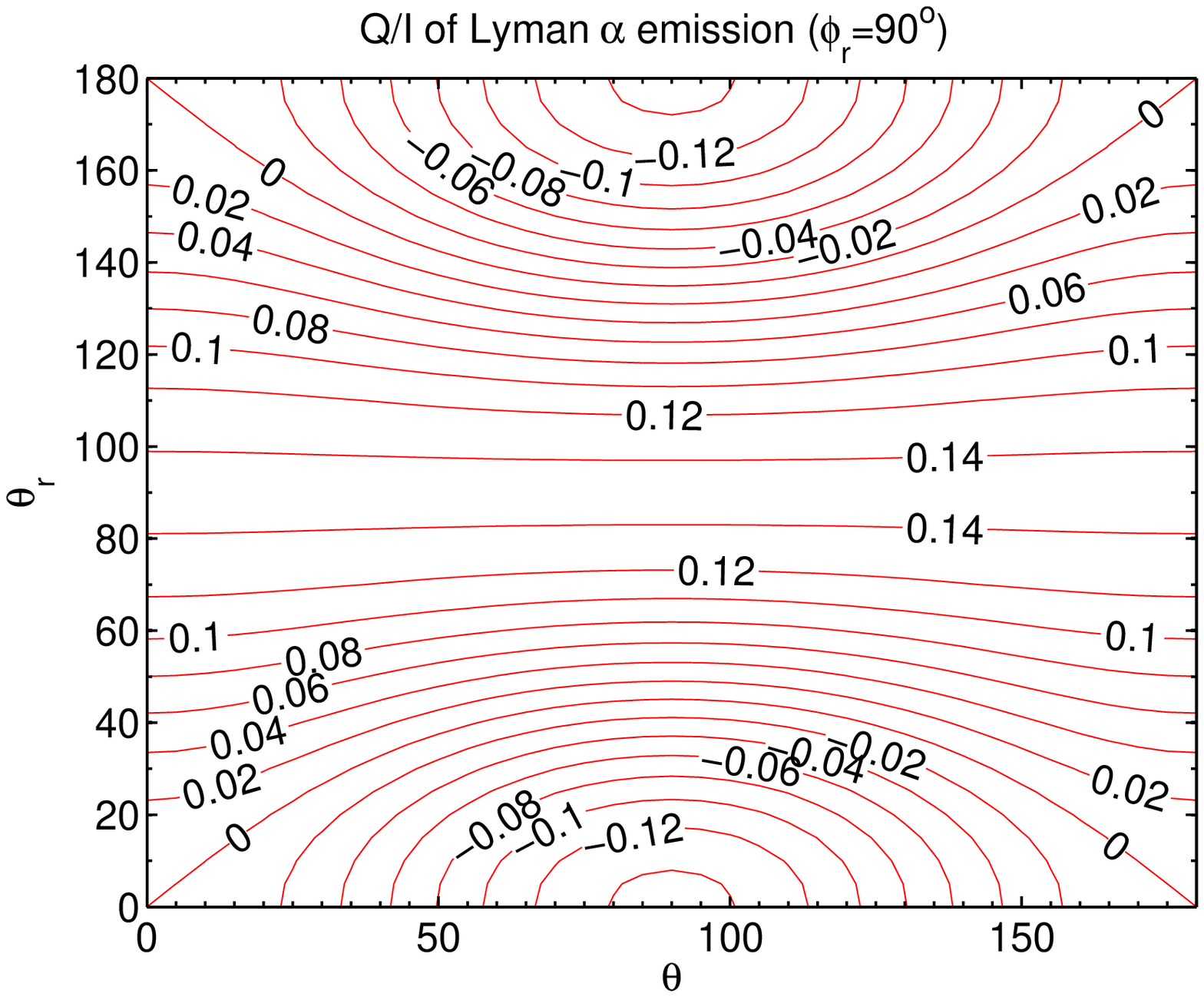}
\includegraphics[%
  width=0.33\textwidth,
  height=0.22\textheight]{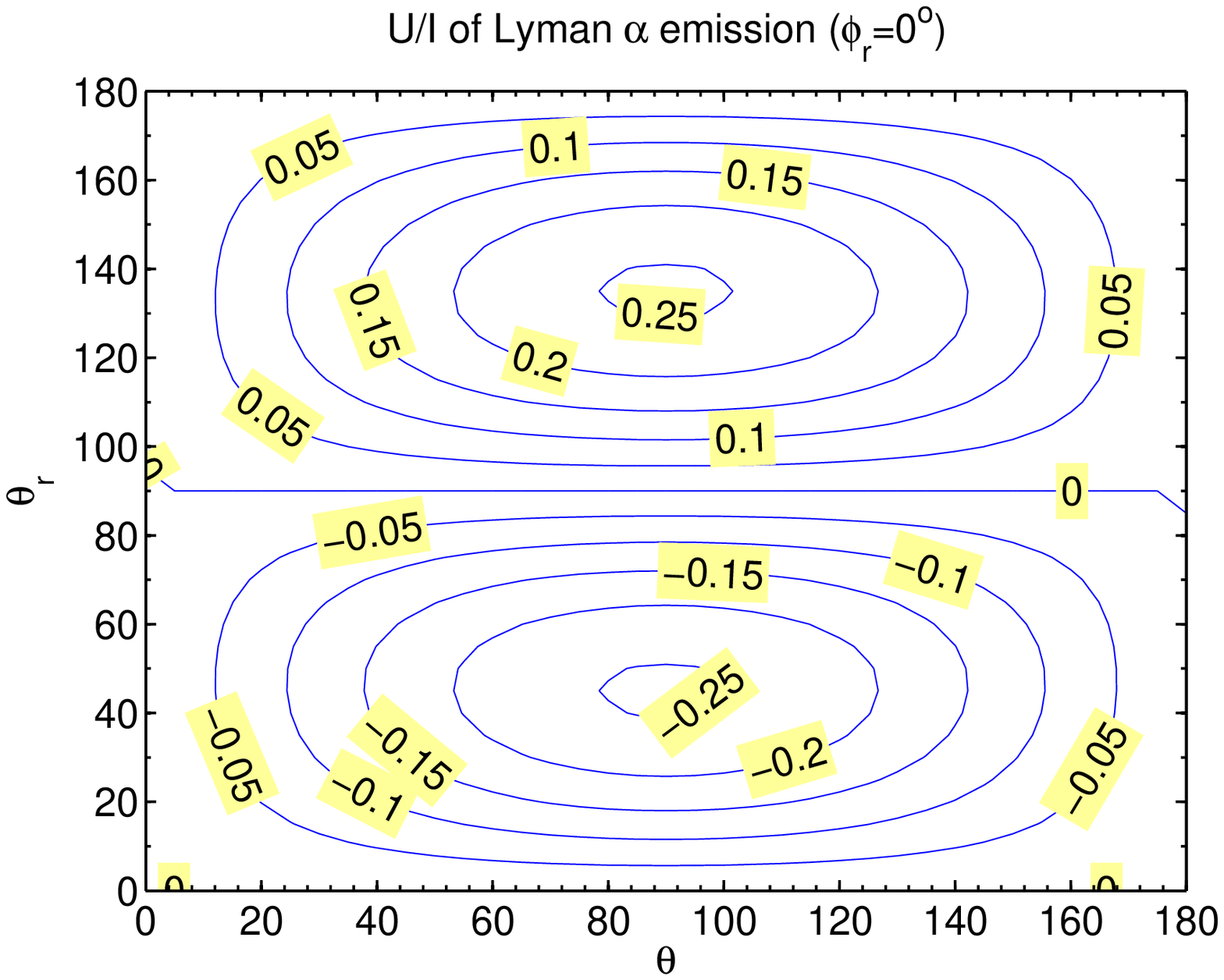}
\includegraphics[%
  width=0.33\textwidth,
  height=0.22\textheight]{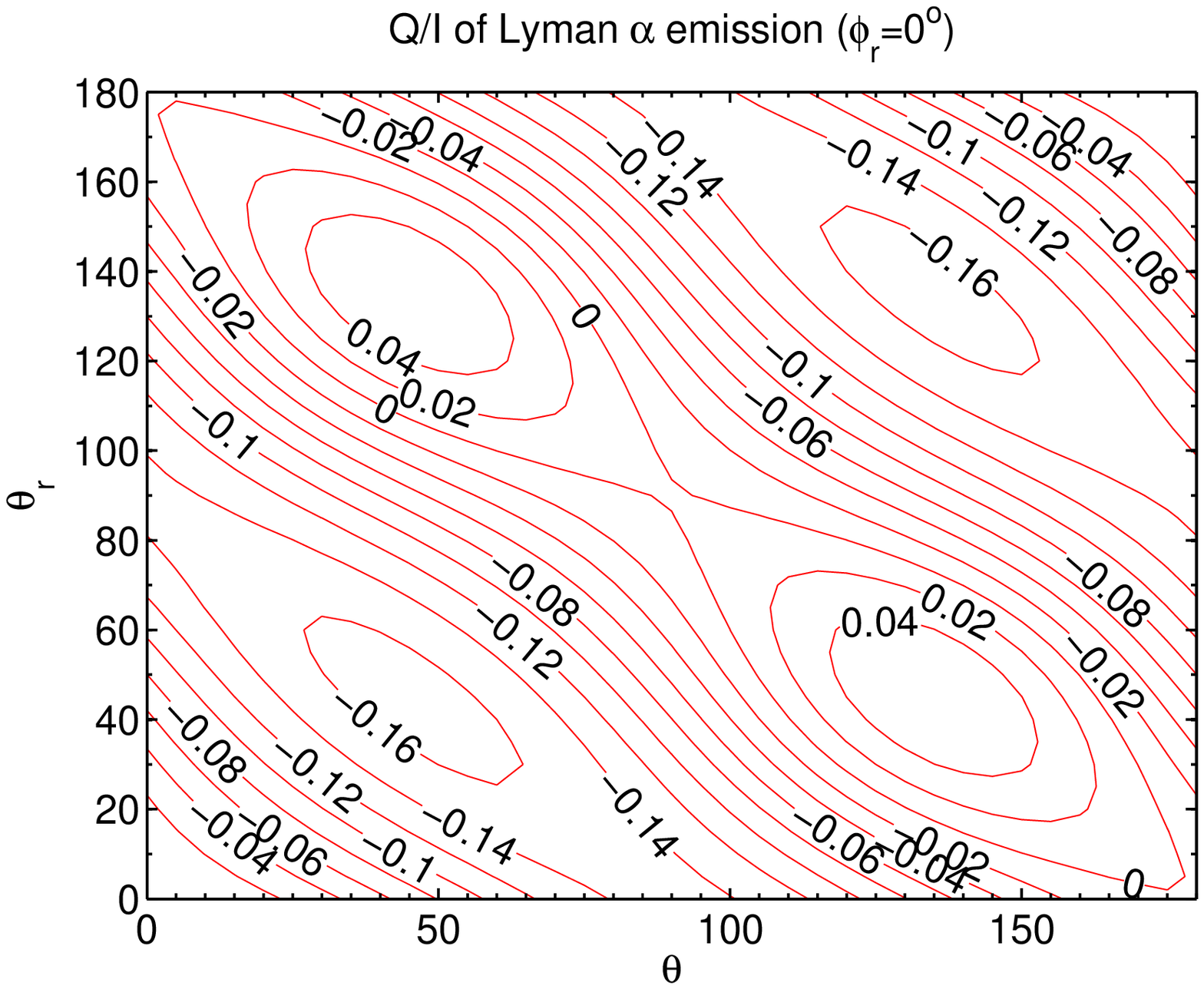}
\caption{Contour graphs of polarization signals of Lyman $\alpha$ emission: {\it left, right}: Q/I, {\it middle}: U/I. Polarization depends on three angles: $\theta_r$, $\theta$ and $\phi_r$ (Fig.\ref{radiageometry}). $\phi_r$ is fixed to $\pi/2$ and 0. At $\phi_r=0$, U=0.}
\label{Hpolcontour}
\end{figure}

The polarization of Lyman $\alpha$ lines are given in Fig.\ref{Hpolcontour}. We see that
compared to Na I, K I, Lyman $\alpha$ line is much more polarized. In general, the more substate the atoms has, the less their polarizability is. As polarized radiation are mostly from those atoms with the largest axial angular momentum, which constitute less percentage in atoms with more sublevels.

\begin{figure}
\includegraphics[%
  width=0.45\textwidth,
  height=0.3\textheight]{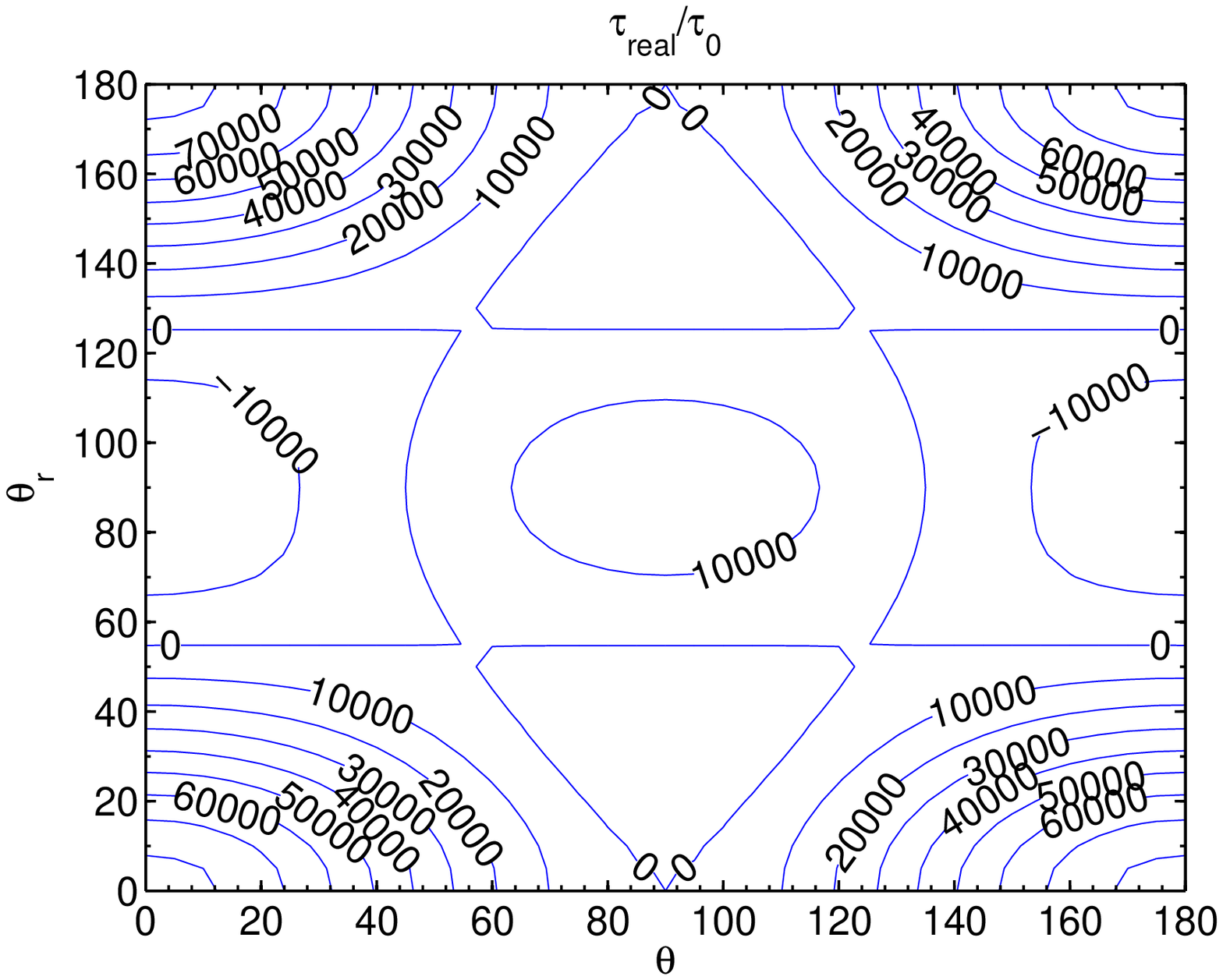}\includegraphics[%
  width=0.45\textwidth,
  height=0.3\textheight]{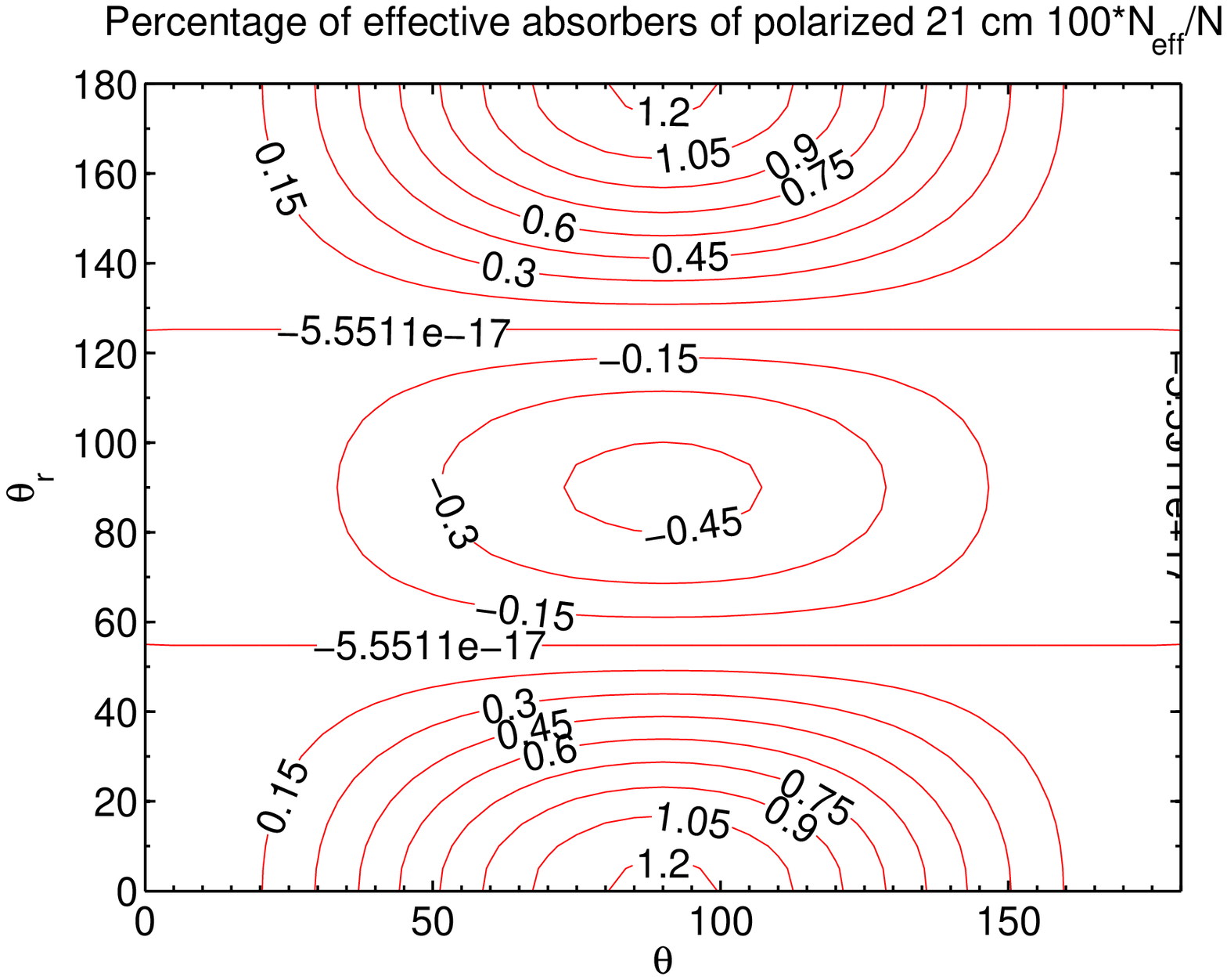}
\caption{The change of optical depth of HI 21 cm due to alignment. {\it Left}: The ratio of the optical depth with alignment taken into account $\tau_{real}$ and the one without alignment counted $\tau_0$. The occupations on the two hyperfine levels on the ground state $F_l=0$ and $F_l=1$ become very close due to the Lyman $\alpha$ pumping. That is why taking into account the alignment component $\rho^2_0$ or not makes a big difference.  {\it Right}: The percentage of effective absorbers of polarized radiation Q of HI 21cm (see Eq.\ref{hyfradio}). Note without alignment $\rho^2_0$, polarization and this quantity would be zeros. If the radiation field has an isotropic component, these results would be reduced by a factor $W_a/W$ (see Eq.\ref{hyfradio},\ref{taurovertau} and the corresponding text).}
\label{H21}
\end{figure}

The alignment on the ground state also causes the change in optical depth $\tau_{21}$ of HI 21cm line, which is a transition between the hyperfine sublevels $F_l=0,1$ (see Fig.\ref{HIPV}{\it left}), 
\bea
\eta_{21}&=&\frac{3}{8\pi}A_m\lambda^2\xi(\nu-\nu_0)n[{\cal J}^0_0\rho^0_0(F_l=0)-\Sigma_{K}{\cal J}^K_0\rho^K_0(F_l=1)/\sqrt{3}]\nonumber\\
&=&\frac{3}{8\pi}A_m\lambda^2\xi(\nu-\nu_0)n\left[\begin{array}{c}\rho^0_0(F_l=0)-(1-1.5\sin^2\theta)\rho^2_0(F_l=1)/\sqrt{2}\\\sqrt{3}/4\sin^2\theta\rho^2_0(F_l=1)\end{array}\right]
\label{hyfradio}
\eea
For comparison, we plot in Fig.\ref{H21} the ratio of ratio of the optical depth with alignment taken into account $\tau_{real}$ and the one without alignment counted $\tau_0$:
\be
\frac{\tau_{real}}{\tau_0}=\frac{[{\cal J}^0_0\rho^0_0(F_l=0)-\Sigma_{K}{\cal J}^K_0\rho^K_0(F_l=1)/\sqrt{3}]}{[{\cal J}^0_0\rho^0_0(F_l=0)-{\cal J}^0_0\rho^0_0(F_l=1)/\sqrt{3}]}\simeq \frac{[-{\cal J}^2_0\rho^2_0(F_l=1)/\sqrt{3}]}{[{\cal J}^0_0\rho^0_0(F_l=0)-{\cal J}^0_0\rho^0_0(F_l=1)/\sqrt{3}]}
\label{taurovertau}
\ee
Since the two hyperfine levels are almost evenly populated in the equilibrium case, the slight change among the populations can make a substantial influence on the transmission of 21cm in the medium (see Fig.\ref{H21}). 

 This may be related to the Tiny-Scale Atomic Structures (TSAS) observed
in different phases of interstellar gas (see Heiles 1997). Besides, as we see from Fig.\ref{H21}, the optical depth can become negative in some cases, indicating the amplification of the 21cm radiation, or maser (see Varshalovich 1967). These effects should be included for HI studies, e.g. Lyman $\alpha$ clouds (Akerman et al. 2005), local bubbles (Redfield \& Linsky 2004), etc. Moreover, polarization can occur due to the alignment. The absorption coefficients for the Stokes parameter Q is given by the second component of Eq.(\ref{hyfradio}) (see also Fig.\ref{H21}{\it right}). According to Eq.(\ref{taurovertau}) the real optical depth is approximately $\propto\rho^2_0\propto {\bar J}^2_0\propto W_a/W$ (see Eq.\ref{irredradia},\ref{Idilu}), where $W_a$ is the anisotropic part of the dilution factor of radiation field (see Van de Hulst 1950 for its expression). Thus if there is an isotropic component in the radiation field, the results will be reduced by a factor of $W_a/W$. Note if photons are scattered multiple times before reaching the atoms, the anisotropy of the radiation field will be diminished and so these effects. Detailed study will be provided elsewhere. 

\subsection{Case of PV}

The overlap of hyperfine structure of upper levels reduces the alignment. In fact, {\bf PV} has the same electron configuration and nuclear spin (see Fig.\ref{HIPV}{\it left}). PV is more aligned as it does not have the overlap on the upper level,
\bea
\left[\begin{array}{c}\rho^0_0(F_l=1)\\\rho^2_0(F_l=1)\end{array}\right]=\varrho^0_0\left[\begin{array}{c}\cos ^4\theta_r+2 \cos ^2\theta_r-19\\
\sqrt{2} \left(1-3 \cos^2\theta_r\right),
\end{array}\right]
\eea 
where $\varrho^0_0=\rho^0_0(F_l=0)/\left(1.876\cos ^4\theta_r+0.289 \cos ^2\theta_r-10.825\right)$. Fig.\ref{HIPV}{\it right} gives the comparison of the density tensor components of H I and PV.

\subsection{Case of NV}

{\bf NV} is also an alkali atom. The nuclear spin of Nitrogen is $I=1$. The total angular
momentum of the ground state thus can be $F=(1\pm1/2)=3/2,/2$ (see Fig.\ref{HIPV}{\it middle}).
Therefore the ground state has totally $(2\times3/2+1)+(2\times1/2+1)=6$
sublevels which enables alignment. 
For NV, the hyperfine splitting 
is much larger than the natural width of the excited state. The smallest hyperfine 
splitting is about $5.6\gamma$, their influence is thus marginal. We obtain its density matrix as follows:
\bea   
\left[\begin{array}{c}\rho^0_0(F_l=3/2)\\\rho^2_0(F_l=3/2)\end{array}\right]=\varrho^0_0\left(
\begin{array}{l}
13.136 \cos^2\theta_r-240.06
   \\
20.90- 62.701 \cos^2\theta_r
\end{array}
\right),
\eea
where $\varrho^0_0=\rho^0_0(F_l=1/2)/(16.62\cos
   ^4\theta_r-1.8 \cos
   ^2\theta_r-167.90)$.

\begin{figure}
\includegraphics[%
  width=0.45\textwidth,
  height=0.3\textheight]{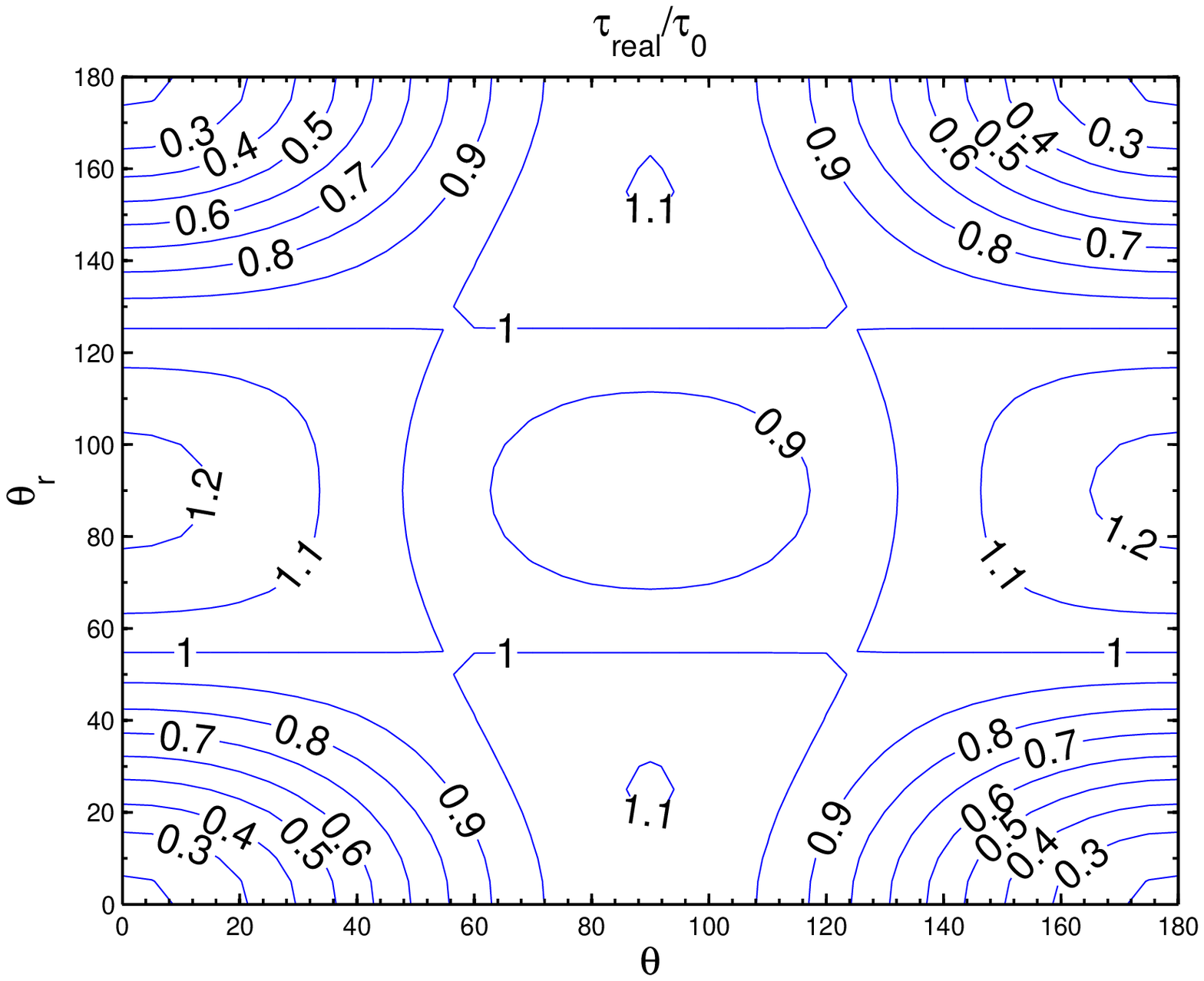}
\includegraphics[%
  width=0.45\textwidth,
  height=0.3\textheight]{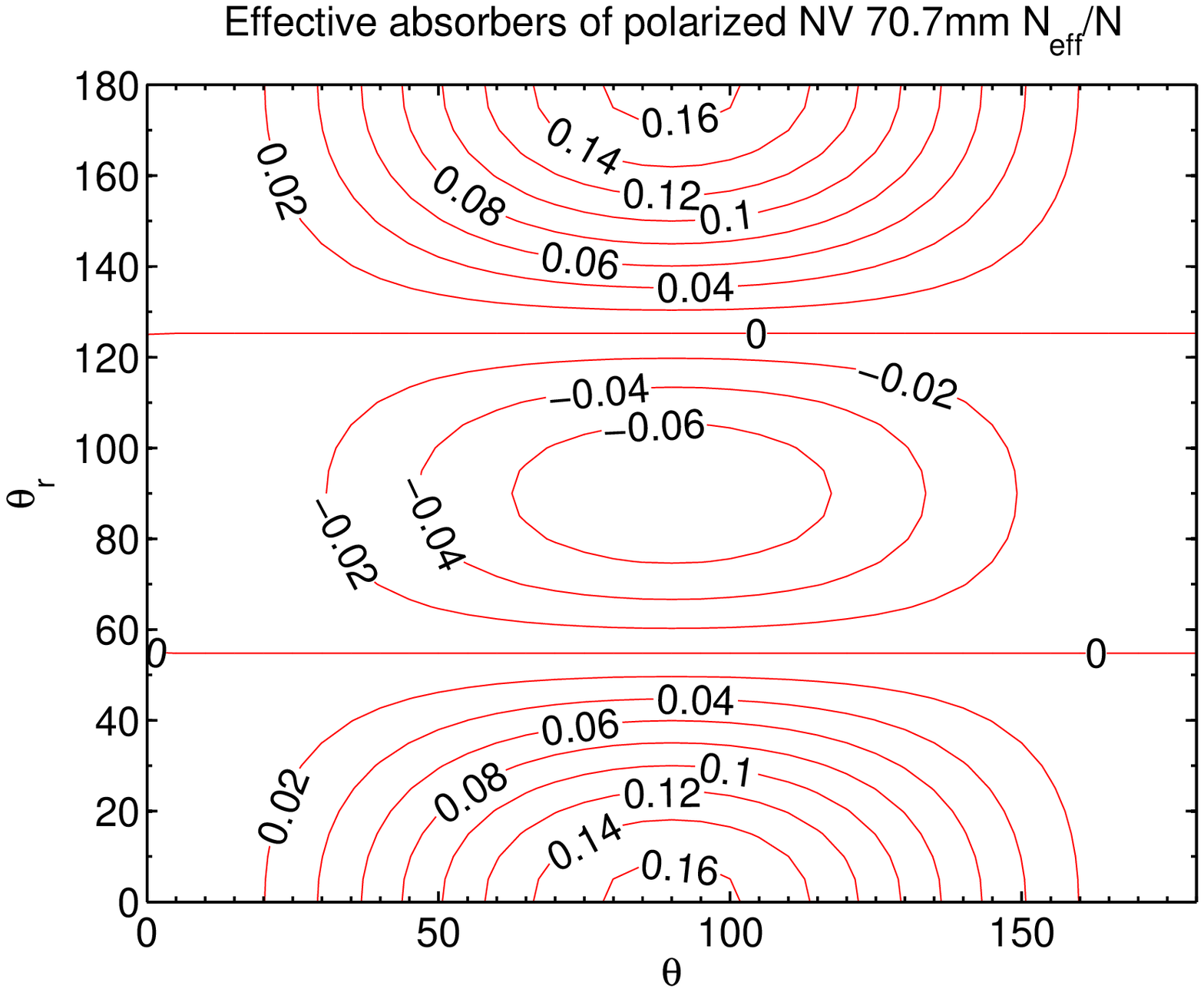}
\caption{The change of optical depth of NV 70.7mm due to alignment. {\it Left}: The ratio of the optical depth with alignment taken into account $\tau_{real}$ and the one without alignment counted $\tau_0$; {\it Right}: The ratio of the density effective absorbers to the total density for polarized radiation Q of NV 70.7mm (see Eq.\ref{hyfradio}). Note without alignment $\rho^2_0$, polarization and this quantity would be zeros.}
\label{NV}
\end{figure}

The hyperfine transition between the two sublevels on the ground state ($\lambda$=70.72mm) has been shown a good tracer of hot rarefied astrophysical plasmas (Sunyaev \& Docenko 2006). Same as HI 21cm, the alignment alters the optical depth (according to Eq.\ref{hyfradio}), which one must take into account when analyzing the NV lines. Further more, polarization appears as a result of the alignment (see Fig.\ref{NV}).
  
\section{More complex atomic species: N I}
\label{NI}
Unlike Alkali atoms, neutral nitrogen is alignable within its fine structure. The ground state of N I is $4S^o_{3/2}$, and the excited state is $4P_{1/2,3/2,5/2}$. The ground state therefore can have four magnetic sublevels with $M=\pm 1/2, 3/2$. Thus hyperfine structure is not a prerequisite for alignment. However, the alignment and resulting polarizations will be miscalculated if we do not include the hyperfine structure. For resonant lines, hyperfine interactions cause substantial precession of electron angular momentum ${\bf J}$ about total angular momentum ${\bf F}$ before spontaneous decay. Thus total angular momentum ${\bf F}$ should be considered and the ${FM_F}$ base must be adopted (Walkup, Migdall \& Pritchard 1982).

Nitrogen has a nuclear spin I=1. Its ground level is thus split into three hyperfine sublevels $F_l=J_l-1, J_l,J_l+1=1/2, 3/2,5/2$. The density tensor $\rho(F_l=5/2)$ has three components with $k=0,2,4$; $\rho(F_l=3/2)$ has two components with $k=0,2$; sublevel $F_l=1/2$ is not alignable and we only need to consider $\rho^0_0$. Solving Eq.(\ref{hypground}), we obtain\footnote{We did the calculation assuming that hyperfine splitting is at least three times the natural line-width and therefore interference term is negligible for the excited state.}
\bea
\rho^{0,2}_0(F_l=3/2)=\varrho^0_0\left(\begin{array}{c} 3\cos
   ^8\theta_r+4 \cos ^6\theta_r-552 \cos
   ^4\theta_r-872\cos
   ^2\theta_r+12325 \\
2\cos ^8\theta_r-72.6 \cos
   ^6\theta_r-235.04\cos ^4\theta_r+1887.5\cos
   ^2\theta_r-600.4\end{array}\right)
\eea
\bea
\rho^{0,2}_0(F_l=5/2)=\varrho^0_0\left(\begin{array}{c}10 \cos
   ^8\theta_r+25 \cos ^6\theta_r-765 \cos
   ^4\theta_r-1017 \cos
   ^2\theta_r+15087 \\
6\cos ^8\theta_r-99.2 \cos
   ^6\theta_r-335.8 \cos ^4\theta_r+4089.6 \cos
   ^2\theta_r-1322.3
\end{array}
\right)
\eea
where $\varrho^0_0=\rho^0_0(F_l=1/2)/(0.2 \cos
   ^{10}\theta_r-2 \cos ^8\theta_r-8.6\cos
   ^6\theta_r-217.5 \cos ^4\theta_r-727.6 \cos
   ^2\theta_r+8733.1)$.

\begin{figure}
\includegraphics[%
  width=0.22\textwidth,
  height=0.25\textheight]{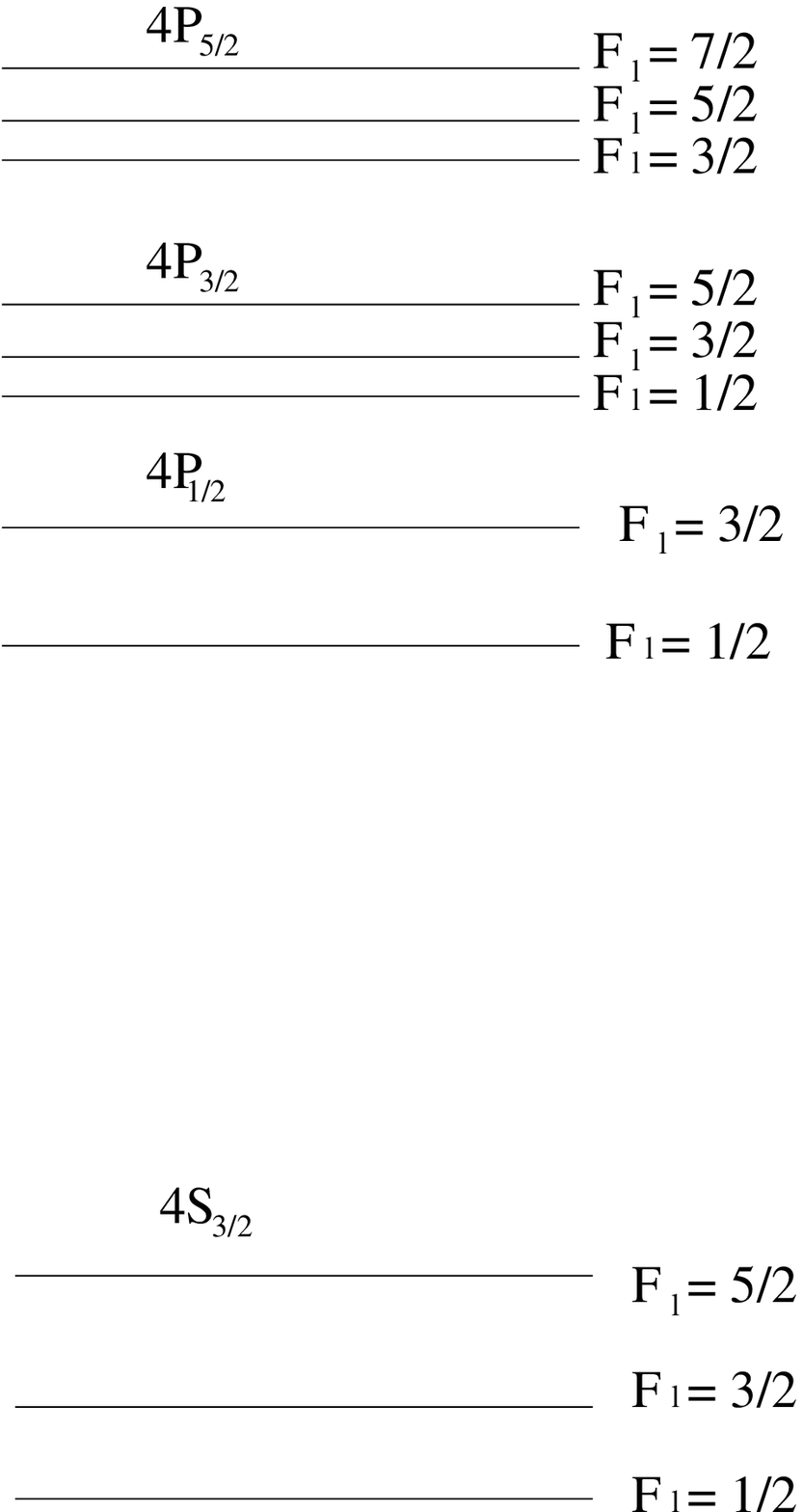}
\includegraphics[%
  width=0.38\textwidth,
  height=0.25\textheight]{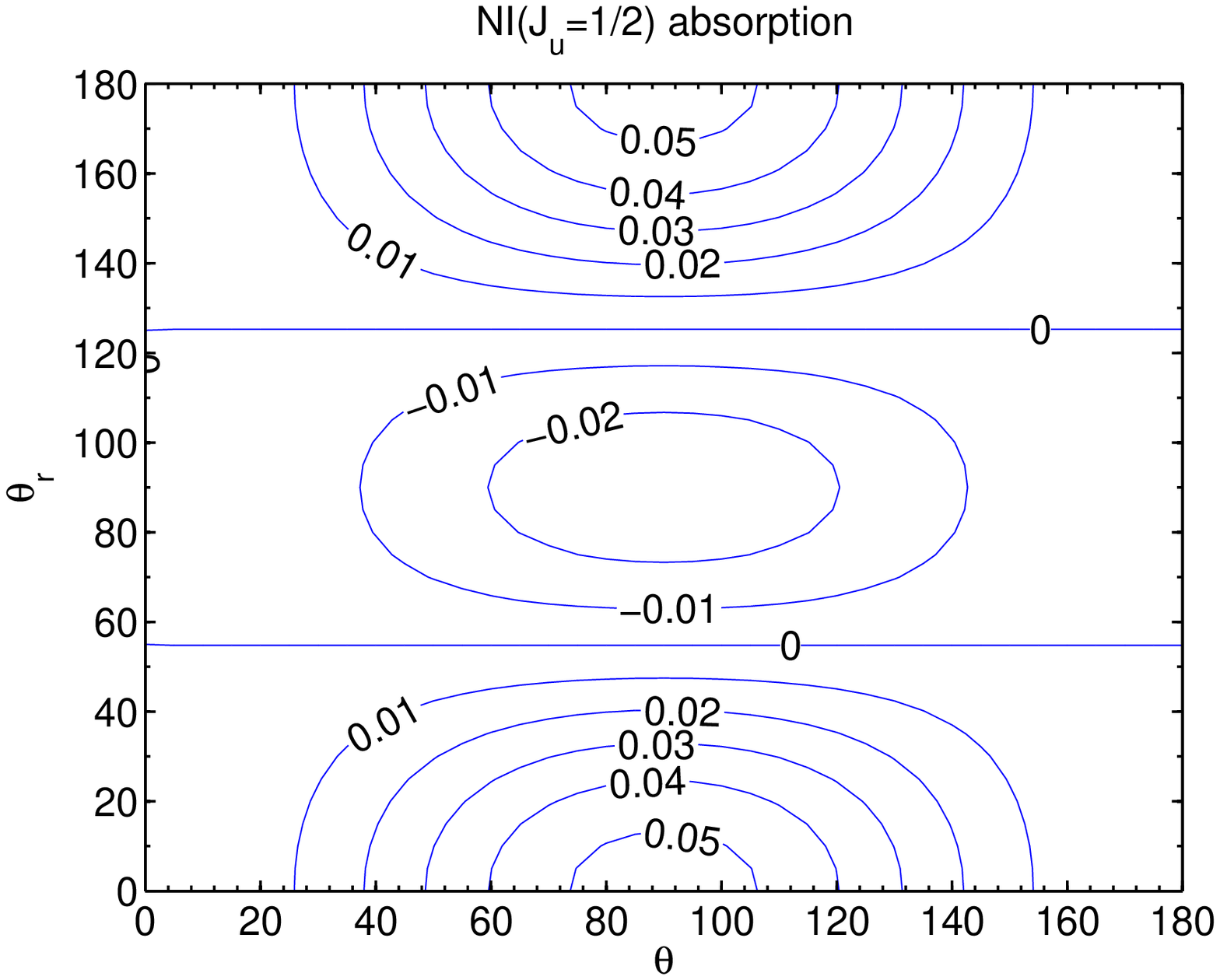}
\includegraphics[%
  width=0.38\textwidth,
  height=0.25\textheight]{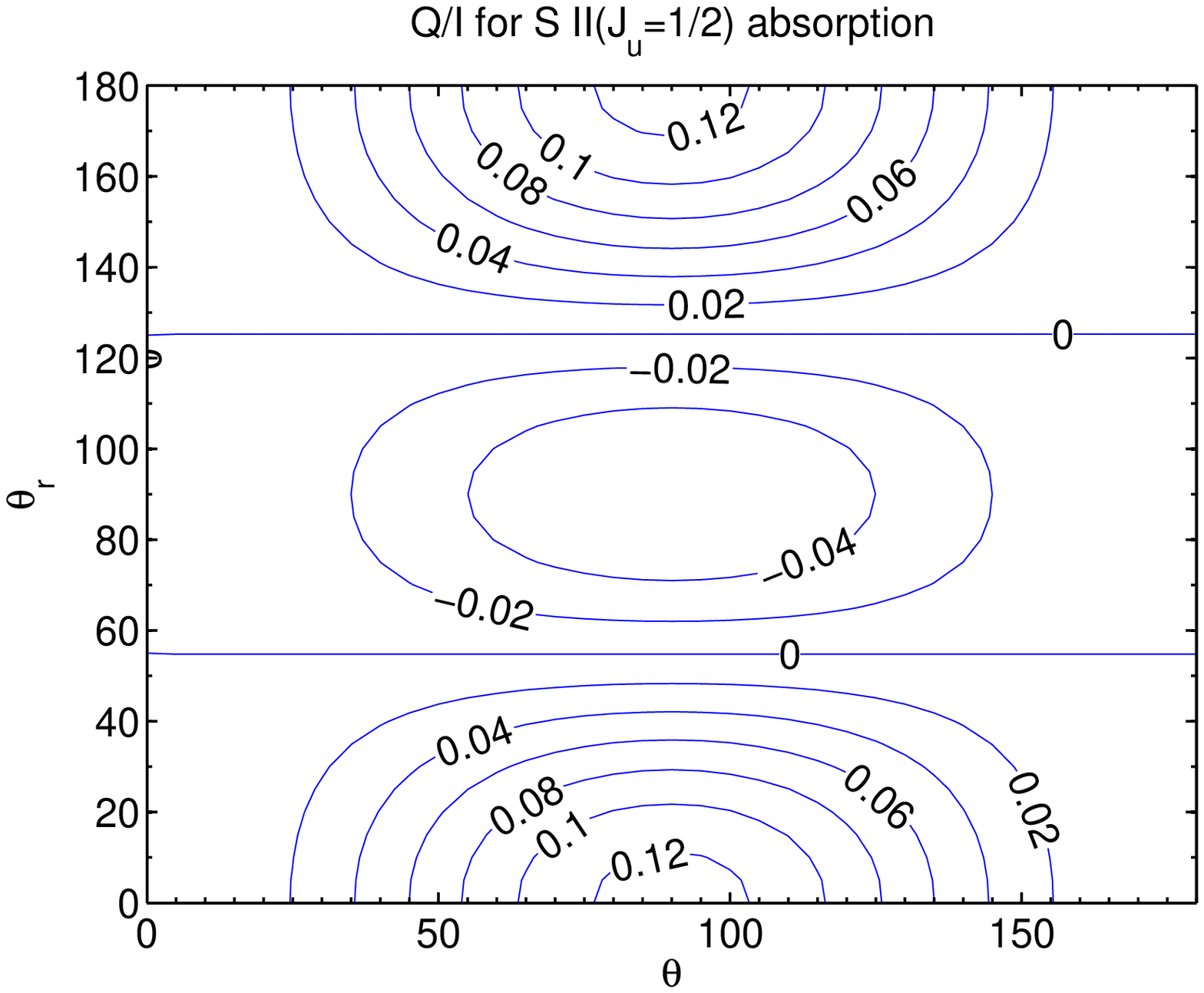}
\caption{{\it Left}:the schematic of N I hyperfine levels;  and the contour graphs of N I ({\it middle}) and S II ({\it right}) degree of
polarization. 
SII does not have nuclear spin. The degree of
polarization is determined by the dipole component of density matrix $\sigma^2_0(\theta_r)$ and the direction of observation $\theta$ (Eqs.\ref{NIpol}). N I and S II have the same electron configuration. However, N I has a nuclear spin while S II does not. More sublevels are allowed in the hyperfine structure of N I, and the alignment is thus reduced and so is the degrees of polarization of N I line compared to that of S II line.}
\label{NIS2pol}
\end{figure}

Since N I has a $J_l=3/2>1$, absorption from N I can be polarized unlike Alkali species (see \S\ref{alkali}). For optically thin case, the polarization produced by absorption through optical depth $\tau=\eta_0d$ is (see Eq.\ref{genericabs})
\be
\frac{Q}{I\tau}\simeq-\frac{\eta_1}{\eta_0}=\frac{1.5\sin^2\theta w^2_{J_lJ_u}\Sigma_{F_l} (-1)^{F_l-J_u}\Upsilon(F_l,2)\rho^2_0(F_l)}{\Sigma_{F_l} (-1)^{F_l-J_u}\left[\sqrt{2}\Upsilon(F_l,0)\rho^0_0(F_l)+\Upsilon(F_l,2)\rho^2_0(F_l)(1-1.5\sin^2\theta)w^2_{J_lJ_u}\right]}
\label{NIpol}
\ee
where $\Upsilon=\left\{\begin{array}{ccc}J_l&J_l&K\\F_l&F_l&I\end{array}\right\}$, $I_0$ is the intensity of background source. We neglect emission here. For a generic case, where the background source (e.g. QSO) is polarized and optical depth is finite, we can obtain in the first order approximation
\bea
I&=&(I_0+Q_0)e^{-\tau(1+\eta_1/\eta_0)}+(I_0-Q_0)e^{-\tau(1-\eta_1/\eta_0)},\nonumber\\
Q&=&(I_0+Q_0)e^{-\tau(1+\eta_1/\eta_0)}-(I_0-Q_0)e^{-\tau(1-\eta_1/\eta_0)},\nonumber\\
U&=&U_0e^{-\tau}, V=V_0e^{-\tau},
\eea
in which ${I_0, Q_0, U_0, V_0}$ are the Stokes parameters of background source. The polarizations of the absorption to $J_u=1/2$ is illustrated in Fig.\ref{NIS2pol}. Note that if the incident light is polarized in a different direction with alignment,
{\it circular polarization} can be generated due to dephasing though it is an 2nd order effect. Consider a background source with a nonzero Stokes parameter $U_0$ shining upon atoms aligned in $Q$ direction\footnote{To remind our readers, The Stokes parameters Q represents the linear polarization along ${\bf e}_1$ minus the linear polarization along ${\bf e}_2$; U refers to the polarization along $({\bf e_1+e_2})/\sqrt{2}$ minus the linear polarization along $({\bf -e_1+e_2})/\sqrt{2}$ (see Fig.\ref{nzplane}{\it right}).}. The polarization will be precessing around the direction of alignment and generate a $V$ component representing a circular polarization 
\be
\frac{V}{I\tau}\simeq \frac{\kappa_Q}{\eta_I}\frac{U_0}{I_0}=\frac{\psi_\nu}{\xi_\nu}\frac{\eta_Q}{\eta_I}\frac{U_0}{I_0}
\ee
where $\kappa$ is the dispersion coefficient, associated with the real part of the refractory index, whose imaginary part corresponds to the absorption coefficient $\eta$. $\psi$ is the dispersion profile and $\xi$ is the absorption profile.

\begin{figure}
\includegraphics[%
  width=0.45\textwidth,
  height=0.3\textheight]{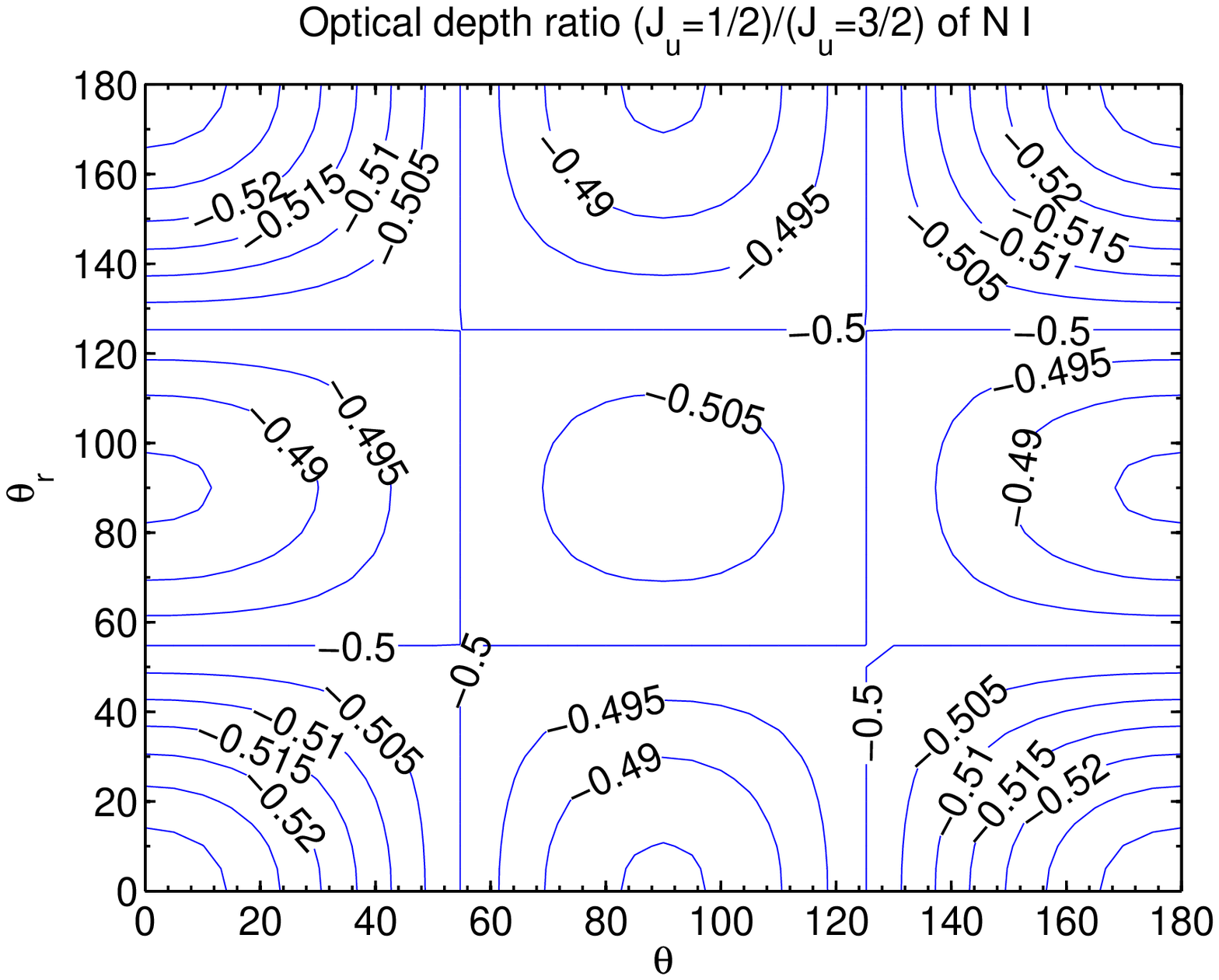}
\includegraphics[%
  width=0.45\textwidth,
  height=0.3\textheight]{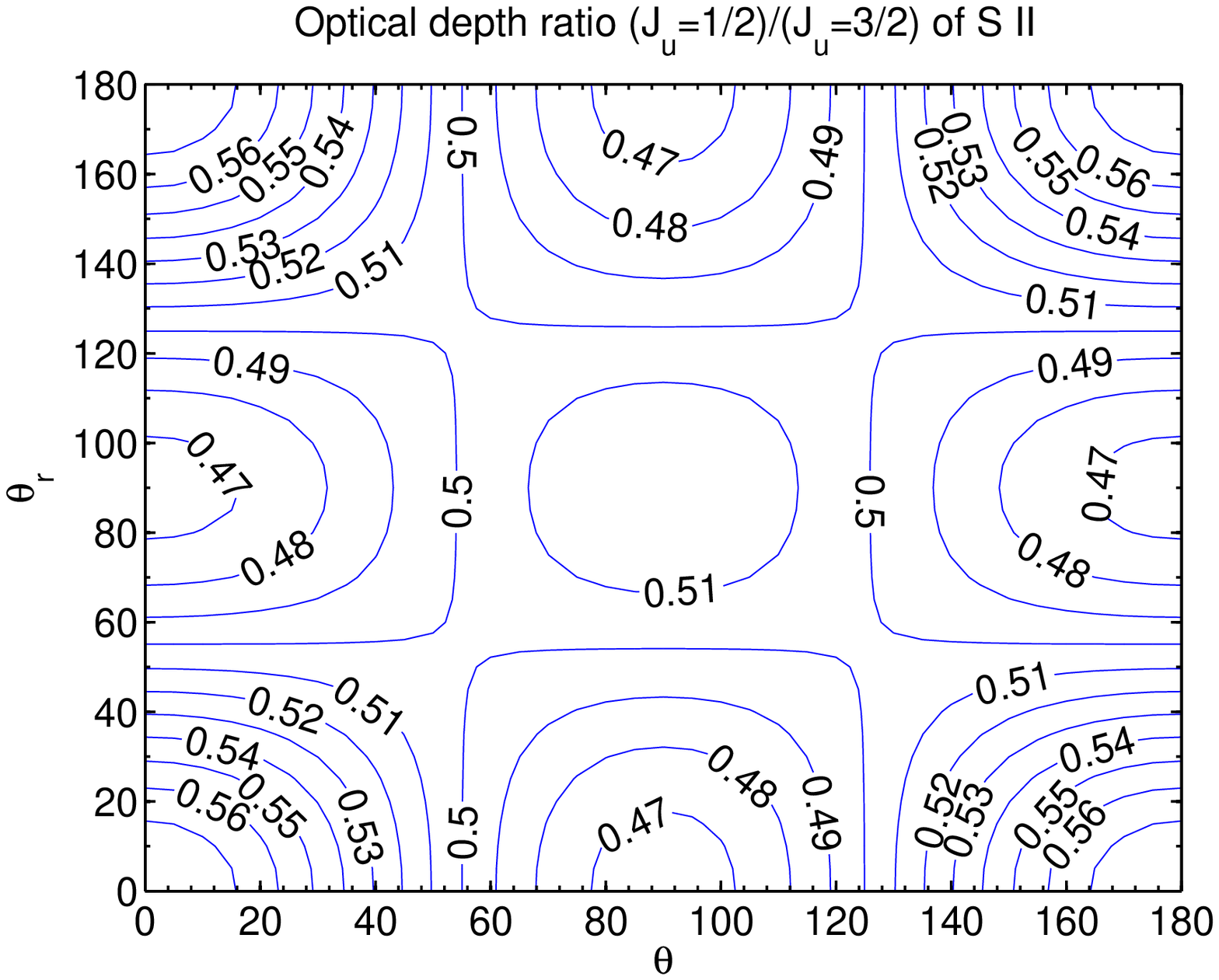}
\caption{The contour graphs of line ratios of N I and S II. The difference of their ratios is only due to the hyperfine structure of N I.}
\label{NIS2ratio}
\end{figure}

Optical depth also varies with the alignment (Fig.\ref{NIS2ratio}). The generic expression of the line ratio of a multiplet is given by
\be
\frac{\tau_1}{\tau_2}=\frac{\Sigma_{F_l} (-1)^{F_l-J_{u1}}\left[\sqrt{2}\Upsilon(F_l,0)\rho^0_0(F_l)+\Upsilon(F_l,2)\rho^2_0(F_l)(1-1.5\sin^2\theta)w^2_{J_lJ_{u1}}\right]}{\Sigma_{F_l} (-1)^{F_l-J_{u2}}\left[\sqrt{2}\Upsilon(F_l,0)\rho^0_0(F_l)+\Upsilon(F_l,2)\rho^2_0(F_l)(1-1.5\sin^2\theta)w^2_{J_lJ_{u2}}\right]}
\label{NIratio}
\ee

Similar to the cases without hyperfine structure (Paper I), absorption is determined by only two angles, $\theta_r$ and $\theta$. Among them, $\theta_r$ determines the ground state alignment and $\theta$ dependence occurs from the direction of observation. We also see the Van-Vleck effect in Fig.\ref{NIS2pol}. Specifically, the polarization is either $\parallel$ or $\perp$ to the magnetic field in the plane of sky; the switch happens at Van-Vleck angle $\theta_r=54.7^o$.

In Fig.\ref{NIS2pol},\ref{NIS2ratio}, we plot together the polarizations and optical depth ratios for N I and S II, result for which 
is taken from Paper I. As we know, N I and S II have the exactly the
same term. The difference between them arises from the hyperfine structure of N I. In other words, if we do not take into account the hyperfine structure of N I, it would be polarized exactly the same way as that of SII. And this can be tested observationally. As we explained above, it is usually true that the more complex the structure is, the less polarized the line is. It can also be interpreted by the nature of hyperfine interactions. Hyperfine interactions causes the precession of electron angular momentum in the field generated by the nuclear spin. Thus similar to the case of magnetic mixing, the polarization is reduced. 

\section{Dilution along line of sight}

For absorption lines, there are atoms far from any pumping source along the line of sight. These atoms are not aligned, and we need to take into account the averaging along line of sight. Different components of the atomic density matrix are modulated by the pumping differently. For atoms far from a source, their dipole component of density matrix $\rho^2_0$ is zero. The zero order term $\rho^0_0$, representing total occupation of a level, however, only changes with alignment by $\lesssim 10\%$. As a first order approximation, we thus can ignore the variation of $\rho^0_0$ due to the pumping and adopt a step function for $\rho^2_0$
\bea
\rho^2_0(r)=\left\{\begin{array}{rl}\rho^2_0(aligned),& {\rm for ~\,r<r_c}\\
0,& {\rm for\,~r>r_c}\end{array}\right..
\eea
where $r_c$ is the distance from a pumping source where the collisional transition rate becomes equal to the optical pumping rate. In this case, we only need to multiply $\rho^2_0$ in Eq.(\ref{hyfradio},\ref{NIpol},\ref{NIratio}) by $N_a/N_{tot}$, the ratio of alignable column density to the total column density along the line of sight. Accordingly the contours (Fig.\ref{H21}-\ref{NIS2ratio}) do not change apart from their amplitudes because the dependencies on $\theta_r,\theta$ only appear in the terms containing $\rho^2_0$. By overlapping the polarization contour maps and the line ratio contour map, we attain the angles $\theta_r$ and $\theta$. Then insert these values into either Eq.(\ref{NIpol}) or Eq.(\ref{NIratio}), we get the ratio $N_a/N_{tot}$. More precise results can be obtained by making iterations. From this ratio, we learn also the conditions in the vicinity of the pumping source. Combining different atomic species, we can make a tomography of magnetic field as the ratio $N_{a}/N_{tot}$ varies with each atomic species. As an example, let us consider the alignment of H I by an O-type star (outside H II region), for which we know the collisional transition rate is $C_{10}/n_H=3.3\times 10^{10}{\rm cm}^3 {\rm s}^{-1}$. By equating it with the Lyman $\alpha$ pumping rate $B{\bar J}^0_0$, we can get $r_c\simeq 15pc$ in cold neutral medium where we adopt $n_H=30{\rm cm}^3{\rm s}^{-1}$. If the star is 100pc away from us, then the dilution factor along line of sight would $\int^{15pc}_{r_s} n_H dr/\int^{100pc}_{r_s} n_H dr\sim 6$, where $r_s$ is the size of Str{\" o}mgren sphere.   

It is also possible that there are multiple independent pumping sources along the line of sight. These situations, however, can be easily identified for diffuse interstellar medium.

\section{Discussion}

\subsection{Hyperfine splittings}

We have considered alignment of atoms with nuclear spin in this paper. For these atoms, it is the total angular momentum, ${\bf F=J+I}$ that should be considered. The prerequisite for alignment in this case is $F>1/2$. There are two 
categories. Atoms like Alkali atoms, Al II, Cu II, etc. would not be 
alignable without hyperfine structure since their electron angular momentum 
in ground state is $J_l<1$. Another category of atoms are alignable within 
fine structure, e.g., N I, Cl I,II,III, etc. However, calculations of
the alignment and polarization would render erroneous results
 if one does not take into account the hyperfine structure of the 
species. This is because for resonant lines, the hyperfine interaction 
time-scale is shorter than that of resonant scattering. 

The first category of atoms above, i.e. with $J_l<1$, cannot produce any 
polarization in absorptions even the ground state is aligned within the hyperfine structure. As we explained in \S\ref{NI}, polarization only reduces due to hyperfine interaction. For Alkali atoms, the absorption is not polarized in the frame of fine structure. Taking into hyperfine structure does not make a difference in this case to the absorptions. However, the
polarization of emission will be affected by the alignment. We
note, that although  the 
 ground level alignment is not a prerequisite for polarization of 
emission for every line, it does affect the 
polarization of emission. 

We discussed a few examples of elements from our list in
Table~1. They were chosen on the basis of astrophysical importance as we 
see it. For instance, sodium D lines are very pronounced ones and easy to 
measure. It is also important for studies of comet wakes, as we discuss
in the paper. 
More calculations should be done in future in relation to particular
astrophysical objects under study.

\subsection{Polarization of absorption lines}

We studied polarization of absorption resulting from alignment of atoms
 with nuclear spin and thus with
hyperfine structure. The degree of polarization is reduced compared to 
atomic species of the same electron configurations but without hyperfine 
structure. The direction of polarization, however, has the same pattern, namely, either $\parallel$ or $\perp$ to the magnetic field on the plane of sky (Paper I). And the switch between the two cases happens at the 
Van-Vleck angle $\theta_r=54.7^o$. In fact, this should be applicable to 
all absorption lines (including molecular lines) regardless of their different structures as long as the following conditions are satisfied. First, pumping light and background light are unpolarized; and then it is in the magnetic realignment regime, namely, magnetic precession is faster than the photon-excitation rate. 

This fact is very useful practically. It means that even we do not have an exact prediction and precise measurement of the degree of polarization of the absorption lines. We can have a 2D mapping of magnetic field on the plane of sky (the angle $\phi_B$ in Fig.\ref{radiageometry}{\it right}) within an accuracy of $90^o$ once we observe their direction of polarizations. In this sense, it has some similarity with the Goldreich-Kylafis effect although it deals with radio emission lines.

For absorption lines, there is inevitably dilution along line of sight, which adds another dependence on the ratio of alignable column density and total column density $N_a/N_{tot}$. For different species, this ratio is different. The ratio should be close to 1 for highly ionized species which only exist near radiation sources. The same is true
 for the absorption from metastable state (see Paper I). 
Combining different species (with different $N_a/N_{tot}$), it is possible to acquire a tomography of the magnetic field {\it in situ}. To
extend the technique we shall present elsewhere calculations for
more atoms with metastable states. 

An additional effect that we consider briefly in \S 6 is the generation of circularly polarized light when linear polarized light passes through aligned atoms. This is a new
interesting effect the implications of which we intend to explore elsewhere.

\subsection{Polarization of emission lines}

In Paper I we dealt with absorption lines of species with fine
structure. This paper deals with absorption and emission
of both the atom species with hyperfine structure and hyperfine plus 
fine structure.

The work on emission of atoms with hyperfine structure can be traced
back in time. Studies
of alignment of neutral sodium in laboratory alignment 
pioneered more that half a 
century ago by Brossel, Kastler \& Winter (1952), Hawkins (1955),
and Kastler (1956). 
These experiments revealed that sodium atoms
can be efficiently aligned in laboratory conditions if atomic beams are 
subjected to anisotropic resonance radiation flux. However, the 
calculations that we aware of were not satisfactory. For instance,
the classical treatment of radiation adopted in Hawkins (1955)  
does not provide the correct measures of the alignment and polarization.
Moreover, the calculations were limited by the particular geometry
of equipment used. On the contrary, our astrophysical applications
require us to
get predictions for arbitrary angles between the line of sight and
magnetic field as well as for arbitrary angles between the illumination
source and magnetic field. 

In the paper we have considered the situation that atoms are subjected to the 
flow of photon that excite transitions at the rate $\tau^{-1}_R=B_{lu}{\bar J}^0_0$ which is
smaller than the Larmor precession rate $\nu_L$, but larger than the rate of disalignment due to collisions $\tau^{-1}_c$. For the cold
gas with 30 hydrogen atoms per cubic cm, the characteristic range over which the atoms can be aligned by an O star is $\lesssim 15$pc for HI due to spin exchange collisions (with rate $C_{10}/n_H=3.3\times 10^{-10}{\rm cm}^3{\rm s}^{-1}$). 

Compared to absorption lines, emission lines are more localized and therefore the dilution along the line of sight can be neglected. The disadvantage of emission lines compared to absorption lines is that the direction of the polarization of emission lines has a complex dependence on the direction of the magnetic field 
and the illusion light. Therefore, the use of
emission lines is more advantageous when combined with other measurements. 

The change of the optical depth is another important consequence of atomic alignment. Such effect can be important for HI as was first discussed by Varshalovich (1967). However, the actual calculations that
take into account the magnetic realignment in ubiquitous astrophysical
magnetic fields are done, as far as we know, only in this paper.
 It might happen that the variations of the optical depth caused by
alignment can be related to the Tiny-Scale Atomic Structures (TSAS) observed
in different phases of interstellar gas (see Heiles 1997). Similar
effects may be present for Lyman alpha clouds and other objects.

\subsection{Implications}

Atomic alignment opens a new channel of information about the physical
properties of the various medium, including environments of circumstellar
regions, AGN and interstellar medium. In particularly, the topology
of magnetic fields that are so important for these environments can be
revealed. What is unique for this new window is the possibility of
obtaining information about 3D directions of magnetic field. Combining
different
emission and absorption lines it seems possible to restore the entire
structure of the region, that would include full 3D information, including
the information about the position of illuminating stars.

We have done calculations for a number of representative atomic species (see Table~1). These emission and/or absorptions from these atomic species, are important lines seen in different astrophysical environments. Indeed, there are much more atomic lines that can be studied the same way as we did here. The particular choice of atoms to use depends both on the instruments
available and the object to be studied. For instance, Sodium D lines have been observed in interplanetary medium, including comet tail, and a Jupiter's moon
Io.  Many lines from alignable species have been observed in QSOs and AGNs (see Verner, Barthel \& Tytler 1994) and they can be used to study the magnetic field {\it in situ}. Needless to say, this technique can be used to any interstellar medium near an emitting source, H II regions, circumstellar discs, supernovae, etc.
For intergalactic gas, Lyman $\alpha$, H I 21 cm, NV and other radio lines, one also should be aware of the fact that alignment also changes the optical depth and the emissivity of the medium. And with our quantitative predictions of the alignment, one can also attain the information of magnetic field in the medium.

 A detailed discussion of alignment of atoms in different conditions 
corresponding to 
circumstellar regions (see Dinerstein, Sterling \& Bowers 2006), AGN 
(see Kriss 2006), Lyman alpha clouds (see Akerman et al. 2005), interplanetary space (see Cremonese et al. 2002), the local bubble
 (see Redfield \& Linsky 2004), etc., will be given elsewhere. 
Note, that for interplanetary studies,
one can investigate not only spatial, but also temporal variations
of magnetic fields. This can allow cost effective way of studying
interplanetary magnetic turbulence at different scales.

Taken together this paper and Paper I provide examples of treating both 
aligned absorbing and emitting species. For the sake of simplicity, we have
not considered a few effects that can affect observations.
For instance, we did not consider effects of the finite telescope resolution
for the emission from the regions that have finite curvature of magnetic field 
or random magnetic field component. Such effects are  well known and
described in the existing polarimetric literature (see Hildebrand 2000). 
For absorption, however, the effect of telescope finite diagram is negligible 
if we study the absorption from a point source. In addition, we considered
emission from optically thin medium, which justifies our neglect for the 
radiative transfer. 

In the present paper we considered the alignment of NaI atoms subjected
to a limited number of scattering events. However, our approach to describe 
time-dependent alignment is general and may be easily applied to other species.
This may be particularly important for studies of transient phenomena using
our technique.

As the resolution and sensitivity
of telescopes increases, atomic alignment will be capable to 
probe the finer structure of astrophysical magnetic fields including those in
the halo of accretion disks, stellar winds etc. Space-based polarimetry should
provide a wide variety of species to study magnetic fields with. 

\section{Summary}
In this paper we calculated the alignment of various atomic species 
having hyperfine structure and
quantified the effect of magnetic fields on alignment. As the result,
we obtained linear polarization that is expected for both scattering
and absorption.  
We have shown that

\begin{itemize}

\item Atomic alignment of atoms and ions with
hyperfine structure of levels happens as the result of interaction 
of the species with anisotropic flow of photons.

\item Atomic alignment affects the polarization state of the 
scattered photons as well as the polarization state of the
absorbed photons. The degree of polarization is influenced
by mixing caused by Larmor
precession of atoms in external magnetic field. This allows a new
way to study magnetic field direction in diffuse medium using polarimetry.

\item The degree of polarization depends on the species under study. Atoms with more levels exhibit, in general, less degree
of alignment. More importantly, it depends on the angle between
the direction of
the pumping light and the observational direction with respect to 
the magnetic field embedded in the medium. 

\item The direction of polarization 
depends  on the direction of  the anisotropic radiation and 
observation in respect
to the magnetic field acting upon at atom, if the polarization of 
scattered light is concerned. 

\item The direction of polarization
is either parallel or perpendicular to magnetic field, if the polarization of absorbed light is studied. The switch between
the two options happens at the Van-Vleck angle between the direction 
of magnetic field and the pumping radiation. 

\item The intensity ratio of scattered lines or absorption lines are also 
influenced by magnetic realignment and therefore also carry the 
information about the direction of magnetic field.

\item Absorption and emission of species along the line of sight
away from the pumping sources interferes with the detected signal,
e.g. influence the degree of the measured polarization and the ratio
of absorption lines. This effect, however, can be accounted for 
iteratively.  

\item If the light incident on the aligned atoms is linearly polarized,
as this is a typical case of QSOs,
circular polarization gets present in the transmitted light.

\item Atomic alignment is an effect that is present for a variety of
species and for different terms of the same atom.
Combining the polarization information as well
as using line intensity ratio data allows  to improve
the precision in mapping of magnetic fields and get
insight into the environments of the aligned atoms.

\item A steady-state alignment is achieved when after many scattering
events. For a limited number of the scattering events
the alignment depends on this number, i.e. ``time-dependent''. 
While the direction of polarization are the same for 
both cases, the degree of polarization increases with the number of scattering until the steady-state alignment is reached.

\item Time variations of magnetic field in interplanetary plasma should
result  in time variations of degree of polarization thus providing
a tool for interplanetary turbulence studies.

\end{itemize}

\begin{acknowledgments}
We are grateful to Jungyeon Cho for  
supplying us with MHD simulations of a comet wake. We thank the anonymous referee for his/her valuable suggestions. Helpful comments from P. Hall are acknowledged. HY is supported by CITA and the National Science and Engineering Research Council of Canada. AL is supported by the NSF Center for Magnetic Self-Organization
in Astrophysical and Laboratory Plasmas and the NSF grant AST 0098597.
\end{acknowledgments}

\appendix

\section{Radiative transitions}

For spontaneous emission from a hyperfine state $JIF'M'$ to another hyperfine state $J'IF,M$, the transition probability per unit time is
\be
a=\frac{64\pi^4e^2a_0^2\nu^3}{3hc^3}\sum_q |<JIFM| V_{q} |J'F'M'>|^2
\label{smalla}
\ee
where $V_q^{i,o}={\bf r\cdot e}_q^{i,o}$ is the projection of dipole moment 
along basis vector ${\bf e}_q$ of radiation, ${\bf e_{\pm}}= (\mp{\bf \hat x}-{\bf \hat y})/\sqrt{2}$, 
${\bf e}_0={\bf \hat z}$. 

For hyperfine lines, in the case of weak interaction in which 
neighboring fine levels do not interact, the electrical dipole matrix element for transition from a hyperfine sublevel $F',M'_F$ of upper level $J'$ to $F,M_F$ of lower level $J$ is given by
\begin{eqnarray}
&R^q_{FF'}&=<IJFM_F|V_q|IJ'F'M'_F>=(-1)^{F'+M_F-1}\left(\begin{array}{ccc}F &1& F'\\-M_F &q& M'_F\end{array}\right)<JIF||V_q||J'I'F'>\nonumber\\
&=&(-1)^{F'+M_F-1}\sqrt{[F,F']}\left(\begin{array}{ccc}F &1& F'\\-M_F &q& M'_F\end{array}\right)\left\{\begin{array}{ccc}
J_l &I &F\\F'& 1& J'\end{array}\right\}<J||V_q||J'>,
\label{hypfine}
\end{eqnarray}
where $M_F$ is quantum number corresponding to 
the projection of the total angular momentum $F$, $I$ corresponds to the nuclei spin. The matrix with big "\{ \}" represents 6-j or 9-j symbol, depending on the size of the matrix. When more than two angular momentum vectors are coupled, there is more than one way to add them up and form the same resultant. 6-j (or 9-j) symbol appears in this case as a recoupling coefficient describing transformations between different coupling schemes.

\section{Irreducible density matrix}
\label{density}
We adopt the irreducible tensorial formalism for performing the calculations (see also Paper I). The relation between irreducible tensor and the standard density matrix of atoms is 
\be
\rho^K_Q(F,F')=\sum_{MM'}(-1)^{F-M}(2K+1)^{1/2}\left(\begin{array}{ccc}
 F & K & F'\\
-M & Q & M'\end{array}\right)<FM|\rho|F'M'>.
\label{irreducerho}
\ee
For photons, their generic expression of irreducible spherical tensor is:
\be
J^K_Q=\sum_{qq'}(-1)^{1+q}[3(2K+1)]^{1/2}\left(\begin{array}{ccc}
 1 & 1 & K\\
q  & -q' & -Q\end{array}\right)J_{qq'},
\label{irreduce}
\ee

\section{Effect of collisions}
\label{collision}
Collisions can cause transitions among the hyperfine sublevels and reduce the ground state alignment. In the regime where collisions are not negligible, the equilibrium equation for the ground state Eqs.(\ref{lowevolution}) should be modified to include the collisions.
For the ground hyperfine level,
\bea
\dot{\rho^k_0}(F^0_l)&=& \sum_{F'_l|_{E'>E}}p_k[J_l]C(F'_l\rightarrow F^0_l)\rho^k_0(F'_l)+\sum_{J_u,F_u,F'_u}p_k[J_u]A(J_u\rightarrow J_l)\rho^k_q(F_u, F'_u) \nonumber\\
&-&\left[\sum_{F'_l|_{E<E'}}\delta_{kk'}C(F^0_l\rightarrow F'_l)+D^k+\sum_{J_uF_uk'}(\delta_{kk'}B_{lu}\bar{J}^0_0+s_{kk'}B_{lu}\bar{J}^2_0 )\right]\rho^{k'}_{0}(F^0_l),
\eea

For other hyperfine levels on the ground state,
\bea
\dot{\rho^k_0}(F_l)&=&  -\sum_{F'_l|E>E'}C(F_l\rightarrow F'_l)\rho^k_0(F_l)+\sum_{F'_l|_{E'>E}}p_k[J_l]C(F'_l\rightarrow F_l)\rho^k_0(F'_l)+\sum_{J_u,F_u,F'_u}p_k[J_u]A(J_u\rightarrow J_l)\rho^k_0(F_u, F'_u) \nonumber\\
&-&\left[\sum_{F'_l|_{E<E'}}\delta_{kk'}C(F_l\rightarrow F'_l)+D^k+\sum_{J_uF_uk'}(\delta_{kk'}B_{lu}\bar{J}^0_0+s_{kk'}B_{lu}\bar{J}^2_0)\right]\rho^{k'}_{0}(F_l)+\sum_{F'_l|_{E'<E}}[J_l]p_{k'}C(F'_l\rightarrow F_l)\rho^{k}_0(F'_l),
\eea
where C is the collisional transition rate among the hyperfine levels on the ground state, $D^k$ is the depolarizing rate due to elastic collisions (see Landi Degl'Innocenti \& Landolfi 2004). E and E' are the energies of hyperfine levels $F_l$ and $F'_l$ respectively.

\section{From observational frame to magnetic frame}
\label{euler}
In real observations, the line of sight is fixed, and the direction of the magnetic field is unknown. Thus a transformation is needed from the observational frame to the theoretical frame where the magnetic field is the quantization axis. This can be done by two Euler rotations as illustrated in Fig.\ref{radiageometry}.  In the original observational coordinate (xyz) system, the direction of radiation is defined as the z axis, and the direction of magnetic field is characterized by polar angles $\theta_B$ and $\phi_B$. First we rotate the whole system by an angle $\phi_B$ about the z-axis, so as to form the second coordinate system $x'y'z'$. The second rotation is from the $z(z')$ axis to the $z''$ axis by an angle  $\theta_B$ about the $y'$-axis.  Mathematically, the two rotations can be fulfilled by multiplying rotation matrices,
\bea
\left[\begin{array}{ccc}
\cos\theta_B&0& -\sin\theta_B \\ 0 &1&0\\ \sin\theta_B &0& \cos\theta_B \end{array}\right]
\left[\begin{array}{ccc}
\cos\phi_B & \sin\phi_B & 0 \\-\sin\phi_B&\cos\phi_B & 0\\0 & 0&1 \end{array}\right]=
\left[\begin{array}{ccc}
\cos\theta_B \cos\phi_B& \cos\theta_B \sin\phi_B & -\sin\theta_B\\
	-\sin\phi_B &		\cos\phi_B			& 0\\
	\sin\theta_B \cos\phi_B& \sin\theta_B \sin\phi_B&	\cos\theta_B \end{array}\right].
\eea

\end{document}